\documentclass[aps,prd,nofootinbib]{revtex4-2}
\usepackage{fontspec}
\usepackage{amsmath}
\usepackage{amsfonts}
\usepackage{amssymb}
\usepackage{makeidx}
\usepackage{graphicx,epsfig}
\usepackage{color}
\usepackage{subfigure}
\usepackage{ulem}
\usepackage{multirow}
\usepackage{epstopdf}
\usepackage{dcolumn}
\usepackage{comment}
\usepackage{caption}
\usepackage{subcaption}
\usepackage{epstopdf}
\usepackage{dcolumn}
\usepackage{slashed}
\usepackage{makecell}
\usepackage{url}
\usepackage{bm}
\usepackage[section]{placeins}
\usepackage[colorlinks=true, citecolor=blue, urlcolor = blue, linkcolor= red, bookmarks=true]{hyperref}
\newcommand{\beq}{\begin{equation}}
	\newcommand{\eeq}{\end{equation}}
\newcommand{\bea}{\begin{eqnarray}}
	\newcommand{\eea}{\end{eqnarray}}

\begin{document}
	
	\title {Relativistic corrections to   heavy quarkonium $+\gamma$ production at super Z factory}
	
	\author{ Xiao-Peng Wang $^{(a)}$}
	\author{Guang-Zhi Xu $^{(a)}$}
    \author{ Yi-Jie Li $^{(a)}$}
	\email{ liyijie@lnu.edu.cn}
	\author{Kui-Yong Liu $^{(b,a)}$}
	\email{liukuiyong@lnu.edu.cn}
	\affiliation{ {\footnotesize (a)~School of Physics, Liaoning University, Shenyang 110036, China}\\
		{\footnotesize (b)~School of Physics and Electronic Technology, Liaoning Normal University, Dalian 116029, China}}

	\date{\today}

\begin{abstract} 
In the framework of nonrelativistic quantum chromodynamics, we study the $\mathcal{O}(v^2)$ relativistic corrections to heavy‑quarkonium production associated with a photon at future electron‑positron colliders operated in the Z‑factory mode. S‑, P‑, and D‑wave heavy quarkonia as well as the mixed state $\psi(3770)$ are considered. We comprehensively investigate tree‑level $s$-channel processes for all states, together with $t$-channel contributions that apply only to ${}^3L_1$ states. It is found that the relativistic corrections substantially reduce the LO cross sections for charmonium states by roughly $20\%$–$70\%$, while the reduction is milder for bottomonium, at the level of $10\%$–$30\%$. Event yields for all studied channels are calculated, and sizable event yields are expected for $\eta_c$, $J/\psi$, $\psi(2S)$, $\Upsilon$, and $\psi(3770)$, especially at FCC‑ee. 
\end{abstract}

	\maketitle
	
	\section{Introduction} \label{introduce}
	Heavy quarkonium, as a bound state consisting of a heavy quark and its antiquark, provides a unique and fertile ground for studying the strong interaction. These systems involve distinct energy scales: the heavy quark mass $m_Q$, which governs the perturbative creation of the heavy quark pair; the heavy quark momentum $m_Qv$ ($v$ is the typical heavy quark velocity in the quarkonium rest frame); and the typical heavy quark kinetic energy $m_Qv^2$, which characterizes the nonperturbative hadronization of the $Q\bar{Q}$ into physical quarkonium. This separation of scales makes quarkonium an ideal laboratory to probe the interplay between perturbative and non-perturbative aspects of quantum chromodynamics (QCD). The nonrelativistic QCD (NRQCD) effective field theory~\cite{Bodwin:1994jh} provides the modern theoretical framework for treating quarkonium production and decay in a rigorous and systematic manner. When the NRQCD framework is confronted with experimental data at the B factory~\cite{Belle:2002tfa,Belle:2004abn,BaBar:2005nic}, it is found that, in addition to the QCD corrections~\cite{Zhang:2005cha,Gong:2007db}, the relativistic effects are also essential~\cite{He:2007te,Bodwin:2007ga}. 
	Moreover, the relativistic corrections are found to be significant for heavy quarkonium production at  Z factory\cite{Trunin:2024ibx,Wang:2025sbx,Wang:2026gso}.

	Exclusive heavy‑quarkonium production in association with a photon at $e^+e^-$ colliders is a topic of current interest.
	On the experimental side, compared with hadron colliders, the well‑controlled beam energy and polarization of the incoming beams, together with a cleaner hadronic background, give $e^+e^-$ colliders unique advantages.
	On the theoretical side, in contrast to double‑heavy‑quarkonium production in $e^+e^-$ annihilation, one can probe more subtle properties of heavy quarkonia for processes with only one heavy quarkonium in the final state. It has also been observed that the single‑heavy‑quarkonium production cross‑section exceeds that of the double‑heavy‑quarkonium process by one order of magnitude at B‑factories and by several orders of magnitude at the Z‑pole factory  \cite{Chen:2013mjb,Sun:2014kva,Liao:2021ifc,Wang:2025sbx}.
	Specifically, the exclusive production of charmonium associated with a photon has been carried out  at BESIII \cite{BESIII:2014uzr,BESIII:2017nty,BESIII:2021yal} and  B factory \cite{BaBar:2003waz,Belle:2015hcs,Belle:2018jqa}. 
	Theoretical studies including the leading-order (LO) cross section~\cite{Chung:2008km}, next-to-leading-order (NLO) QCD corrections~\cite{Li:2009ki,Sang:2009jc,Chao:2013cca}, and next-next-to-leading-order (NNLO) QCD corrections~\cite{Chen:2017pyi,Yu:2020tri,Sang:2020fql,Li:2025pbt} have been investigated for the process $e^+e^-\rightarrow \eta_c(\chi_{c0,1,2})+\gamma$ within the NRQCD framework.
	The relativistic corrections to these processes were also explored \cite{Sang:2009jc,Fan:2012dy,Li:2013nna,Xu:2014zra,Brambilla:2017kgw}. 
	Alternative theoretical frameworks have also been adopted to investigate these processes. The quasi-real electron method has been applied to explore the initial-state radiation (ISR) production of $J/\psi(\Upsilon)$ \cite{Benayoun:1999hm}. The light cone formalism has been utilized to analyze the production mechanism of the $e^+e^-\rightarrow \eta_c(\chi_{c0,1,2})+\gamma$ process \cite{Shifman:1980dk,Jia:2008ep,Braguta:2010mf}. Furthermore, the Bethe-Salpeter equation approach has been employed to study both the $e^+e^-\rightarrow \eta_c(\chi_{c0,1,2})+\gamma$ production \cite{Bhatnagar:2022wgg,Guleria:2024rac,Bhatnagar:2024ghr} and the ISR production of $J/\psi(\Upsilon)$ \cite{Bhatnagar:2025keo}.
	
	Besides the existing  $e^+e^-$ colliders, the future $e^+e^-$ colliders, such as CEPC \cite{CEPCStudyGroup:2018ghi}, FCC \cite{Agapov:2022bhm}, and ILC \cite{ILC:2013jhg}, would also provide good platforms to study heavy quarkonium processes. Additionally, based on their high energies and unprecedented integrated luminosities $\mathcal{L}\sim10^{34-36}cm^{-2}s^{-1}$ in Z factory mode, a super Z factory was proposed \cite{Ma.JP}. At the Z factory energy, Chang et al. \cite{Chang:2010am} investigated the  $e^+e^-\rightarrow H+\gamma$  processes using the Bethe-Salpeter method and emphasized the ISR effect. Refs. \cite{Chen:2013mjb,Chen:2013itc,Sun:2014kva} calculated $e^+e^-\rightarrow H+\gamma$ production up to NLO in $\alpha_s$. Liao et al. evaluated the excited heavy quarkonium production at LO in $\alpha_s$ and $v^2$\cite{Liao:2021ifc}. The $\eta_Q+\gamma$ production at NLO in $\alpha_s$ and resummation of large logarithms $\ln\frac{s}{m^2}$ at next-to-leading-logarithmic accuracy were performed in Ref. \cite{Chung:2019ota}. Recently, Trunin \cite{Trunin:2024ibx} has calculated the resummation of relativistic corrections to heavy quarkonium$+\gamma$ production. In this work, we will further study the relativistic corrections to $e^+e^-\rightarrow H+\gamma$  production at the super Z factory.
	
	Beyond conventional quarkonium physics, high-precision $e^+e^-$ collider measurements also enable the exploration of exotic quarkonium-like hadrons (the ``XYZ'' states).
	In the last two decades, a great number of XYZ states have been discovered \cite{Guo:2017jvc,Olsen:2017bmm,Brambilla:2019esw,Chen:2022asf}, although many of them do not match the predictions based on conventional hadron model\cite{Godfrey:1985xj,Barnes:2005pb}. 
	In this context, excited conventional quarkonium spectra offer a useful benchmark for exotic‑hadron studies, including D‑wave states. 
	For the lowest lying (n=1) D-wave heavy quarkonia, the 
	$\psi_1(1^3D_1)$ is known as 
	$\psi(3770)$\cite{Rapidis:1977cv}. The X(3823), observed at Belle \cite{Belle:2013ewt} and BESIII \cite{BESIII:2015iqd}, was identified as 
	$\psi_2(1^3D_2)$ \cite{Deng:2016stx,Li:2022qhg,Li:2023zig}. The  observed 
	X(3842) \cite{LHCb:2019lnr} can be a good candidate for 
	$\psi_3(1^3D_3)$\cite{Yu:2019kcr,Li:2023cpl}. The Belle Collaboration also searched for 
	$\eta_{c2}(1^1D_2)$ in B decays \cite{Belle:2020esr} and 
	$e^+e^-\rightarrow\eta_{c2}+\gamma$
	\cite{Belle:2021cjk}, but no significant signal was found. For D-wave bottomonium, only the 
	$^3D_2$
	has been seen in radiative decay from 
	$\Upsilon(3S)$ \cite{CLEO:2004npj,BaBar:2010tqb}. The assignments of the conventional D-wave heavy quarkonia to the measured resonance states are important for the verification of potential models and QCD. The higher-order QCD corrections to D-wave heavy quarkonium related processes have been studied within the NRQCD framework \cite{He:2008xb,Fan:2009cj,Fan:2011aa,Chen:2015zta,Dong:2013qw,Yang:2020pyh}, and the relativistic effects were considered for them \cite{Wang:2013nya,Wang:2016hbl,Li:2016cou,Li:2022qhg,Du:2024gna} using the Bethe-Salpeter equation, which implies that currently the studies of higher-order corrections to processes involving D-wave heavy quarkonia are relatively few. Therefore, we will also take the D-wave  quarkonia into consideration.
	
	Motivated by the above considerations, we present a comprehensive study of NLO relativistic corrections to heavy‑quarkonium production via $e^+e^-\rightarrow H(Q\bar Q)+\gamma$ at the future super Z‑factory within the NRQCD framework. Our calculation covers S‑wave, P‑wave, and D‑wave heavy quarkonium states and includes both s‑channel and t‑channel tree‑level diagrams. 
	The detailed processes include: $e^+e^-\rightarrow  \eta_c+\gamma, e^+e^-\rightarrow  J/\psi+\gamma,  e^+e^-\rightarrow  h_c+\gamma, e^+e^-\rightarrow  \chi_{c0,1,2}+\gamma, e^+e^-\rightarrow  \eta_{c2}+\gamma, e^+e^-\rightarrow \psi_{1,2,3}+\gamma, e^+e^-\rightarrow  \eta_b+\gamma, e^+e^-\rightarrow \Upsilon+\gamma,  e^+e^-\rightarrow  h_b+\gamma, e^+e^-\rightarrow  \chi_{b0,1,2}+\gamma, e^+e^-\rightarrow  \eta_{b2}+\gamma, e^+e^-\rightarrow \Upsilon_{1,2,3}+\gamma$.

	The remainder of this paper is structured as follows. Sec. \ref{solution} describes the theoretical framework and the computational method. The input parameters and the LDMEs used in our calculations are detailed in Sec. \ref{input}. A comprehensive analysis of both the s‑channel and t‑channel contributions is then presented in Sec. \ref{res}, where we compute the LO and NLO($v^2$) relativistic correction (differential) cross sections, and also discuss event generation and the associated uncertainties. Finally, Sec. \ref{sumary} provides a brief conclusion.

	\section{Formula and Method} \label{solution}

	\subsection{NRQCD Factorization and Relativistic Corrections}
	Within the NRQCD effective‑field theory\cite{Bodwin:1994jh},  the production and decay of heavy quarkonium are factorized into  the short-distance coefficients (SDCs) and long-distance matrix elements (LDMEs), the former can be calculated perturbatively, and the latter are obtained by lattice calculations or phenomenological models, or determined by fitting experimental data. The LDMEs are related to the four-fermion operators, characterized by the velocity $v$ of the heavy quark in the meson rest frame.
	The  production cross sections of heavy quarkonium in process $e^+e^-\rightarrow H +\gamma$   can be factorized as\cite{Bodwin:1994jh,Ko:1996xw}
	\bea
	\sigma(e^+e^-\rightarrow H+\gamma)=\sum_{n}[\frac{F(n)}{m_Q^{d_{\mathcal{O}(n)}-4}}\langle\mathcal{O}^H(n)\rangle+ \frac{G(n)}{m_Q^{d_{\mathcal{P}(n)}-4}}\langle\mathcal{P}^H(n)\rangle ]
	\eea
	Here $n$ denotes the quantum‑number configuration ${}^{2S+1}L_J^{[1,8]}$ of the intermediate $Q\bar{Q}$ state, whose quantum numbers can differ from those of the physical quarkonium state $H$.
	$\mathcal{O}^H(n)$ and $\mathcal{P}^H(n)$ are the LO and $\mathcal{O}(v^2)$ four‑fermion operators pertaining to the transition $Q\bar{Q}(n)\rightarrow H$, with dimensions $d_{\mathcal{O}(n)}$ and $d_{\mathcal{P}(n)}$, respectively. $d_{\mathcal{O}(n)}=6,8,10$ and $d_{\mathcal{P}(n)}=8,10,12$ correspond to the $S$‑, $P$‑, and $D$‑wave $Q\bar{Q}$ intermediate states, respectively.
	$F(n)$ and $G(n)$ are the corresponding SDCs.

	The relevant four-quark operators for the production of $H=\eta_Q$, $J/\psi$($\Upsilon$), $h_Q$,...,$\psi_1/\Upsilon_1$, $\psi_2/\Upsilon_2$, $\psi_3/\Upsilon_3$ are defined as\cite{Bodwin:1994jh,Ko:1996xw,He:2009bf}	
	{\allowdisplaybreaks
		\begin{align}
			\mathcal{O}^{\eta_Q}(^1S_0^{[1]})&=\chi^{\dagger}\psi(a^{\dagger}_{\eta_Q}a_{\eta_Q})\psi^{\dagger}\chi,\\
			\mathcal{P}^{\eta_Q}(^1S_0^{[1]})&=\frac{1}{2}\big[\chi^{\dagger}\psi(a^{\dagger}_{\eta_Q}a_{\eta_Q})\psi^{\dagger}\big(-\frac{i}{2}\overleftrightarrow{\boldsymbol{D}}\big)^2\chi+\text{h.c.}\big],\\
			\mathcal{O}^{J/\psi/\Upsilon}(^3S_1^{[1]})&=\chi^{\dagger}\sigma^i\psi(a^{\dagger}_{J/\psi}a_{J/\psi})\psi^{\dagger}\sigma^i\chi,\\
			\mathcal{P}^{J/\psi/\Upsilon}(^3S_1^{[1]})&=\frac{1}{2}\big[\chi^{\dagger}\sigma^i\psi(a^{\dagger}_{J/\psi}a_{J/\psi})\psi^{\dagger}\sigma^i\big(-\frac{i}{2}\overleftrightarrow{\boldsymbol{D}}\big)^2\chi+\text{h.c.}\big],\\
			\mathcal{O}^{h_Q}(^1P_1^{[1]})&=\chi^{\dagger}\big(-\frac{i}{2}\overleftrightarrow{D}^i\big)\psi(a^{\dagger}_{h_Q}a_{h_Q})\psi^{\dagger}\big(-\frac{i}{2}\overleftrightarrow{D}^i\big)\chi,\\
			\mathcal{P}^{h_Q}(^1P_1^{[1]})&=\frac{1}{2}\big[\chi^{\dagger}\big(-\frac{i}{2}\overleftrightarrow{D}^i\big)\big(\frac{i}{2}\overleftrightarrow{\boldsymbol{D}}\big)^2\psi(a^{\dagger}_{h_Q}a_{h_Q})\psi^{\dagger}\big(-\frac{i}{2}\overleftrightarrow{D}^i\big)\chi+\text{h.c.}\big],\\
			&\ldots\nonumber\\
			\mathcal{O}^{\psi_1/\Upsilon_1}(^3D_1^{[1]})&=\frac{3}{5}\chi^{\dagger}K^{i} \psi(a^{\dagger}_{\psi_1/\Upsilon_1}a_{\psi_1/\Upsilon_1})\psi^{\dagger}K^{i}\chi,\\
			\mathcal{P}^{\psi_1/\Upsilon_1}(^3D_1^{[1]})&=\frac{3}{10}\big[\chi^{\dagger}K^{i}\big(\frac{i}{2}\overleftrightarrow{\boldsymbol{D}}\big)^2 \psi(a^{\dagger}_{\psi_1/\Upsilon_1}a_{\psi_1/\Upsilon_1})\psi^{\dagger}K^{i}\chi+\text{h.c.}\big],\\
			\mathcal{O}^{\psi_2/\Upsilon_2}(^3D_2^{[1]})&=\frac{1}{6}\chi^{\dagger}K^{ij} \psi(a^{\dagger}_{\psi_2/\Upsilon_2}a_{\psi_2/\Upsilon_2})\psi^{\dagger}K^{ij}\chi,\\
			\mathcal{P}^{\psi_2/\Upsilon_2}(^3D_2^{[1]})&=\frac{1}{12}\big[\chi^{\dagger}K^{ij}\big(\frac{i}{2}\overleftrightarrow{\boldsymbol{D}}\big)^2 \psi(a^{\dagger}_{\psi_2/\Upsilon_2}a_{\psi_2/\Upsilon_2})\psi^{\dagger}K^{ij}\chi+\text{h.c.}\big],\\
			\mathcal{O}^{\psi_3/\Upsilon_3}(^3D_3^{[1]})&=\frac{1}{9}\chi^{\dagger}K^{ijk} \psi(a^{\dagger}_{\psi_3/\Upsilon_3}a_{\psi_3/\Upsilon_3})\psi^{\dagger}K^{ijk}\chi,\\
			\mathcal{P}^{\psi_3/\Upsilon_3}(^3D_3^{[1]})&=\frac{1}{18}\big[\chi^{\dagger}K^{ijk}\big(\frac{i}{2}\overleftrightarrow{\boldsymbol{D}}\big)^2 \psi(a^{\dagger}_{\psi_3/\Upsilon_3}a_{\psi_3/\Upsilon_3})\psi^{\dagger}K^{ijk}\chi+\text{h.c.}\big],
		\end{align}
	}
	where $\chi$ and $\psi$ are the Pauli spinors describing antiquark creation and quark annihilation, respectively. $\sigma$ is the Pauli matrices and $\boldsymbol{D}$ is the gauge-covariant derivative with $\overleftrightarrow{\boldsymbol{D}}=\overrightarrow{\boldsymbol{D}}-\overleftarrow{\boldsymbol{D}}$. $\overleftrightarrow{D}^{(i}\sigma^{j)}$ is used as the notation for the symmetric traceless component of a tensor:
	$\overleftrightarrow{D}^{(i}\sigma^{j)}=(\overleftrightarrow{D}^{(i}\sigma^{j)}+\overleftrightarrow{D}^{(j}\sigma^{i)})/2-\overleftrightarrow{D}^k\sigma^k\delta^{ij}/3$. The notations $\boldsymbol{K}$ are $ {K}^i=\sigma^jS^{ij}$, $K^{ij}=\epsilon^{ikl}\sigma^lS^{jk}+\epsilon^{jkl}\sigma^lS^{ik}$, $K^{ijk}=\sigma^iS^{jk}+\sigma^jS^{ki}+\sigma^kS^{ij}-\frac{2}{5}\sigma^l(\delta^{jk}S^{il}+\delta^{ki}S^{jl}+\delta^{ij}S^{kl})$, with $S^{ij}=(-\frac{i}{2})^2(\overleftrightarrow{D}^i\overleftrightarrow{D}^j-\frac{1}{3}\overleftrightarrow{D}^2\delta^{ij})$.
	
	The SDCs $F(n)$ and $G(n)$ can be calculated perturbatively through the matching conditions for short-distance $Q\bar{Q}$ production between perturbative QCD and NRQCD.
	\bea
	\sigma[e^+e^-\rightarrow Q\bar{Q}(n)+\gamma]|_{\text{pert QCD}}= [\frac{F(n)}{m_Q^{d_{\mathcal{O}(n)}-4}}\langle0|\mathcal{O}_n^{Q\bar{Q}}(n)|0\rangle+\frac{G(n)}{m_Q^{d_\mathcal{P}(n)-4}}\langle0|\mathcal{P}_n^{Q\bar{Q}}(n)|0\rangle]_{\text{pert NRQCD}}		
	\label{matching}
	\eea	
	
	In the perturbative QCD calculation, the four-momenta of quark and antiquark inside the $Q\bar{Q}$ pair in an arbitrary frame can be related to those in the meson rest frame via an appropriate Lorentz boost matrix $L$ as,
	\bea
	p_Q&=&\frac{1}{2}p_1+q=L(\frac{1}{2}p_r+q_r),\cr
	p_{\bar{Q}}&=&\frac{1}{2}p_1-q=L(\frac{1}{2}p_r-q_r)
	\eea  
	where $p_1$ and $q$ are the total momentum of the heavy quarkonium and the relative momentum between the constituent quark and antiquark, respectively, satisfying $p_1^2=4E_q^2$ and $p_1\cdot q=0$. Their rest‑frame expressions read $p_r^{\mu}=(2E_q,0), q_r^{\mu}=(0,\boldsymbol{q})$, with $E_q=\sqrt{m_Q^2+\boldsymbol{q}^2}$.
	$|\boldsymbol{q}|=m_Qv$, where $v$ is relative velocity of quarks in heavy quarkonium rest frame.
	
	The full scattering amplitude can be expressed as
	\bea
	\mathcal{M}(e^+e^-\rightarrow Q\bar{Q}(n)+\gamma)=\sqrt{\frac{m_Q}{E_q}}A(E_q,q)
	\eea 
	where the factor $\sqrt{m_Q/E_q}$ compensates the relativistic normalization of the $Q\bar{Q}$
	pair and 
	\bea
	A(E_q,q)=\sum_{L_{z}S_{z}}\sum_{s_{Q}s_{\bar{Q}}}\sum_{kl}\langle s_{Q};s_{\bar{Q}}|S,S_{z}\rangle\langle LL_{z};SS_{z}|JJ_{z}\rangle\langle3,k;\bar{3},l|1\rangle\mathcal{A}\left(e^+e^-\rightarrow Q_{k}(p_{Q};s_{Q})+\bar{Q}_{l}(p_{\bar{Q}};s_{\bar{Q}})+\gamma(p_2)\right)
	\label{x1}
	\eea
	with $\mathcal{A}\left(e^+e^-\to Q_{k}(p_{Q};s_Q)+\bar{Q}_{l}(p_{\bar{Q}};s_{\bar{Q}})+\gamma(p_2)\right)$ denoting the amplitude for the process producing a $Q\bar{Q}$ pair together with a photon, and $\langle 3,k;\bar{3},l|1\rangle=\delta_{kl}/\sqrt{N_c}$ for projecting the $Q\bar{Q}$ pair onto the color‑singlet state. 
	The Dirac spinor product $v(p_{\bar{Q}})\bar{u}(p_Q)$ is projected onto the specific ${}^{2S+1}L_J$ quarkonium state via the standard spinor projection method. 
	The Lorentz-covariant projection operators for spin-singlet and spin-triplet states in an arbitrary frame are given respectively by~\cite{Bodwin:2002cfe}.
	\bea
	\mathbb{P}(00)=	\sum_{s_Qs_{\bar{Q}}}v(p_{\bar{Q}})\bar{u}(p_Q)\langle s_Q,s_{\bar{Q}}|0,0\rangle&=&
	\frac{1}{\sqrt{2}(E_q+m_Q)}(-\slashed{p}_{\bar{Q}}+m_Q)\gamma_5\frac{\slashed{p}_1+2E_q}{4E_q}(\slashed{p}_Q+m_Q),\\
	\mathbb{P}(1S_z)=\sum_{s_Q s_{\bar{Q}}}v(p_{\bar{Q}})\bar{u}(p_Q)\langle s_Q,s_{\bar{Q}}|1,S_z\rangle&=&
	\frac{1}{\sqrt{2}(E_q+m_Q)}(-\slashed{p}_{\bar{Q}}+m_Q)\slashed{\epsilon}\frac{\slashed{p}_1+2E_q}{4E_q}(\slashed{p}_Q+m_Q).
	\label{Proj1}
	\eea 
	where $\epsilon^\mu$ is the polarization vector of the spin triplet state and the Dirac spinors are normalized as $\bar{u}u=-\bar{v}v=2m_Q$.

	To expand Eq. (\ref{x1}) as a single series in $q$, it is convenient to define
	\bea
	A_{\alpha_{1}\alpha_{2}\cdots\alpha_{m}}\equiv\frac{\partial^m A(E_q,q)}{\partial q^{\alpha_1}\partial q^{\alpha_2}\cdots\partial q^{\alpha_m}}\Big|_{q=0},\quad A_0\equiv A(E_q,q)\Big|_{q=0},
	\eea
	so that 
	\bea
	A(E_q,q)=A_0+q^{\alpha_1}A_{\alpha_1}+\frac{1}{2}q^{\alpha_1}q^{\alpha_2}A_{\alpha_1\alpha_2}+\cdots
	\eea
	For S-wave and D-wave production, only even powers of $q_i$ contribute, and for P-wave only odd powers of $q_i$ contribute. To extract the high-order corrections in $v^2$, we need to further decompose the high-rank tensor products into the irreducible representations and retain the nonzero terms. And we have the following replacements 
	\begin{equation}
		\label{eq:qavg}
		\begin{aligned}
			\left.
			\begin{aligned}
				q^{\alpha_1}q^{\alpha_2}
				&\rightarrow \frac{|\boldsymbol{q}|^2}{3}\Pi^{\alpha_1\alpha_2},
			\end{aligned}
			\right.
			&& \text{S-wave},\\[8pt]
			\left.
			\begin{aligned}
				q^{\alpha_1}
				&\rightarrow |\boldsymbol{q}|\epsilon_{L_z}^{\alpha_1}, \\
				q^{\alpha_1}q^{\alpha_2}q^{\alpha_3}
				&\rightarrow \frac{|\boldsymbol{q}|^3}{5}\Pi^{\alpha_1\alpha_2\alpha_3},
			\end{aligned}
			\right\}
			&& \text{P-wave},\\[8pt]
			\left.
			\begin{aligned}
				q^{\alpha_1}q^{\alpha_2}
				&\rightarrow |\boldsymbol{q}|^2\epsilon^{\alpha_1\alpha_2}_{L_z}, \\
				q^{\alpha_1}q^{\alpha_2}q^{\alpha_3}q^{\alpha_4}
				&\rightarrow \frac{|\boldsymbol{q}|^4}{7}\Pi^{\alpha_1\alpha_2\alpha_3\alpha_4},
			\end{aligned}
			\right\}
			&& \text{D-wave}.
		\end{aligned}
	\end{equation}
	where
	\bea 
	\Pi^{\alpha_1\alpha_2}&=&-g^{\alpha_1\alpha_2}+\frac{p_1^{\alpha_1}p_1^{\alpha_2}}{4E_q^2},\\
	\Pi^{\alpha_1\alpha_2\alpha_3}&=&\Pi^{\alpha_1\alpha_2}\epsilon_{L_z}^{\alpha_3}+\Pi^{\alpha_2\alpha_3}\epsilon_{L_z}^{\alpha_1}+\Pi^{\alpha_3\alpha_1}\epsilon_{L_z}^{\alpha_2},\\
	\Pi^{\alpha_1\alpha_2\alpha_3\alpha_4}&=&\Pi^{\alpha_1\alpha_2}\epsilon_{L_z}^{\alpha_3\alpha_4}+\Pi^{\alpha_2\alpha_3}\epsilon_{L_z}^{\alpha_4\alpha_1}+\Pi^{\alpha_3\alpha_4}\epsilon_{L_z}^{\alpha_1\alpha_2}+\Pi^{\alpha_1\alpha_3}\epsilon_{L_z}^{\alpha_2\alpha_4}+\Pi^{\alpha_1\alpha_4}\epsilon_{L_z}^{\alpha_2\alpha_3}+\Pi^{\alpha_2\alpha_4}\epsilon_{L_z}^{\alpha_3\alpha_1}
	\eea
	Up to $\text{NLO}(v^2)$
	we thus have 
	\bea
	\mathcal{M}(e^++e^-\rightarrow Q\bar{Q}(n)+\gamma)&=&
	\begin{cases}
		\sqrt{\frac{m_Q}{E_q}}[A_0+\frac{|\boldsymbol{q}|^2}{6}\Pi^{\alpha_1\alpha_2}A_{\alpha_1\alpha_2}],\text{for S wave}\\
		\sqrt{\frac{m_Q}{E_q}}[|\boldsymbol{q}|\epsilon_{L_z}^{\alpha_1}A_{\alpha_1}+\frac{|\boldsymbol{q}|^3}{30}\Pi^{\alpha_1\alpha_2\alpha_3}A_{\alpha_1\alpha_2\alpha_3}],\text{for P wave}\\
		\sqrt{\frac{m_Q}{E_q}}[\frac{1}{2}|\boldsymbol{q}|^2\epsilon_{L_z}^{\alpha_1\alpha_2}A_{\alpha_1\alpha_2}+\frac{|\boldsymbol{q}|^4}{168}\Pi^{\alpha_1\alpha_2\alpha_3\alpha_4}A_{\alpha_1\alpha_2\alpha_3\alpha_4} ],\text{for D wave}
	\end{cases}
	\label{x2}
	\eea
	In Eq.~(\ref{x2}), the second term within the square brackets for each state gives the explicit $\mathcal{O}(v^2)$ contribution. For the parts of the expression outside these $|\boldsymbol{q}|^n$ factors, $|\boldsymbol{q}|^2$ can be set to zero, i.e., $E_q=m_Q$. By contrast, the first term contains both LO and NLO contributions. Its implicit $|\boldsymbol{q}|^2$ dependence is obtained by expanding $E_q$.
	
	As for the kinematics, 
	the Mandelstam variables can be written as: 
	\bea
	s=(k_1+k_2)^2=(p_1+p_2)^2, 
	t=(p_1-k_1)^2=(p_2-k_2)^2, 
	u=(p_2-k_1)^2=(p_1-k_2)^2 
	\label{mandel}
	\eea
	and they satisfy the  relation 
	\bea
    s+t+u= 4E_q^2
	\label{os}
	\eea
	Notice that $s$ is independent of $\boldsymbol{q}^2$. 
	In the nonrelativistic limit  $\boldsymbol{q}^2\rightarrow 0$, we have $t\rightarrow t_0$, 
	$u\rightarrow u_0$, and 
	$s+t_0+u_0=4m_Q^2$. 
    With  Eqs.~(\ref{mandel},\ref{os}), the Mandelstam variables $t$ and $u$ can be further written as\cite{Xu:2012am}
	
	\begin{equation}
		\begin{aligned}
			t
			&=-\frac{s-4E_q^2}{2}(1-\sqrt{1-\frac{4m_e^2}{s}}\cos\theta)
			=t_0\frac{s-4E_q^2}{s-4m_Q^2}\cr
			u
			&= -\frac{s-4E_q^2}{2}(1+\sqrt{1-\frac{4m_e^2}{s}}\cos\theta)
			=u_0\frac{s-4E_q^2}{s-4m_Q^2}
			\label{rcrep}
		\end{aligned}
	\end{equation}
	where $\theta$ is the angle between $k_1$ (electron) and $p_1$ (heavy quarkonium). 
The $t$- and $u$-channel contributions, for which the squared amplitude contains the $1/(tu)$ term, exhibit collinear singularities at $\theta\to0,\pi$ in the limit $m_e\to0$, analogous to the forward singularity in Møller scattering. We retain only the electron mass in the factor multiplying $\cos\theta$ within our expressions for $t$ and $u$ to regularize these collinear singularities, while all other electron-mass corrections are neglected.

	We are now in a position to present the master formulas for the SDCs, 
	\bea
	\frac{F(n)}{m_Q^{d_{\mathcal{O}(n)}-4}}&=&\frac{1}{8s}\int d\Phi_{20}\left[\sum|M|^2\right]_{E_q\to m_Q},\\
	\frac{G(n)}{m_Q^{d_{\mathcal{P}(n)}-4}}&=&\frac{1}{8s}\int d\Phi_{20}\left[\sum(f_{ps}*|M|^2+|N|^2)\right]_{E_q\to m_Q},
	\eea
	with 
	\bea
	|M|^2&=&
	\begin{cases}
		|A_0|^2,  &\text{for S-wave}\\
		\epsilon_{L_z}^{*\alpha_1'}\epsilon_{L_z}^{\alpha_1}A^*_{\alpha_1'}A_{\alpha_1},  &\text{for P-wave}\\
		\frac{1}{4}\epsilon_{L_z}^{*\alpha_1'\alpha_2'}\epsilon_{L_z}^{\alpha_1\alpha_2}A^*_{\alpha_1'\alpha_2'}A_{\alpha_1\alpha_2},  &\text{for D-wave}\\
        \frac{1}{2}\epsilon_{L_z}^{\alpha_1\alpha_2}A_0^*A_{\alpha_1\alpha_2},  &\text{for S-D interference}
	\end{cases}
    \label{Msqu}
	\eea
	and 
	\begin{eqnarray}
		|N|^2&=&
		\begin{cases}
			\displaystyle
			\frac{\partial}{\partial E_q}\frac{|A_0|^2}{2E_q}
			+\frac{1}{3}\Pi^{\alpha_1\alpha_2}\operatorname{Re}\bigl[A^*_0\,A_{\alpha_1\alpha_2}\bigr],
			&\hspace{-2em}\text{for S-wave},\\[10pt]
			\displaystyle
			\frac{\partial}{\partial E_q}\biggl[
			\frac{1}{2E_q}\epsilon_{L_z}^{*\alpha_1'}\epsilon_{L_z}^{\alpha_1}
			A^*_{\alpha_1'}A_{\alpha_1}
			\biggr]
			+\frac{1}{15}\epsilon_{L_z}^{*\alpha_1'}\,\Pi^{\alpha_1\alpha_2\alpha_3}\,
			\operatorname{Re}\bigl[A^*_{\alpha_1'}\,A_{\alpha_1\alpha_2\alpha_3}\bigr],
			&\hspace{-2em}\text{for P-wave},\\[12pt]
			\displaystyle
			\frac{\partial}{\partial E_q}\biggl[\frac{1}{8E_q}\epsilon_{L_z}^{*\alpha_1'\alpha_2'}\epsilon_{L_z}^{\alpha_1\alpha_2}
			A^*_{\alpha_1'\alpha_2'}A_{\alpha_1\alpha_2}
			\biggr]
			+\frac{1}{168}\epsilon_{L_z}^{*\alpha_1'\alpha_2'}\,\Pi^{\alpha_1\alpha_2\alpha_3\alpha_4}\,
			\operatorname{Re}\!\biggl[A^*_{\alpha_1'\alpha_2'}\,A_{\alpha_1\alpha_2\alpha_3\alpha_4}\biggr],
			&\hspace{-2em}\text{for D-wave},\\[12pt]
            \displaystyle
            \frac{\partial}{\partial E_q}\biggl[\frac{1}{4E_q}\epsilon_{L_z}^{\alpha_1\alpha_2}A_0^*A_{\alpha_1\alpha_2}\biggr]
            +\frac{1}{168}\Pi^{\alpha_1\alpha_2\alpha_3\alpha_4}\operatorname{Re}\biggl[A_0^*A_{\alpha_1\alpha_2\alpha_3\alpha_4}\biggr]
            +\frac{1}{12}\Pi^{\alpha_1'\alpha_2'}\epsilon_{L_z}^{\alpha_1\alpha_2}\operatorname{Re}\biggl[A_{\alpha_1'\alpha_2'}^*A_{\alpha_1\alpha_2}\biggr],\\
			&\hspace{-4em}\text{for S-D interference}.
		\end{cases}
    \label{Nsqu}
	\end{eqnarray}
	Here, $d\Phi_{20}$ denotes the $\mathcal{O}(v^0)$ two‑body phase‑space measure.
	The symbol $\sum$ stands for summation over all polarizations of the initial and final states. 
	The spin‑averaging factor for the unpolarized $e^+e^-$ initial state is incorporated in the overall factor $1/(8s)$.
	The factor $f_{ps}=r/[m_Q^2(r-1)]$\cite{Li:2013nna}, with $r=4m_Q^2/s$, originates from the relativistic correction to the phase‑space integral.
	To calculate the above squared amplitudes, we also need their corresponding polarization sum relations, which   can be found in Refs~\cite{ Ko:1996xw,He:2009bf}.

	\subsection{Diagrams and amplitudes}
	
	\begin{widetext}	
		\begin{figure*}
			\begin{tabular}{c c }
				\includegraphics[width=0.9\textwidth]{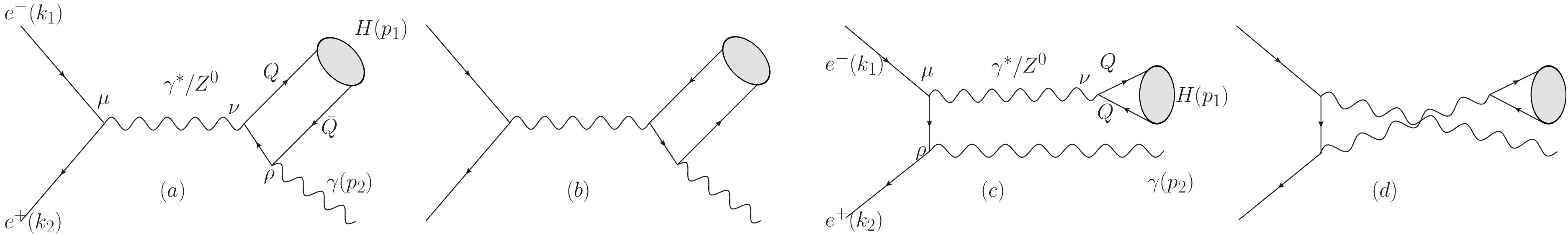}
			\end{tabular}
			\caption{ Feynman diagrams for $e^-(k_1)+e^+(k_2)\rightarrow H(p_1)+\gamma(p_2)$ at the tree level. 
				$H$ stands for heavy‑quarkonium states. Diagrams (a) and (b) correspond to $s$-channel processes, while diagrams (c) and (d) are the $t$-channel (ISR) processes.}
			\label{feynmandia}
		\end{figure*}
	\end{widetext}
	
	There are eight Feynman diagrams for $e^-(k_1)+e^+(k_2)\rightarrow H(Q\bar{Q})(p_1)+\gamma(p_2)$ at the tree level $(\alpha^3)$, which are shown in Fig. \ref{feynmandia}. Diagrams (a) and (b) belong to the s-channel, while diagrams (c) and (d) belong to the t-channel ISR processes. The amplitude defined in Eq.~\eqref{x1} for each individual Feynman diagram reads as 
	\bea
	A&=& \mathcal{L}_{rr'}^{\mu}\mathcal{D}_{\mu\nu}\mathcal{H}^{\nu}
	\eea
	
	The leptonic currents are 
	\begin{equation}\label{eq:lepcur}
		\begin{aligned}
			\mathcal{L}_{rr'(a,b)}^\mu&=\bar{v}_r(k_2)\Gamma_{ee}^{\mu}u_{r'}(k_1)\cr
			\mathcal{L}_{rr'(c)}^{\mu}&=\bar{v}_r(k_2)	\Gamma_{\gamma ee}^{\rho}\frac{\slashed{p}_2-\slashed{k}_2}{(k_2-p_2)^2}\Gamma_{ee}^\mu u_{r'}(k_1)  \cr
			\mathcal{L}_{rr'(d)}^{\mu}&=\bar{v}_r(k_2)\Gamma_{ee}^\mu\frac{\slashed{p}_1-\slashed{k}_2}{(k_2-p_1)^2} 	\Gamma_{\gamma ee}^{\rho}u_{r'}(k_1) 
		\end{aligned}
	\end{equation}
	where the indices $r$ and $r'$ are spin indices for the positron and electron spinors, respectively.
	The vertices $\Gamma^\mu_{ee}$ and $\Gamma^\rho_{\gamma ee}$ read
	\bea
	\Gamma_{ee}^{\mu}&=& 
	\begin{cases}
		-ie\gamma^\mu,  \text{for    } \gamma^* \text{ propagated}  \\
		\frac{-ie}{4\cos\theta_w\sin\theta_w}\gamma^{\mu}(1-4\sin^2\theta_w-\gamma^5), \text{for    } Z^0 \text{ propagated} 
	\end{cases}\\
	\Gamma_{\gamma ee}^{\rho}&=&-ie\gamma^{\rho}\epsilon^*_{\gamma\rho}(p_2)
	\eea
	Here $\epsilon_{\gamma}(p_2)$ is polarization vector of the outgoing photon.
	
	The propagators $\mathcal{D}_{\mu\nu}$ are
	\bea
	\mathcal{D}_{\mu\nu(a,b)}&=& 
	\begin{cases}
		\frac{-ig_{\mu\nu}}{l^2},  \text{for   } \gamma^* \text{ propagated}    \\
		\frac{i}{l^2-m_Z^2+im_{Z}\Gamma_Z} (-g_{\mu\nu}+\frac{l_{\mu}l_{\nu}}{l^2}),\text{for    } Z^0 \text{ propagated} 
	\end{cases}\\
	\mathcal{D}_{\mu\nu(c,d)}&=& 
	\begin{cases}
		\frac{-ig_{\mu\nu}}{p_1^2},  \text{for   } \gamma^* \text{ propagated}    \\
		\frac{i}{p^2_1-m_Z^2+im_{Z}\Gamma_Z} (-g_{\mu\nu}+\frac{p_{1\mu}p_{1\nu}}{p_1^2}),\text{for    } Z^0 \text{ propagated} 
	\end{cases}
	\eea  
	where $l=p_1+p_2$, with $l^2=s$.
	
	The hadronic parts are 
	\bea
	\mathcal{H}^{\nu}_{(a)}&=&i\sqrt{N_c}\text{Tr}[\mathbb{P}(p_1)\Gamma_{Q\bar{Q}}^{\nu}\frac{-\slashed{p}_{\bar{Q}}-\slashed{p}_2+m_Q}{(p_{\bar{Q}}+p_2)^2-m_Q^2}	\Gamma_{\gamma Q\bar{Q}}^{\rho}],\cr
	\mathcal{H}^{\nu}_{(b)}&=&i\sqrt{N_c}\text{Tr}[\mathbb{P}(p_1)	\Gamma_{\gamma Q\bar{Q}}^{\rho}\frac{\slashed{p}_{Q}+\slashed{p}_2+m_Q}{(-p_{Q}-p_2)^2-m_Q^2}\Gamma_{Q\bar{Q}}^{\nu}],\cr
	\mathcal{H}^{\nu}_{(c,d)}&=&i\sqrt{N_c}\text{Tr}[\mathbb{P}(p_1)\Gamma_{Q\bar{Q}}^{\nu}],
	\eea
	where $\sqrt{N_c}$ is the overall color factor for each diagram and the vertices are defined as 
	\bea
	\Gamma_{Q\bar{Q}}^\nu&=&
	\begin{cases}
		iee_Q\gamma^{\nu},      \text{for   } \gamma^* \text{ propagated}\\
		\frac{-ie}{4\cos\theta_w\sin\theta_w}\gamma^\nu(-1+4|e_Q|\sin^2\theta_w+\gamma^5),       \text{for Q=c,    } Z^0 \text{ propagated} 	\\
		\frac{-ie}{4\cos\theta_w\sin\theta_w}\gamma^\nu(1-4|e_Q|\sin^2\theta_w-\gamma^5),       \text{for Q=b,    } Z^0 \text{ propagated} 	
	\end{cases}
	\eea
	and 
	\bea
	\Gamma_{\gamma Q\bar{Q}}^{\rho}=iee_Q\gamma^\rho\epsilon^*_{\gamma\rho}(p_2).
	\eea    	  
	Here $e_Q$ is  the charge fraction of the heavy quark, i.e., $e_Q=2/3$ and $-1/3$ for charm and bottom, respectively. $\mathbb{P}(p_1)$ is the spin projector which will be given in the next subsection.

	\section{Input parameters}
	\label{input}
	
	We take the following input parameters in the calculations:
	\bea
\alpha=1/130.9, ~	m_Z=91.1876~\text{GeV},~	m_e=5.1\times10^{-4}~\text{GeV},	~\Gamma_Z=2.4952~\text{GeV},	\cr
~\sin^2\theta_w=0.2312, 	~\Gamma[J/\psi\rightarrow e^+e^-]=5.55\times10^{-6}~\text{GeV},~\Gamma[\Upsilon\rightarrow e^+e^-]=1.34\times10^{-6}~\text{GeV},\cr
~\Gamma[\psi(3770)\rightarrow e^+e^-]=2.61\times10^{-7}~\text{GeV}, 	~\Gamma[\psi(2S)\rightarrow e^+e^-]=2.33\times10^{-6}~\text{GeV},\cr
~M_{J/\psi}=3.097~\text{GeV},~M_{\Upsilon}=9.460~\text{GeV},~M_{\psi(2S)}=3.686~\text{GeV},~v^2_{c\bar{c}}=0.23,~ v^2_{b\bar{b}}=0.1
\label{values}
\eea
	Here, the Weinberg angle $\theta_w$, electron mass $m_e$, $Z$-boson mass $m_Z$, total $Z$ decay width $\Gamma_Z$, and the vector‑meson leptonic decay widths are taken from the PDG~\cite{ParticleDataGroup:2024cfk}. These leptonic decay widths will be used to calculate the event yields in the following text. For each specific $S$‑, $P$‑, and $D$‑wave production process, $m_Q$ is taken to be approximately half of the corresponding quarkonium mass~\cite{A:2023bxv}. For the LDMEs, we employ their analogs, namely the squared radial wave functions or their $l$-th derivatives evaluated at the origin, collectively denoted as $|R^{(l)}(0)|^2$, which are calculated via the Cornell potential. The specific values of $m_Q$ and $|R^{(l)}(0)|^2$ 
are listed in Table~\ref{inmass}.
	
\begin{table}[htb]
\centering
\caption{Masses (units: $\text{GeV}$) of constituent quarks and the squared radial wave functions at the origin $|R^{(l)}_{nL}(0)|^2$ (units: $\text{GeV}^{3+2l}$)\cite{A:2023bxv}.}
\begin{tabular}{|c|c||c|c||c|c||c|c|}
\hline
\multicolumn{8}{|c|}{\textbf{Charmonium}} \\
\hline
$m_c$ & $|R_{1S}(0)|^2$ & $m_c$ & $|R^{(1)}_{1P}(0)|^2$ & $m_c$ & $|R^{(2)}_{1D}(0)|^2$ & $m_c$ & $|R_{2S}(0)|^2$ \\
\hline
1.50 & 1.4856 & 1.75 & 0.1007 & 1.90 & 0.0159 & 1.84 & 0.7956 \\
\hline
\multicolumn{8}{|c|}{\textbf{Bottomonium}} \\
\hline
$m_b$ & $|R_{1S}(0)|^2$ & $m_b$ & $|R^{(1)}_{1P}(0)|^2$ & $m_b$ & $|R^{(2)}_{1D}(0)|^2$ & & \\
\hline
4.71 & 9.1909 & 4.95 & 1.5663 & 5.07 & 0.6325 & & \\
\hline
\end{tabular}
\label{inmass}
\end{table}

	
	\section{Results and discussion}
	\label{res}
	
	\subsection{s-channel production}

\begin{widetext}
		\begin{figure*}[htbp]
			\begin{tabular}{c c c }
			\includegraphics[width=0.333\textwidth]{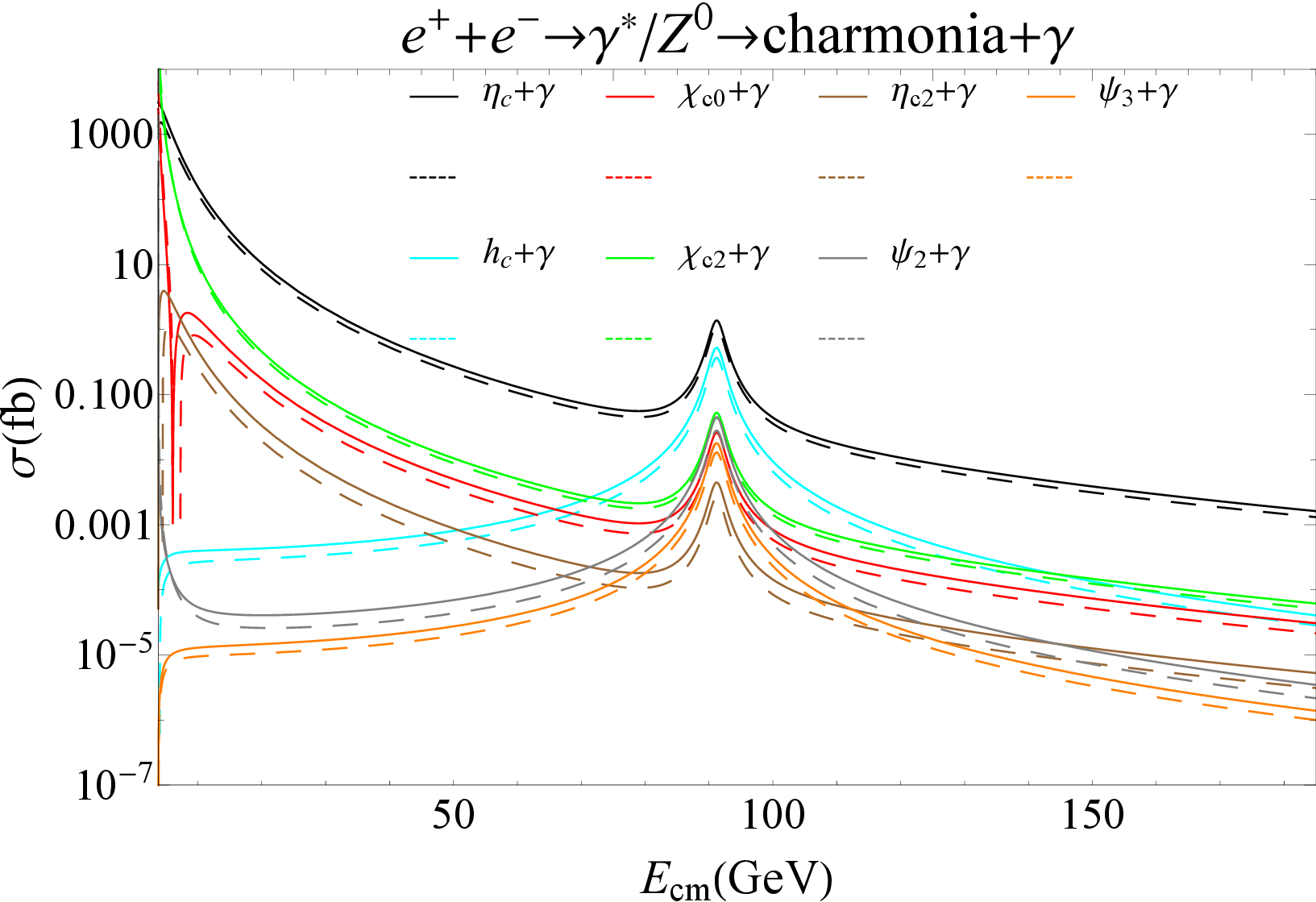}
			\end{tabular}
			\begin{tabular}{c c c }
				\includegraphics[width=0.333\textwidth]{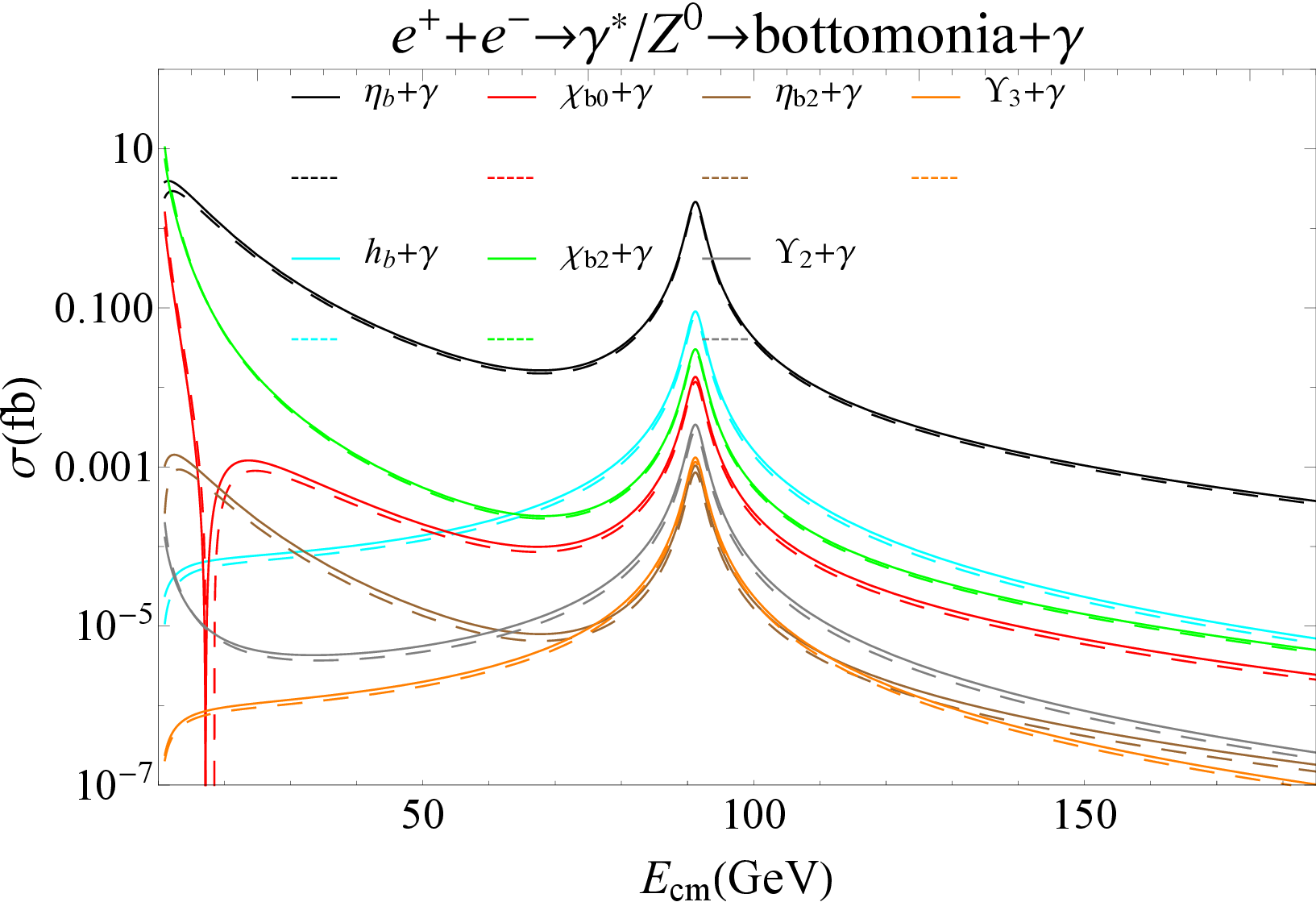}
			\end{tabular}
			\caption{ Cross sections for s-channel heavy quarkonium production as a function of the c.m. energy $(E_{cm}=\sqrt{s})$. The solid and dashed lines denote the LO and NLO($v^2$) results, respectively. }
			\label{Cs}
		\end{figure*}
	\end{widetext}

	In this subsection, we investigate the production of heavy‑quarkonium states allowed in the $s$-channel but forbidden in the $t$-channel due to $J^{PC}$ conservation. These states include ${}^1S_0, {}^1P_1, {}^3P_0, {}^3P_2, {}^1D_2, {}^3D_2, {}^3D_3$. 
	The LO and NLO($v^2$) cross sections as functions of the c.m. energy are presented in Fig.~\ref{Cs}. 
	The cross‑section lines correspond to the sum of contributions from the $\gamma^*$ propagator, the $Z^0$ propagator, and their interference. Contributions mediated by the $\gamma^*$ propagator dominate in the low‑energy region, while those from the $Z^0$ propagator dominate near the $Z^0$ mass owing to $Z^0$ resonance enhancement. 
	A dip is observed for $\chi_{Q0}+\gamma$ production at low energy, originating from the overall suppression factor $(s-12m_Q^2)^2$~\cite{Chen:2013mjb,Liao:2021ifc}. 
Compared with the charmonium case, relativistic corrections for bottomonium production are relatively small because $v^2_{b\bar{b}}$ is small.

	The interference between different mechanisms may yield non‑negligible effects \cite{Sun:2018rgx,Jiang:2018wmv}. The interference between resonant states and the continuum is generally governed by the relative phases between the amplitudes, which can modify both the height and position of resonance peaks \cite{Xu:2015goa,Jia:2026onq}. For the $s$-channel processes, the $\gamma^*$- and $Z^0$-propagated processes correspond to standard electroweak amplitudes, differing only in their propagators and interaction vertices, with no additional relative‑phase difference. We can therefore directly present the relation for the interference cross section in terms of the separate $\gamma^*$‑mediated and $Z^0$-mediated cross sections:
	\begin{equation}
		\sigma_{\rm int} =  R_\gamma\,\sigma_{\gamma^*},\qquad \sigma_{\rm int} =  R_Z\,\sigma_{Z^0},
		\label{eq:int-def-R}
	\end{equation}
	where $\sigma_{\rm int}$, $\sigma_{\gamma^*}$, and $\sigma_{Z^0}$ stand for the interference, $\gamma^*$-mediated, and $Z^0$-mediated cross sections, respectively. 
	And we define two dimensionless interference ratios $R_\gamma$ and $R_Z$,
	\begin{align}\label{interratios}
		R_\gamma
		&= 2\frac{s(s-m_Z^2)}{(s-m_Z^2)^2+m_{Z}^2\Gamma_Z^2}
		\frac{(1-4\sin^2\theta_w)(1-4|e_Q|\sin^2\theta_w)}{|e_Q|(4\cos\theta_w\sin\theta_w)^2},
		\nonumber\\[4pt]
		R_Z
		&= 2\frac{s-m_Z^2}{s}
		\frac{|e_Q|(4\cos\theta_w\sin\theta_w)^2(1-4\sin^2\theta_w)}
		{\big[(1-4\sin^2\theta_w)^2+1\big](1-4|e_Q|\sin^2\theta_w)}.
	\end{align}
	Since the propagators and interaction vertices are identical for different heavy‑quarkonium states with only $s$-channel contributions, $R_\gamma$ and $R_Z$ take the same form for these states at both LO and $\text{NLO}(v^2)$. Note that $R_\gamma$ and $R_Z$ still depend on the heavy‑quark charge $e_Q$, hence their numerical values differ between charmonium and bottomonium.
	The interference cross section exhibits an asymmetric dispersive line‑shape across the $Z^0$ resonance, vanishing at the $Z^0$ pole and leading to a shift of the resonance peak position.

	The numerical cross sections at the 
	$Z^0$ pole for both LO and NLO in 
	$v^2$ are given in Table~\ref{TCS}. Our LO results are in agreement with Refs.~\cite{Chung:2008km, Li:2013nna,Chen:2013mjb,Liao:2021ifc}, and the NLO($v^2$) results for $\gamma^*$-mediated processes are consistent with Refs.~\cite{Sang:2009jc,Li:2013nna}, if the same input parameters are used.

	\begin{table}
		\caption{ Cross sections for the s-channel production of different quarkonia at 
			$\sqrt{s}=m_Z$, via the $\gamma^*$-propagator, the $Z^0$-propagator, and their sum (including the interference), where LO and NLO($v^2$) denote the leading and next-to-leading orders in the 
			$v^2$ expansion, respectively. }
		\begin{tabular}{|c|c c| |c|c c| c c| c c|}
			\hline
			&\multicolumn{2}{c| |}{$\sigma$}&  &\multicolumn{6}{c|}{$\sigma$}\\
			\hline
			propagator&\multicolumn{2}{c| |}{$Z^0$(units: $10^{-2}fb$) }& &\multicolumn{2}{c| }{$\gamma^*$(units: $10^{-4}fb$) }&\multicolumn{2}{c| }{$Z^0$(units: $10^{-2}fb$) }&\multicolumn{2}{c|}{$\text{Total}$(units: $10^{-2}fb$) }\\
			\hline
			&LO&NLO($v^2$)&	 &LO& NLO($v^2$) &LO& NLO($v^2$) &LO & NLO($v^2$) \\
			\hline
			\multicolumn{10}{|c|}{	charmonia }\\
			\hline
			$h_c$&52.51&36.79 &$\eta_c$&246.6&199.3&135.5&109.5&138.0&111.5\\
			\hline
			$\psi_{2}$&4.504&2.805	&$\chi_{c0}$&4.650&3.254&2.555&1.788&2.601&1.820\\
			\hline
			$\psi_{3}$&1.793&1.293 & $\chi_{c2}$&9.424&7.909&5.178&4.345&5.272&4.425\\
			\hline	
			& &	&$\eta_{c2}$&0.799&0.471&0.439&0.259&0.447&0.263\\
			\hline	
			\multicolumn{10}{|c|}{bottomonia}\\
			\hline
			$h_b$&9.020&7.839	  &$\eta_b$&30.08&27.54&215.1&197.0&215.4&197.2\\
			\hline
			$\Upsilon_{2}$&0.340&0.285&$\chi_{b0}$&0.19&0.16&1.355&1.173&1.357&1.175\\
			\hline
			$\Upsilon_{3}$&0.131&0.115 & $\chi_{b2}$&0.422&0.393&3.018&2.810&3.022&2.814\\
			\hline	
			&&&$\eta_{b2}$&0.015&0.012&0.104&0.085&0.104&0.085\\
			\hline
		\end{tabular}
		\label{TCS}
	\end{table}

\begin{widetext}
		\begin{figure*}[htbp]
			\begin{tabular}{c c c }
				\includegraphics[width=0.333\textwidth]{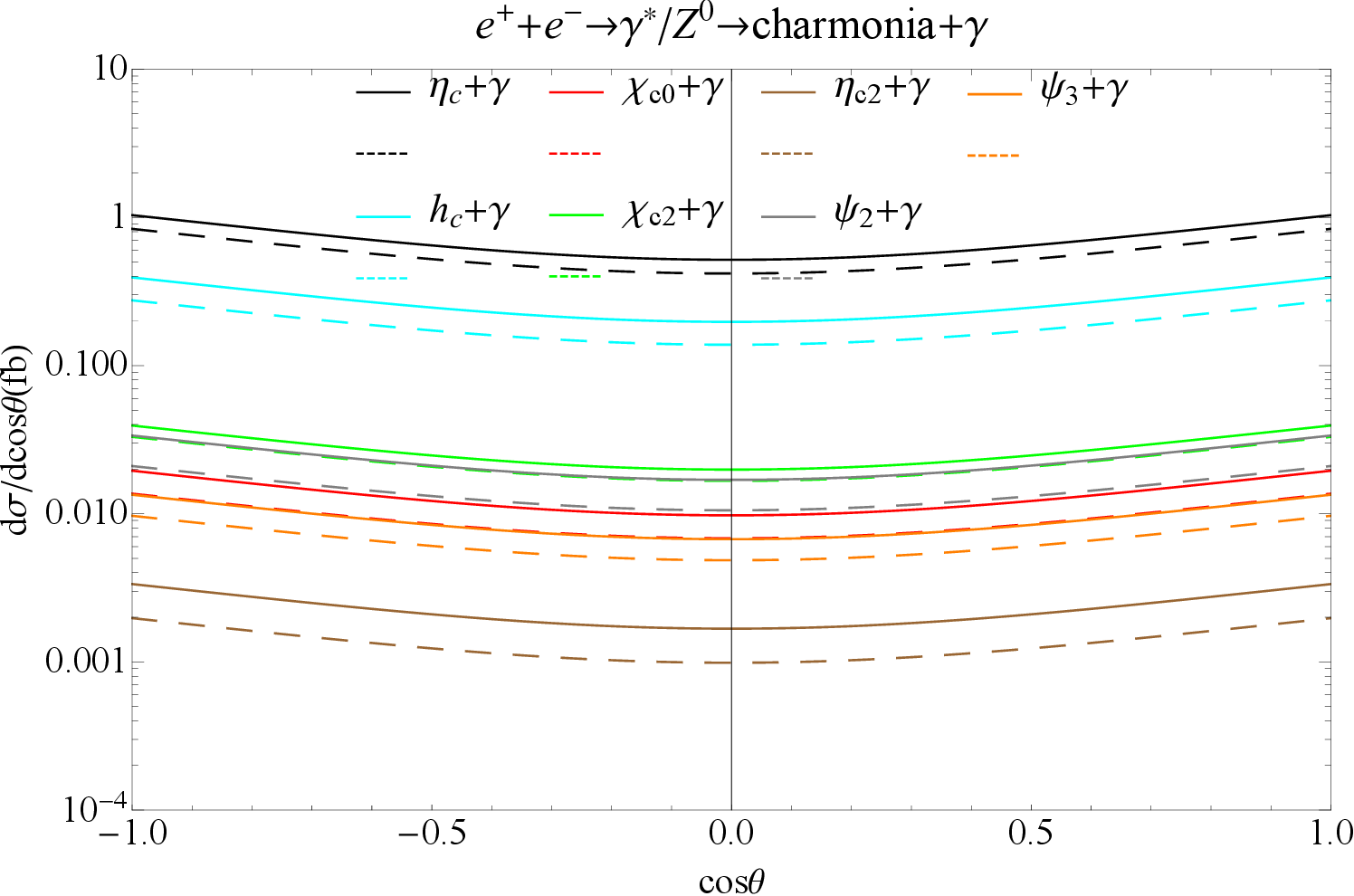}
				\includegraphics[width=0.333\textwidth]{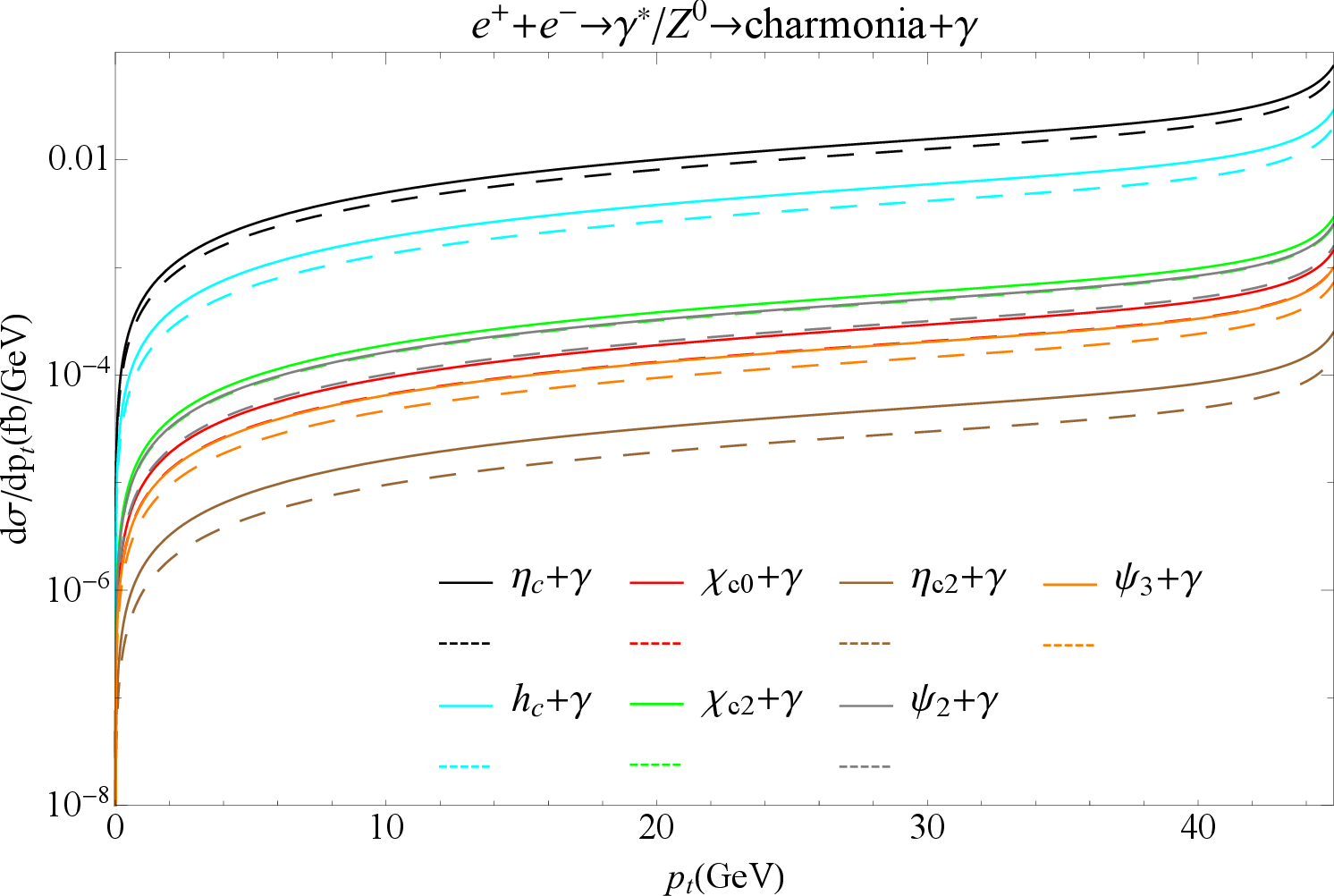}
				\includegraphics[width=0.333\textwidth]{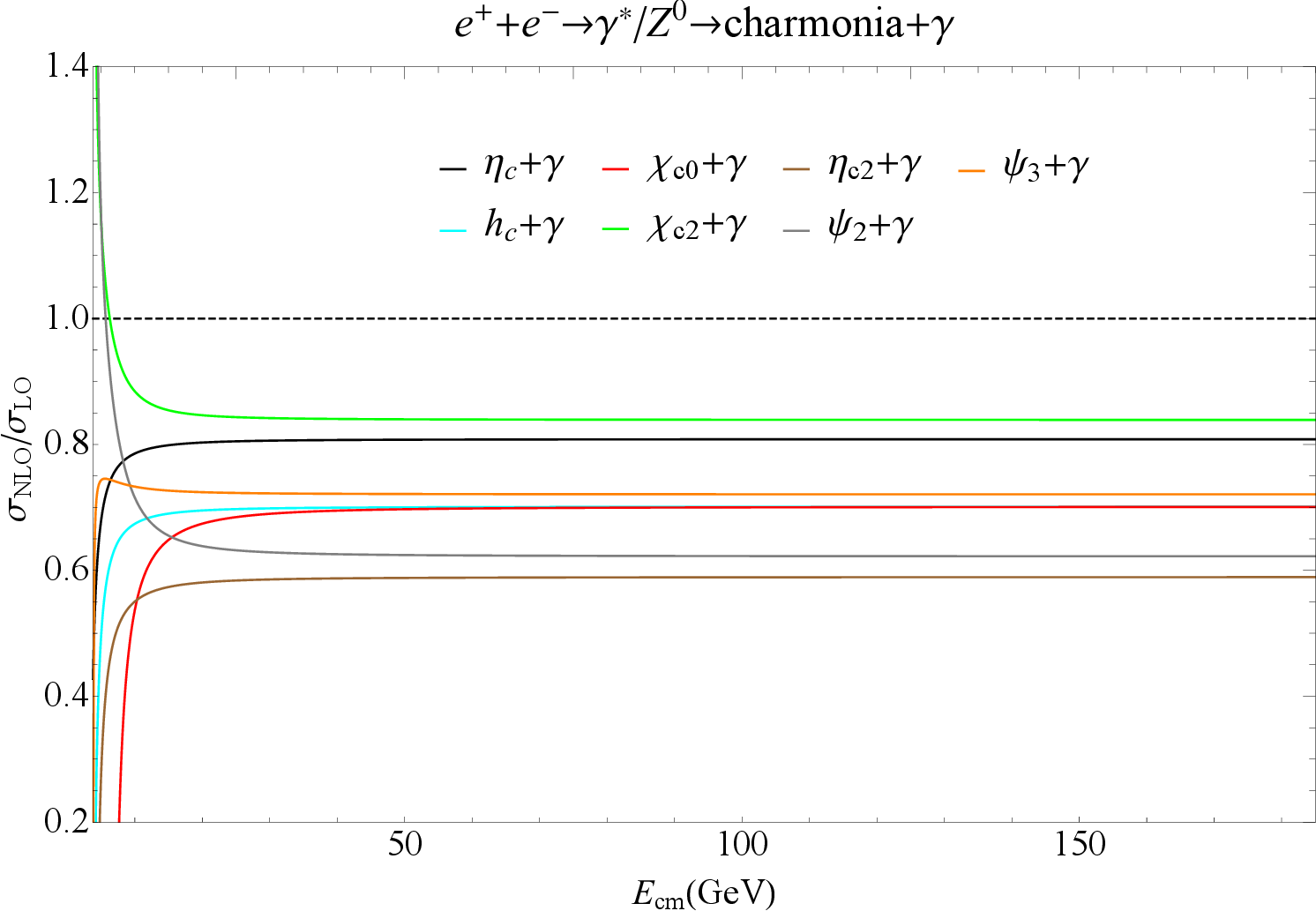}
			\end{tabular}		 
			\begin{tabular}{c c c }
				\includegraphics[width=0.333\textwidth]{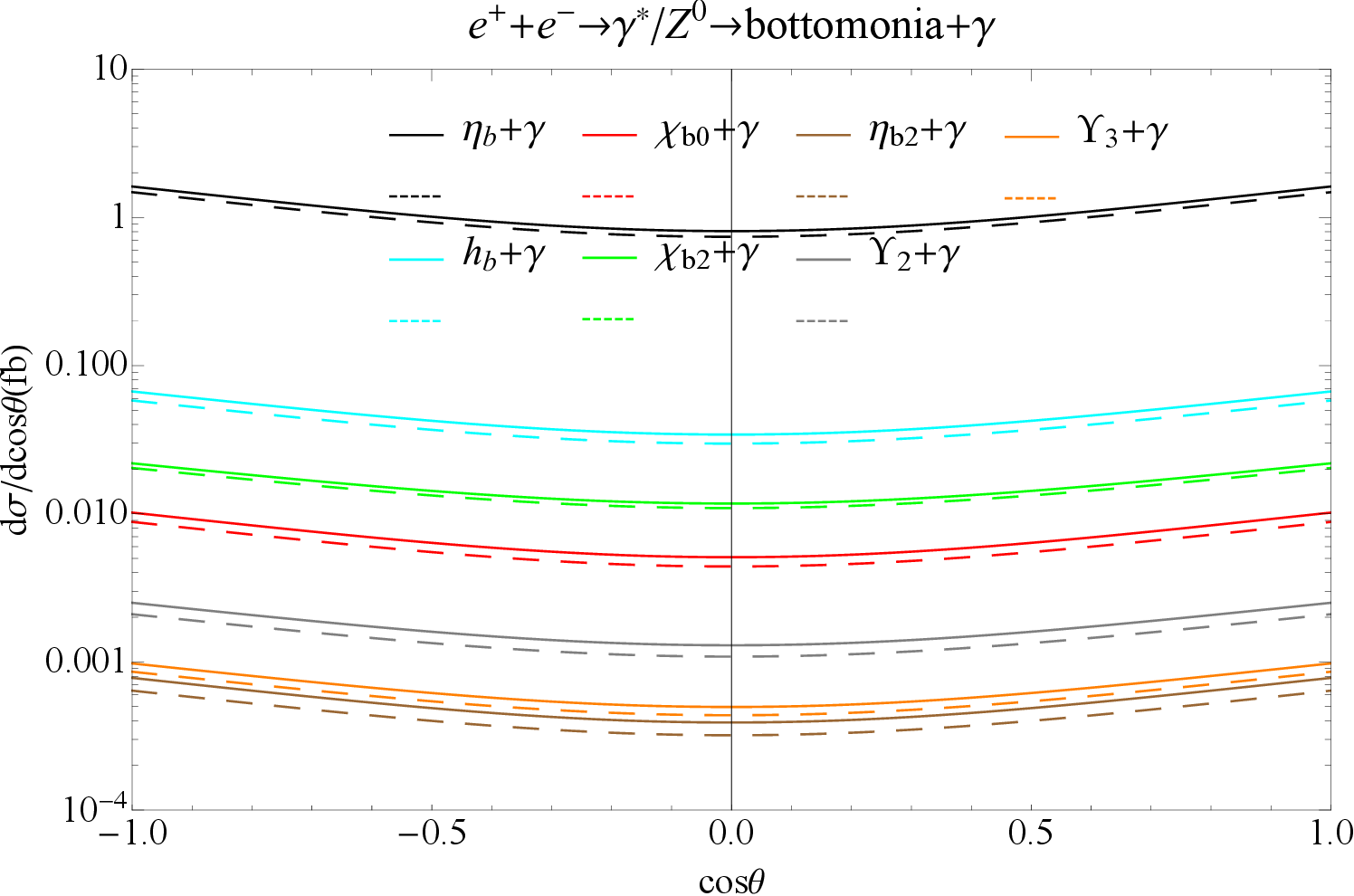}		 
				\includegraphics[width=0.333\textwidth]{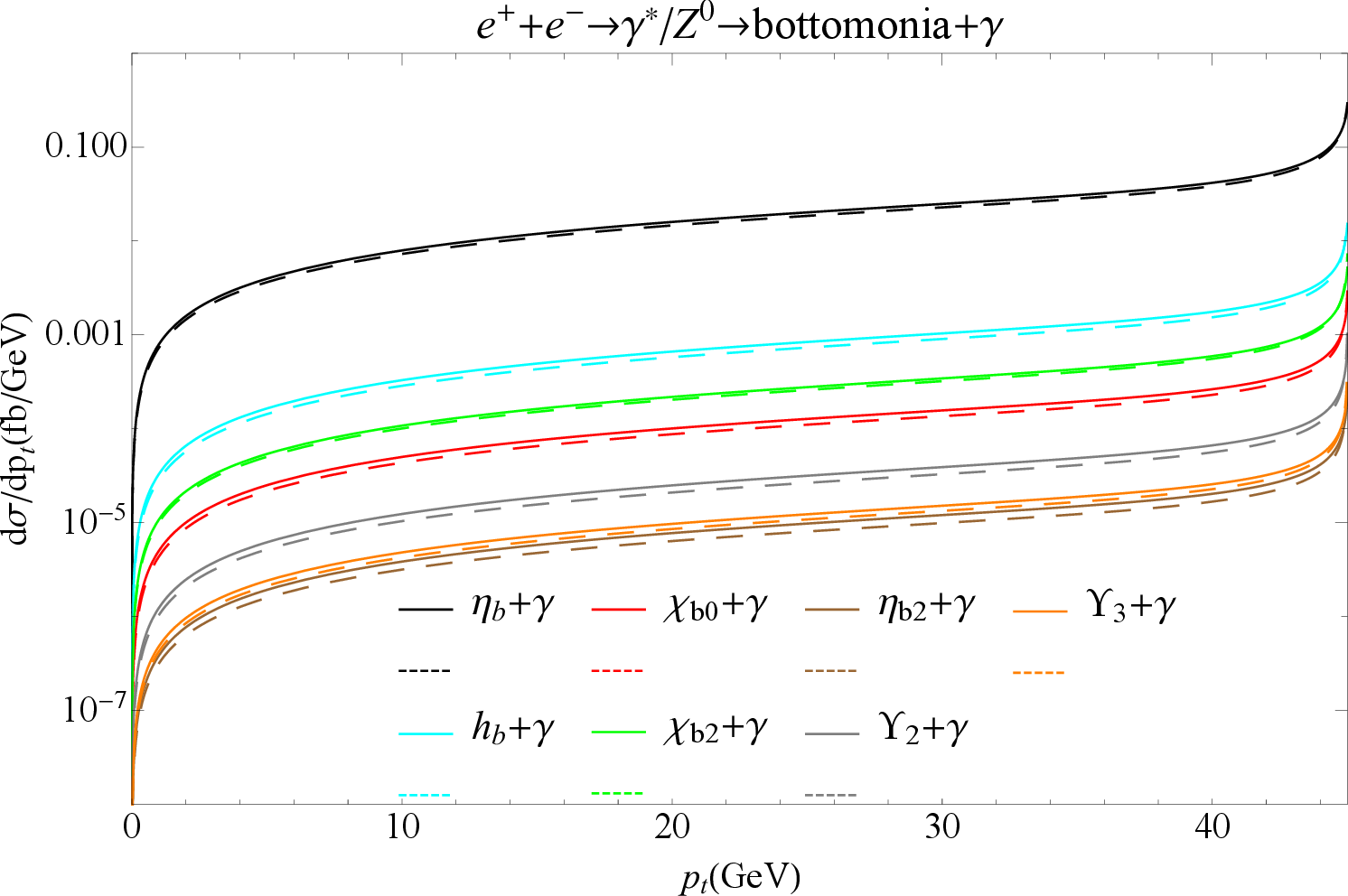}
				\includegraphics[width=0.333\textwidth]{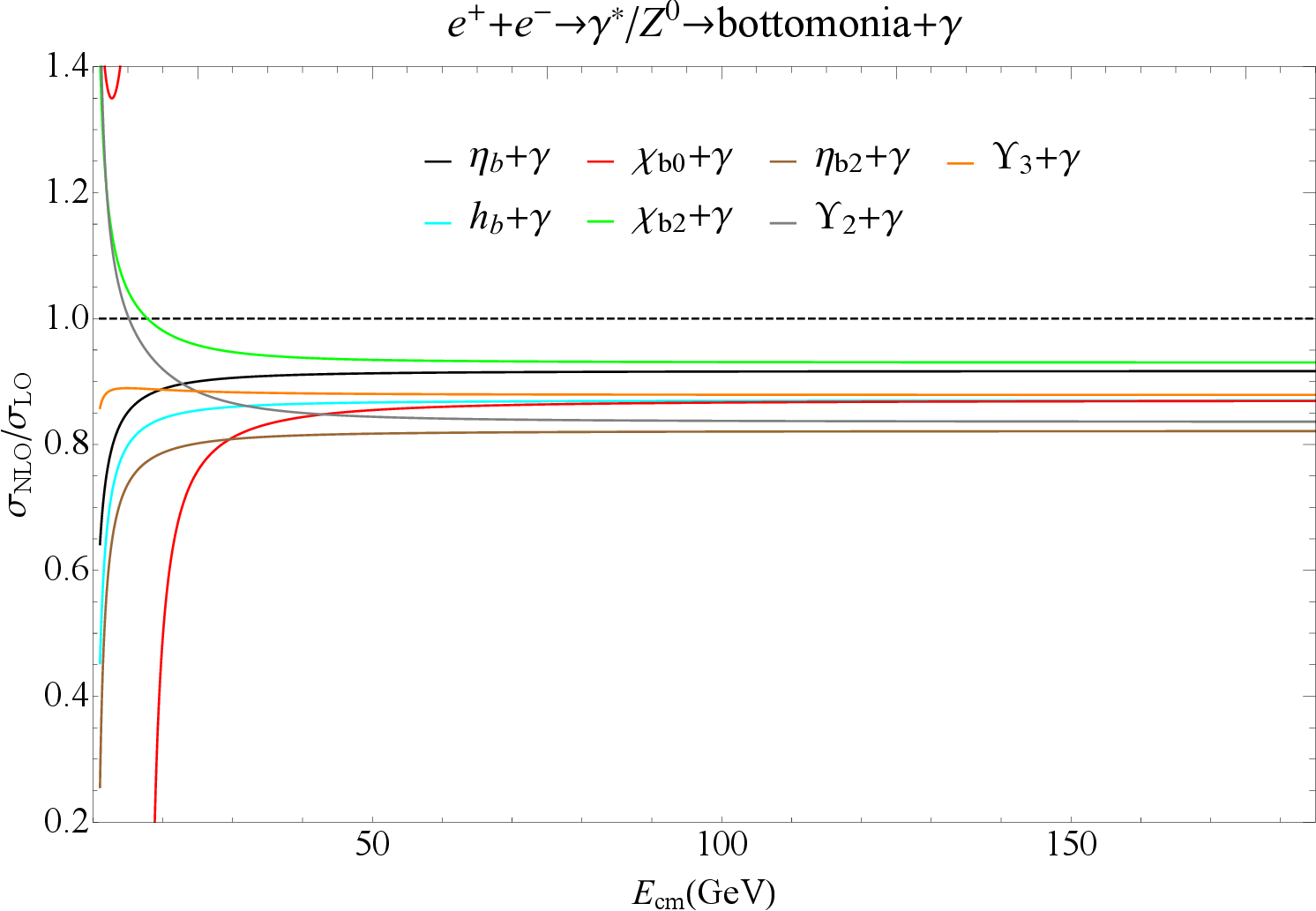}
			\end{tabular}	
			\caption{ Differential cross sections $d\sigma/d\cos\theta$, $d\sigma/dp_t$ at $\sqrt{s}=m_Z$, and the K-factor $\sigma_{NLO(v^2)}/\sigma_{LO}$ versus c.m. energy for different heavy quarkonium (s-channel) production. The solid line represents LO result and the dashed line represents   NLO($v^2$) result. }
			\label{cc}
		\end{figure*}
	\end{widetext}

	The differential cross section $\frac{d\sigma}{dp_t}$ can be written as:
\bea
	\frac{d\sigma}{dp_t}=\left|\frac{d\cos\theta}{dp_t}\right|\frac{d\sigma}{d\cos\theta}=\frac{p_t}{|\boldsymbol{p}_1|\sqrt{\boldsymbol{p}_1^2-p_t^2}}\frac{d\sigma}{d\cos\theta}
	\label{pt1}
\eea
and the three momentum:
\bea
	|\boldsymbol{p}_1|=\frac{\sqrt{\lambda[s,(2E_{q})^2,0^2]}}{2\sqrt{s}}
	\label{pt2}
\eea
	with
	$
	\lambda(x,y,z)=x^2+y^2+z^2-2(xy+xz+yz)
	$.
	Combining $E_q=\sqrt{m_Q^2+\boldsymbol{q}^2}$ and Eqs. (\ref{pt1},\ref{pt2}), we can further get a $\mathcal{O}(v^2)$  expression,
\bea
\frac{d\sigma}{dp_t}=\frac{4p_t}{\sqrt{s}} \frac{(1-r) C + 2 r v^2 D}{(1-r)^2 C^{3/2}}
\frac{d\sigma}{d\cos\theta}+\mathcal{O}(v^4),
\eea
where $r\equiv 4m_Q^2/s$, $C \equiv s(1-r)^2-4p_t^2$, and $D \equiv s(1-r)^2-2p_t^2$.
	The differential cross sections ($d\sigma/d\cos\theta$, $d\sigma/dp_t$) at $\sqrt{s}=m_Z$, as well as the K-factors as functions of the center-of-mass energy, are presented in Fig. \ref{cc}. One can observe that relativistic corrections are more significant for charmonium production, where the K-factors generally range from 0.55 to 0.85, while those for bottomonium vary between 0.80 and 0.95. The ratios of relativistic correction SDCs to LO ones are summarized in Table~\ref{Rratio}. For S-wave and P-wave states, our results agree with the previous calculations in Refs.~\cite{Xu:2012am,Xu:2014zra,Wang:2025sbx,Wang:2026gso}.

	\begin{table}
		\caption{The ratio between the RC SDC and the LO SDC, i.e.  R[n]=G[n]/F[n], for different heavy quarkonium (s-channel)  production  in the high energy limit ($\sqrt{s}\gg m_Z$).}
		\begin{tabular}{|c|c| |c|c|c|c|}
			\hline
			quarkonium state&$ R[n] $&quarkonium state &\multicolumn{3}{c|}{$R[n]$}\\
			\hline
			\hline
			propagator&$Z^0$&propagator&$\gamma^*$&$Z^0$&$\gamma^*/Z^0$\\
			\hline
			\hline
			$h_Q(^1P_1)$&$-\frac{13}{10}$&$\eta_Q (^1S_0)$&$-\frac{5}{6}$&$-\frac{5}{6}$&$-\frac{5}{6}$\\
			\hline
			$\psi_2/\Upsilon_{2} (^3D_2)$&$-\frac{23}{14}$	&$\chi_{Q0} (^3P_0)$&$-\frac{13}{10}$&$-\frac{13}{10}$&$-\frac{13}{10}$\\
			\hline
			$\psi_3/\Upsilon_{3} (^3D_3)$&$-\frac{17}{14}$ &  $\chi_{Q2} (^3P_2)$&$-\frac{7}{10}$&$-\frac{7}{10}$&$-\frac{7}{10}$\\
			\hline
			&	&$\eta_{Q2} (^1D_2)$&$-\frac{25}{14}$&$-\frac{25}{14}$&$-\frac{25}{14}$\\
			\hline
		\end{tabular}
		\label{Rratio}
	\end{table}
	
	\subsection{ Production of ~${}^3L_1$~ states}

	In this subsection, we investigate the production of ${}^3S_1$, ${}^3P_1$, and ${}^3D_1$ states. These states receive non‑zero $t$-channel contributions, corresponding to $\gamma^*$-fragmentation into ${}^3S_1,{}^3D_1$ and $Z^0$-fragmentation into ${}^3S_1,{}^3P_1,{}^3D_1$.
	
	\begin{widetext}
		\begin{figure*}[htbp]	
			\begin{tabular}{c c c }
				\includegraphics[width=0.333\textwidth]{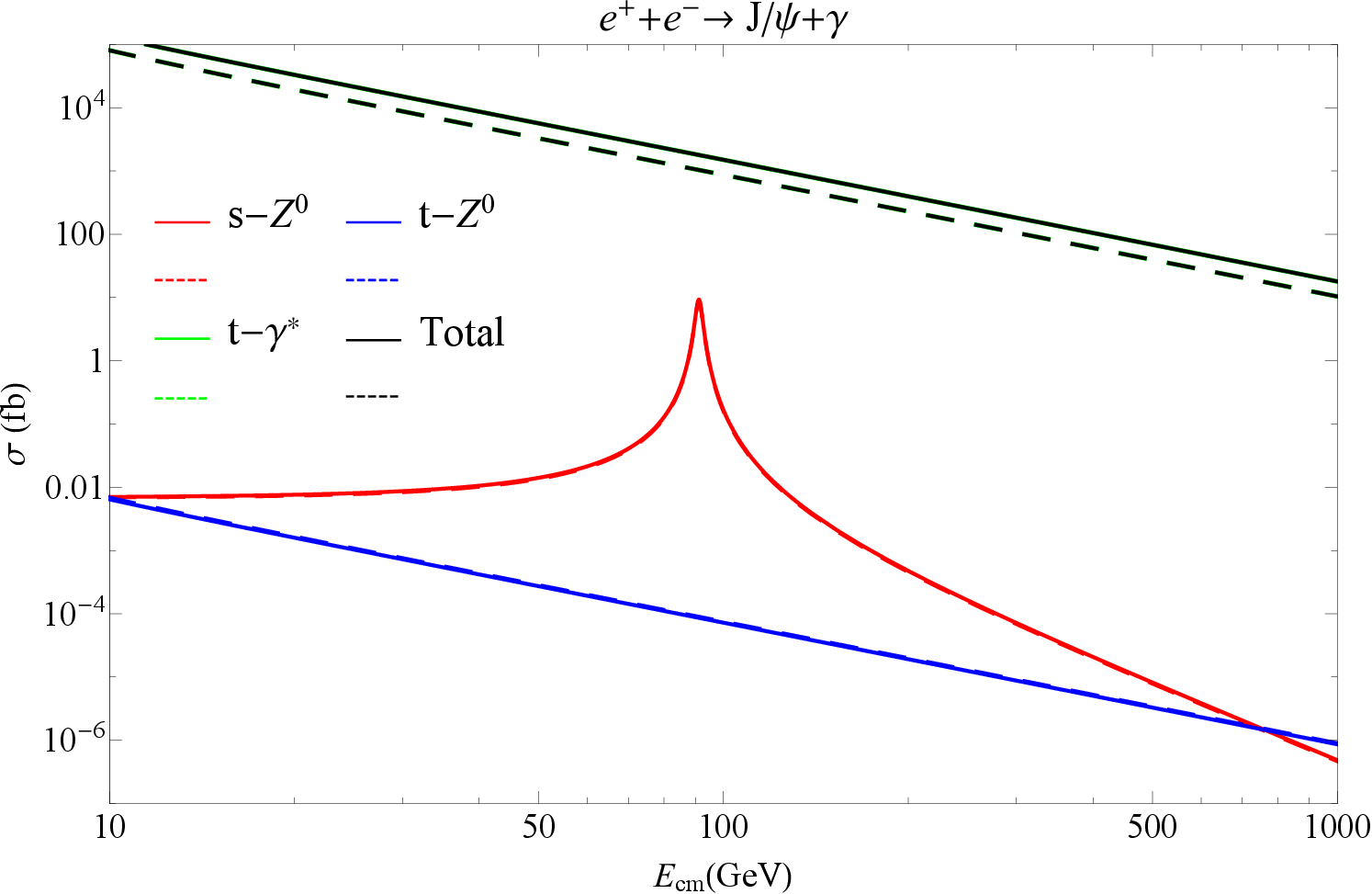}
				\includegraphics[width=0.333\textwidth]{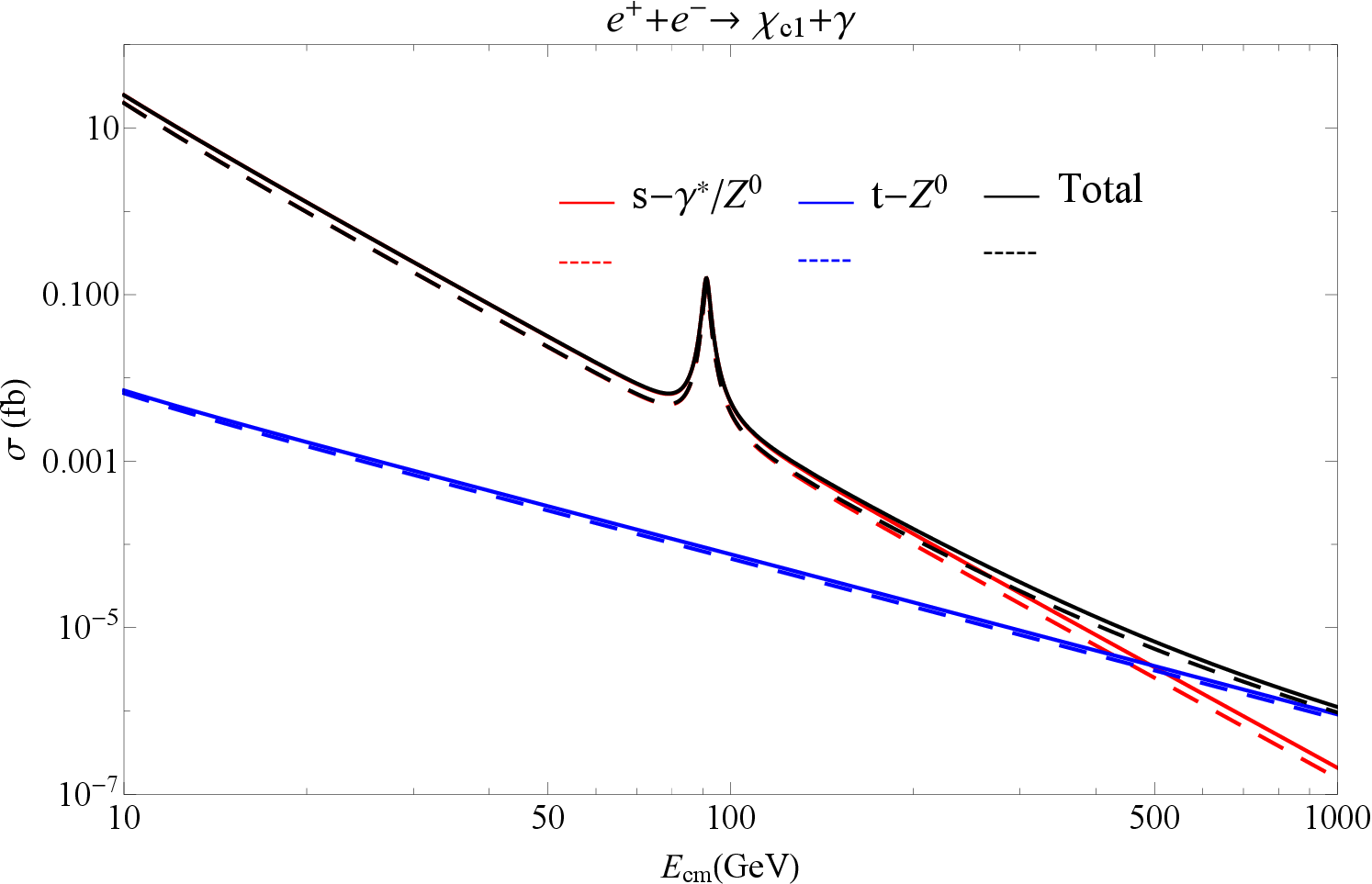}
				\includegraphics[width=0.333\textwidth]{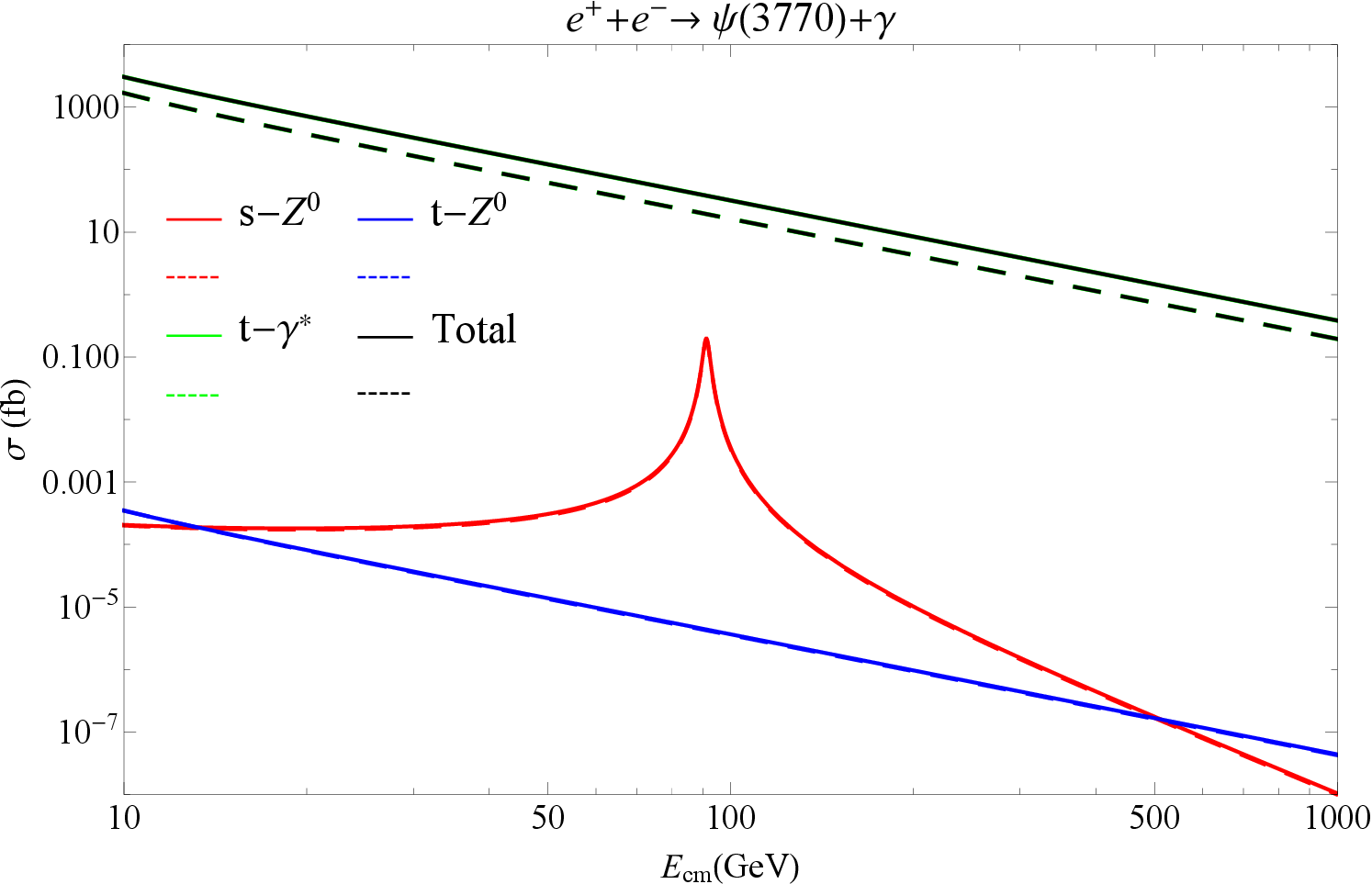}
			\end{tabular}
			\begin{tabular}{c c c }
				\includegraphics[width=0.333\textwidth]{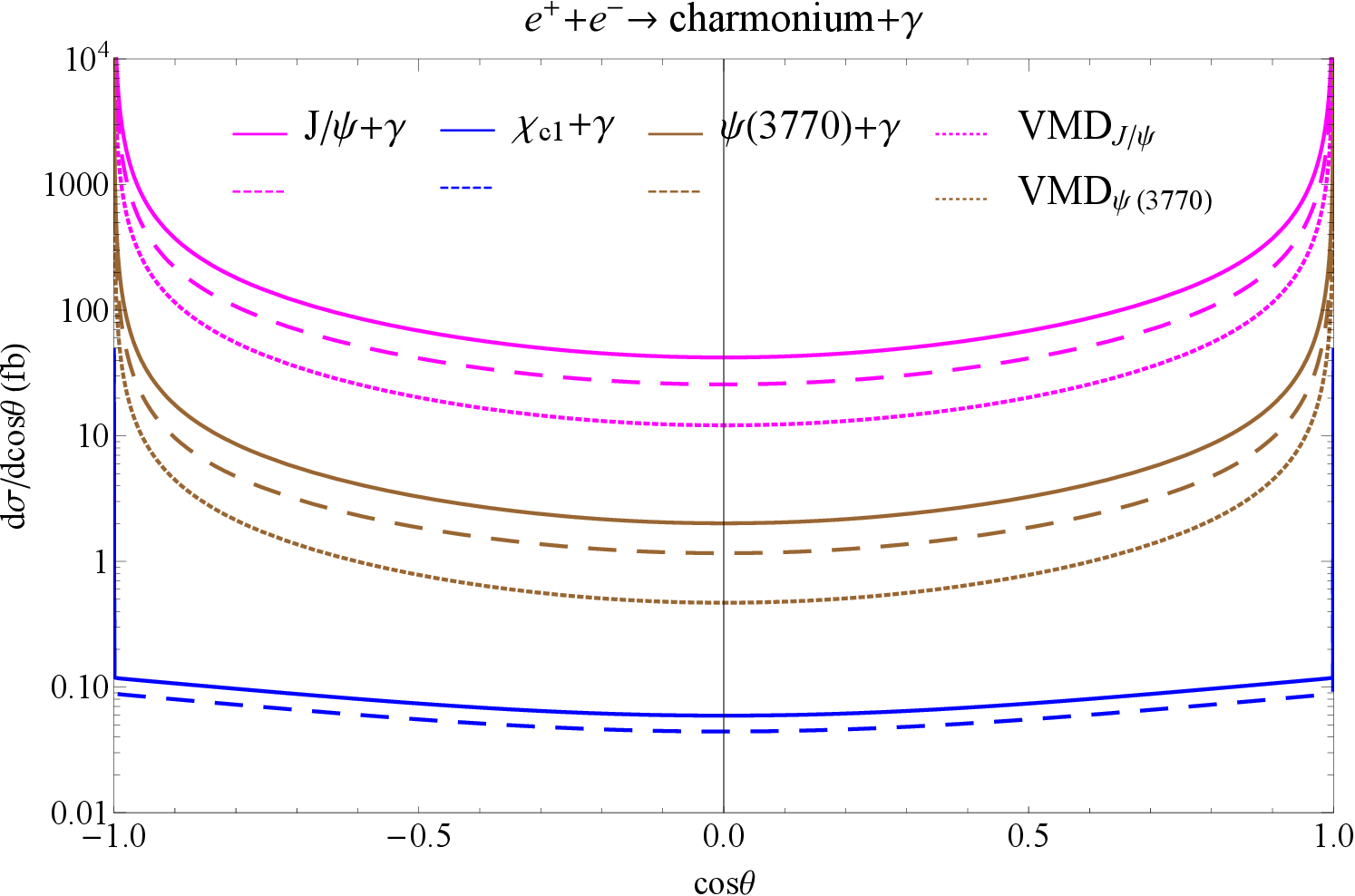}
				\includegraphics[width=0.333\textwidth]{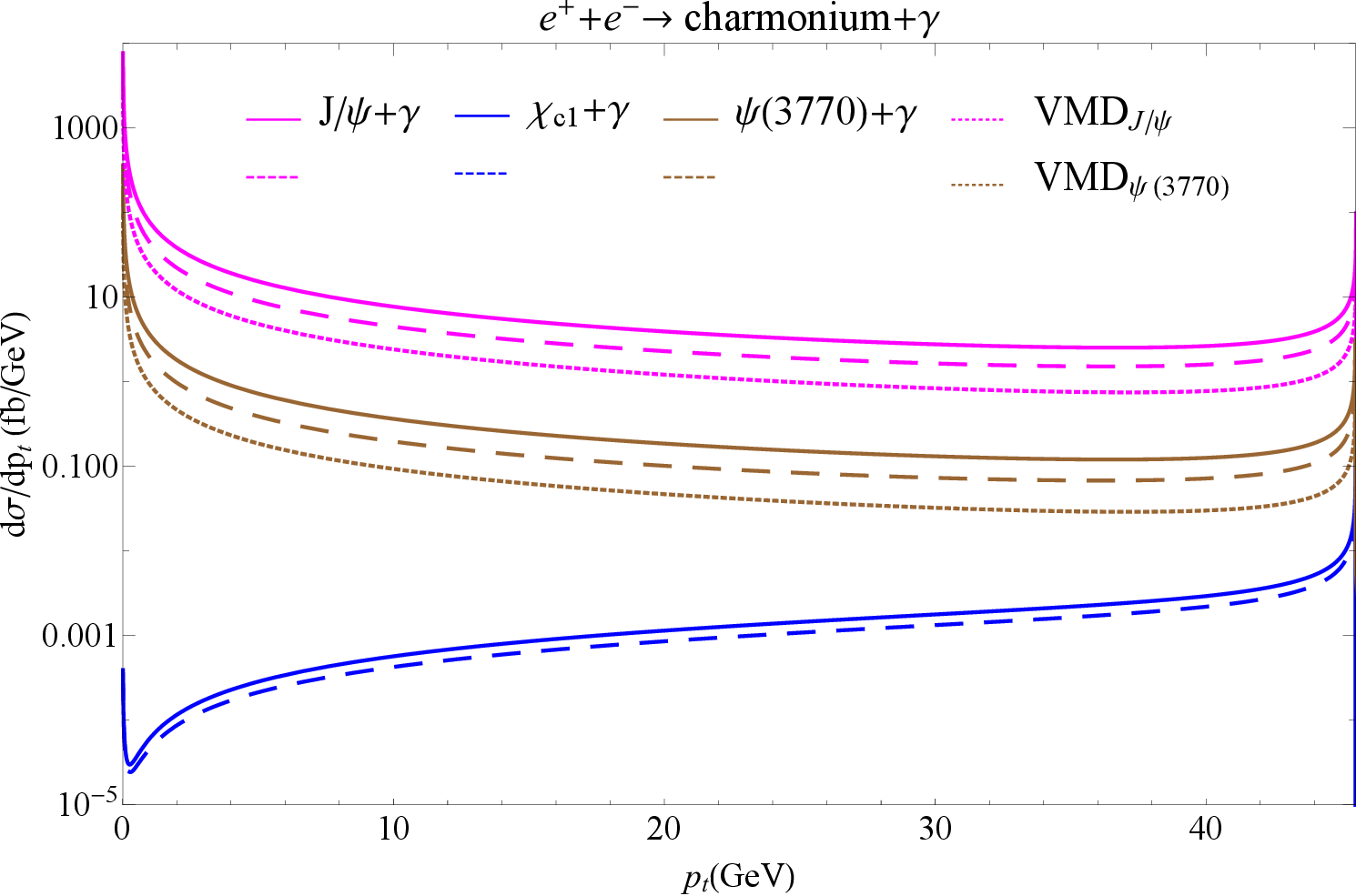}
				\includegraphics[width=0.333\textwidth]{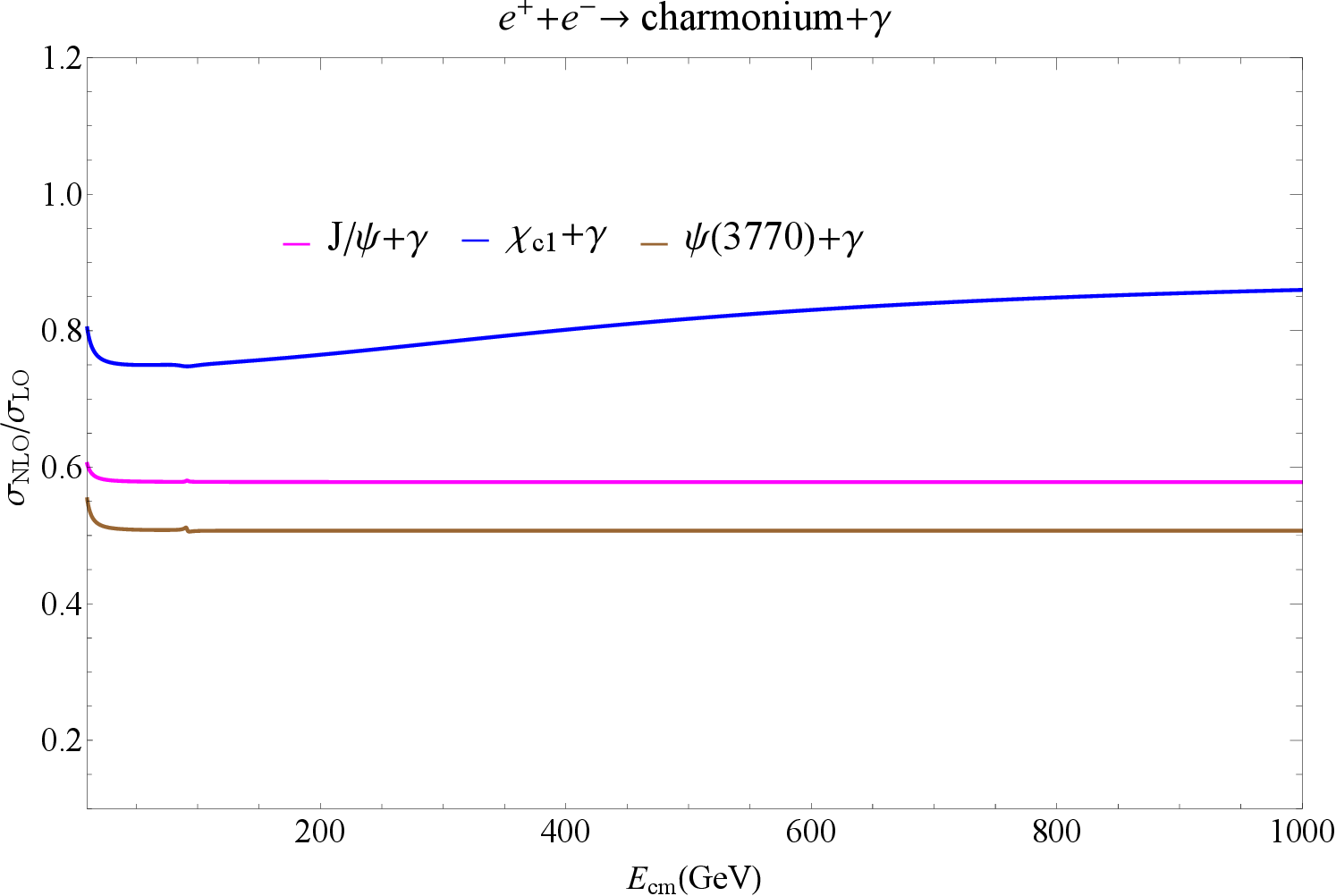}
			\end{tabular}
			\caption{  
$^3L_1$ charmonia production (including both $s$ and $t$-channel contributions). The upper panels show cross sections for different subprocesses, where ``$s$–$Z^0$'' denotes the $s$-channel $Z^0$-propagated contribution, ``$t$–$\gamma^*$'' denotes the $t$-channel $\gamma^*$-fragmentation contribution, and so on. ``Total'' stands for the sum of all channels including their interference terms. The lower panels present the total differential cross sections $d\sigma/d\cos\theta$, $d\sigma/dp_t$ evaluated at $\sqrt{s}=m_Z$, as well as the K‑factor $\sigma_{\text{NLO}(v^2)}/\sigma_{\text{LO}}$. The solid lines correspond to the LO results, the dashed lines to the $\text{NLO}(v^2)$ results, and the dotted lines to the VMD results.
}
			\label{ccc}
		\end{figure*}
	\end{widetext}
	
	\begin{widetext}
		\begin{figure*}[htbp]		
			\begin{tabular}{c c c }
				\includegraphics[width=0.333\textwidth]{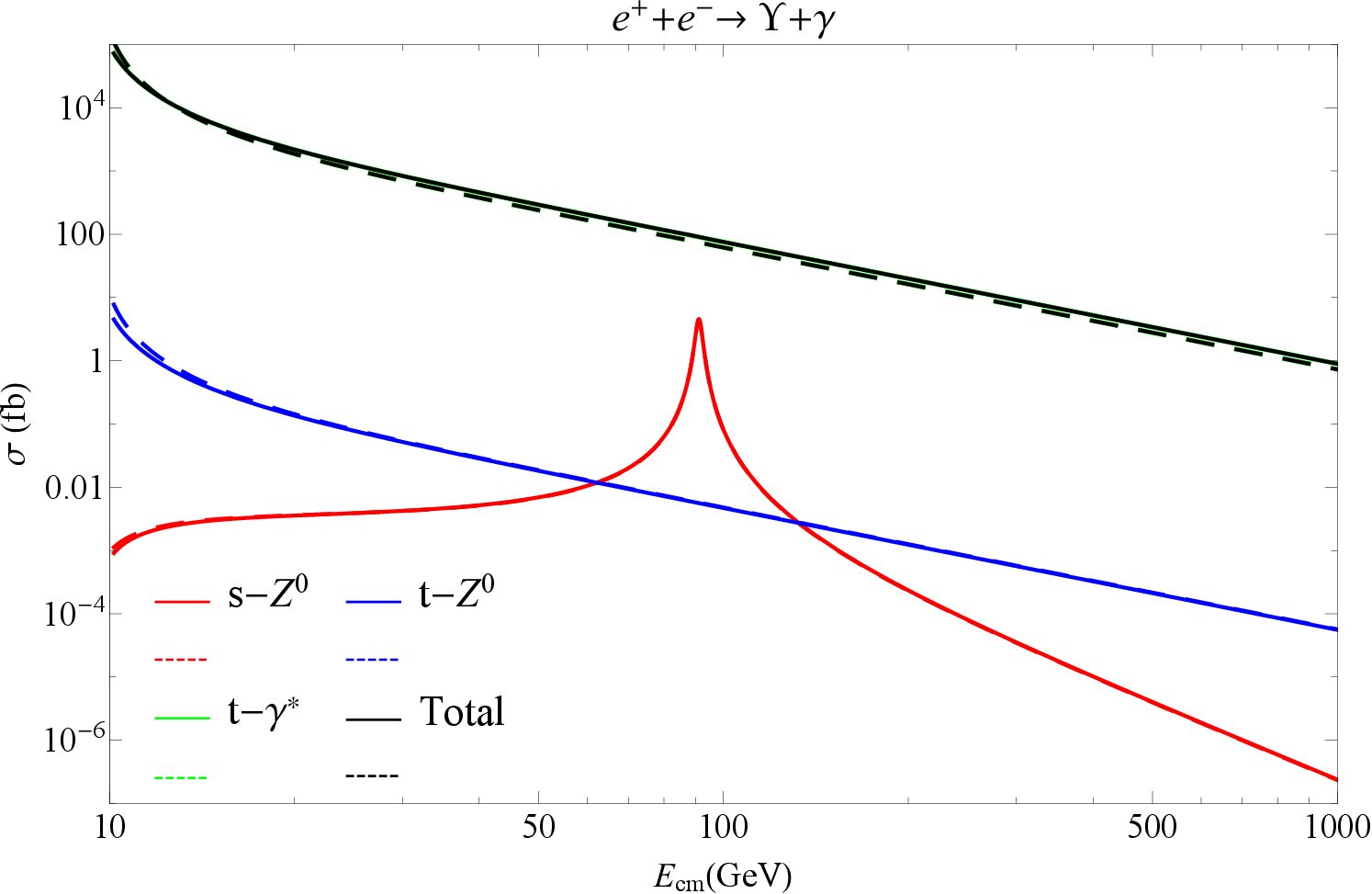}
				\includegraphics[width=0.333\textwidth]{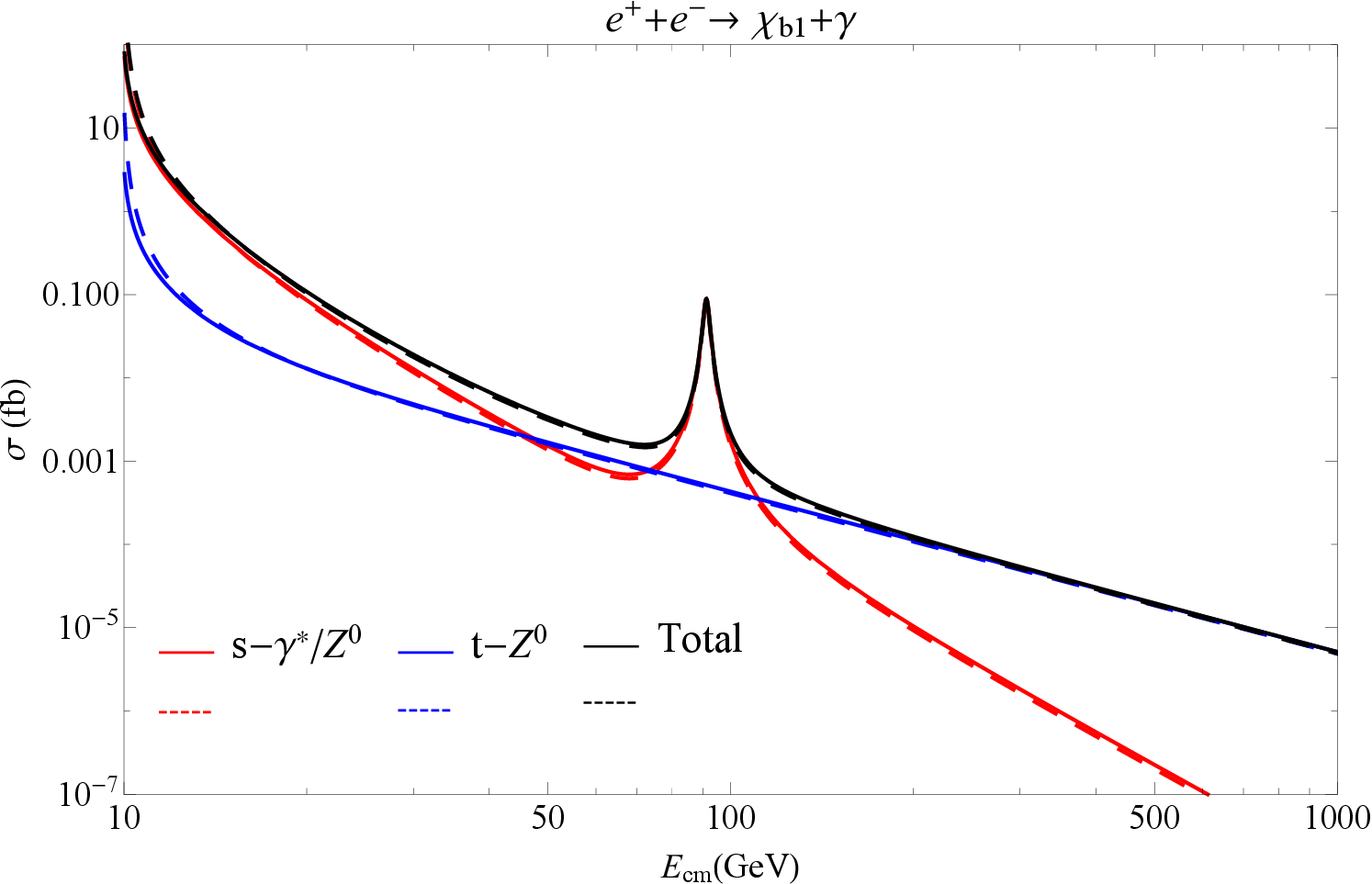}
				\includegraphics[width=0.333\textwidth]{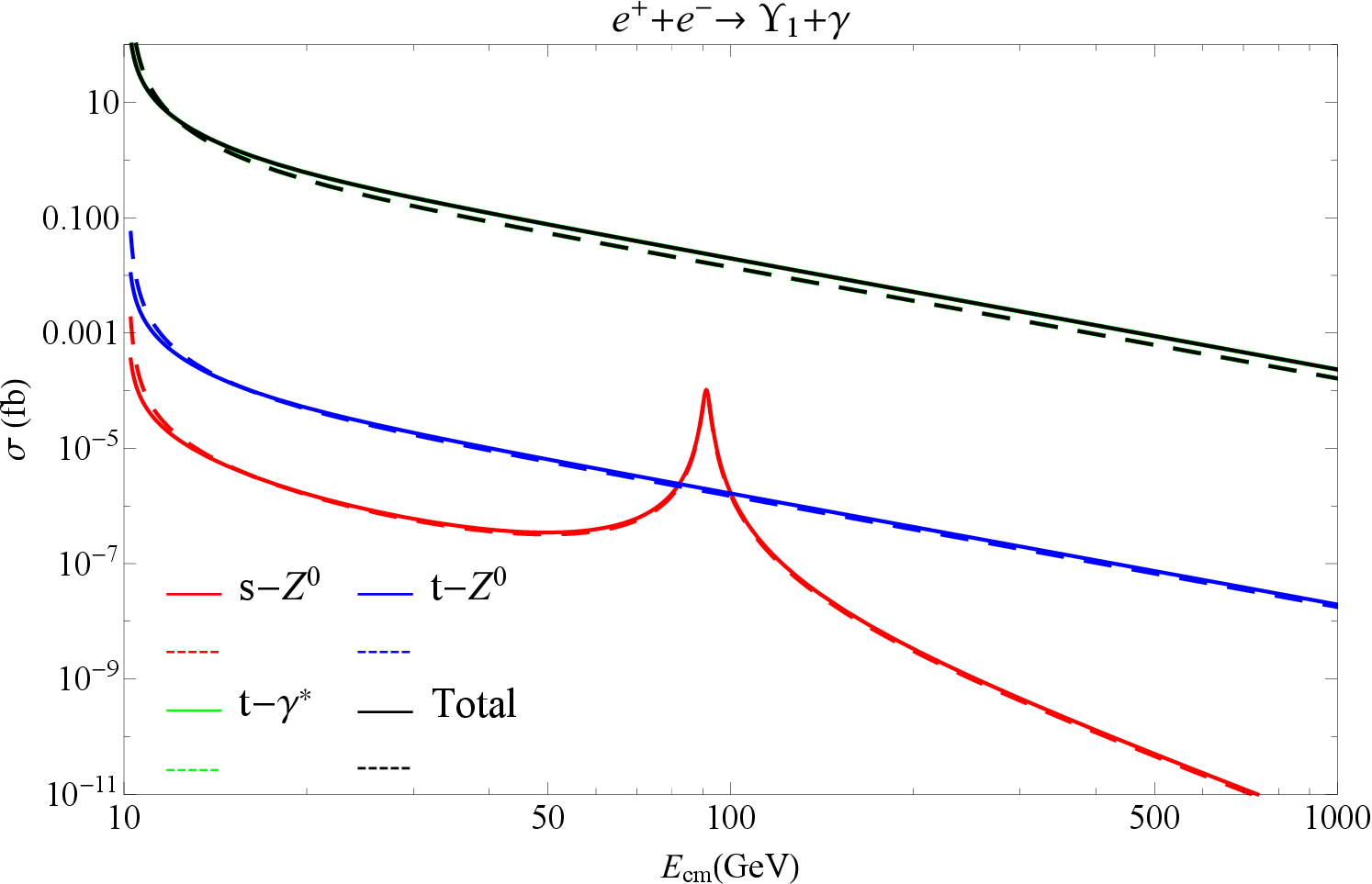}
			\end{tabular}
			\begin{tabular}{c c c }
				\includegraphics[width=0.333\textwidth]{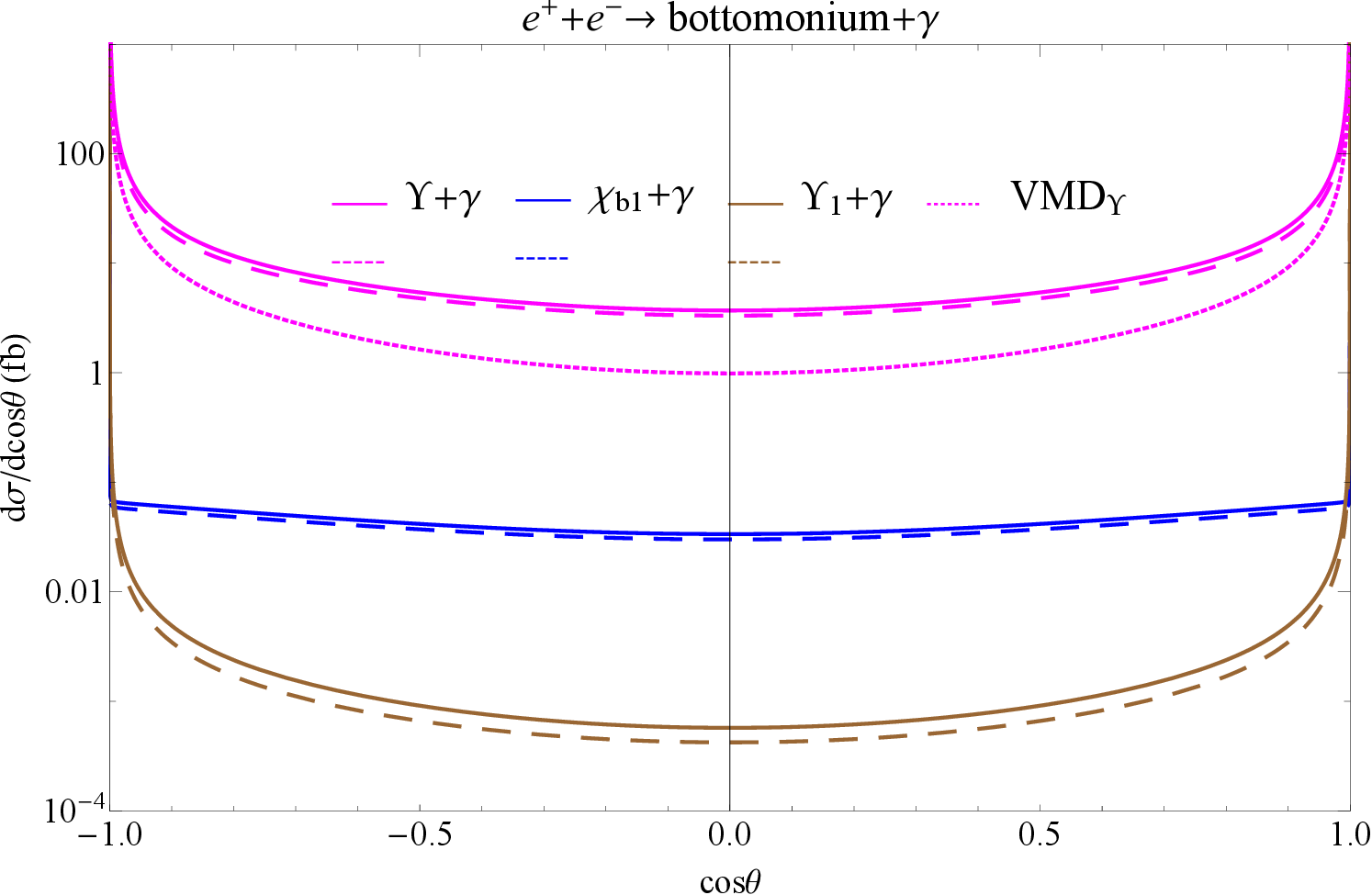}
				\includegraphics[width=0.333\textwidth]{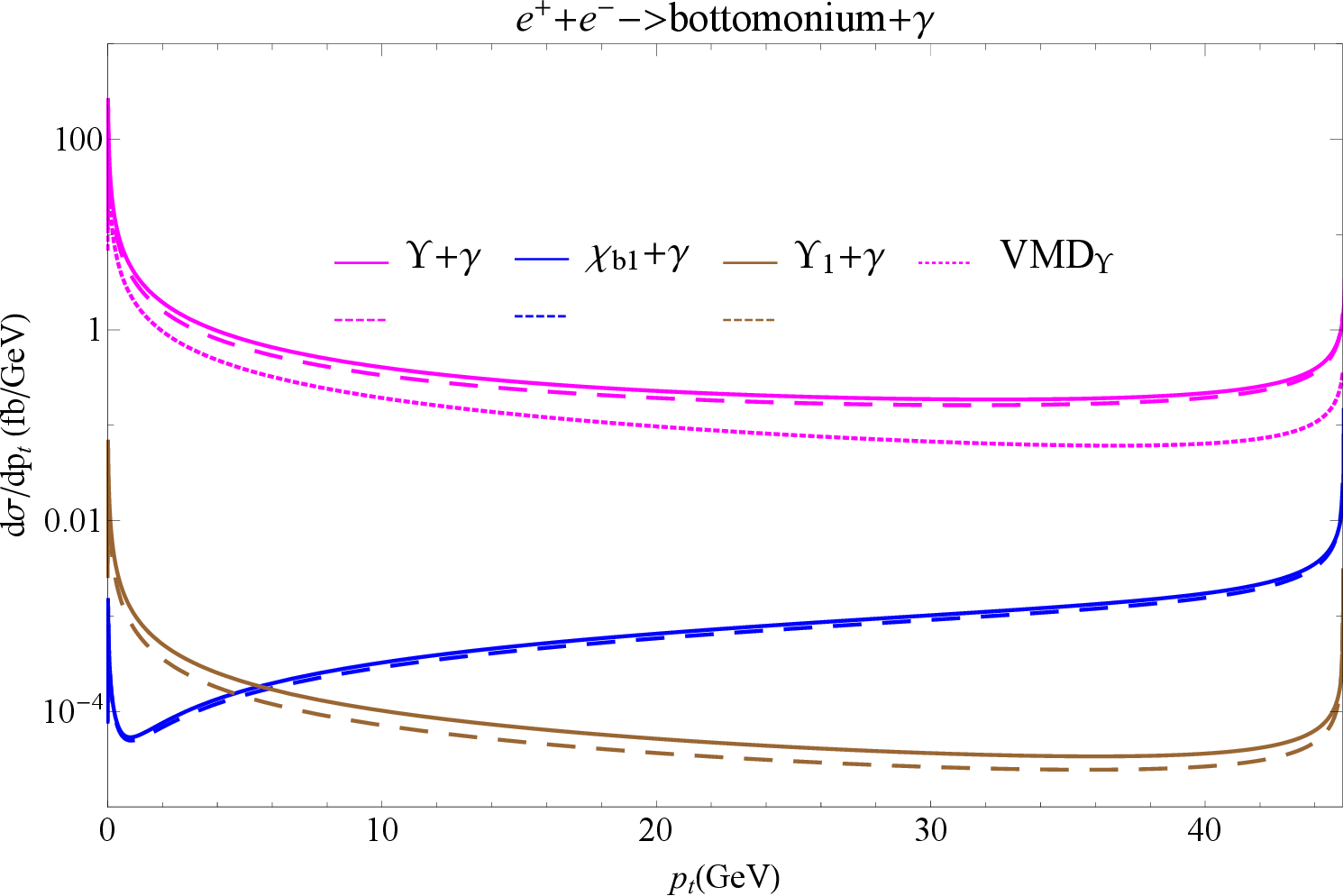}
				\includegraphics[width=0.333\textwidth]{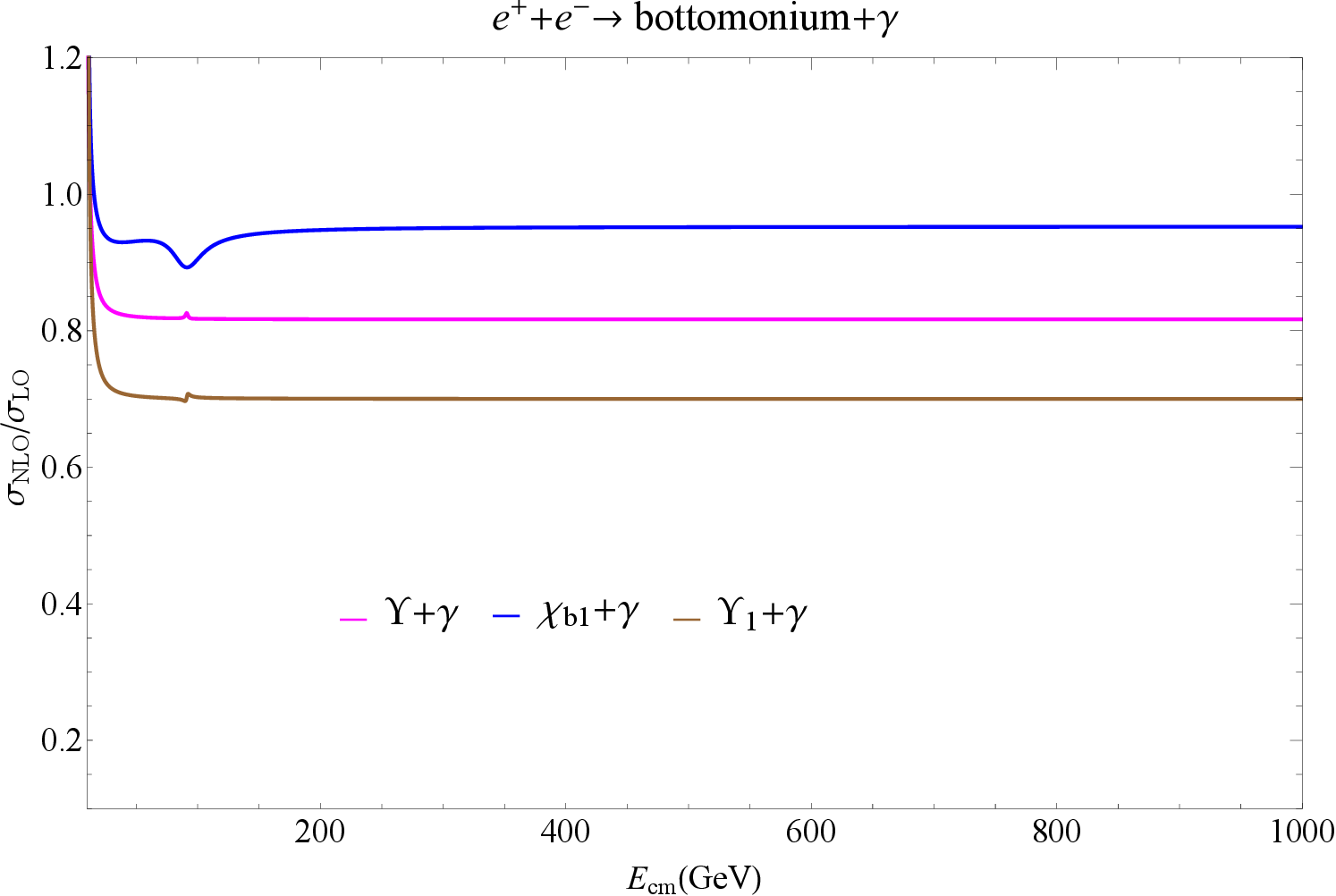}
			\end{tabular}
			\caption{
$^3L_1$ bottomonia production (including both $s$ and $t$-channel contributions). The notation is the same as in Fig.~\ref{ccc}.
}
			\label{bbb}
		\end{figure*}
	\end{widetext}

The total and differential cross sections for ${}^3L_1$ heavy-quarkonium production including both $s$- and $t$-channel contributions are presented in Figs.~\ref{ccc} and~\ref{bbb}.
Here, we do not present results for the pure $^3D_1$ charmonium state $\psi_1$, but instead show those for $\psi(3770)$.
Since the $\psi(3770)$ can be viewed as a $1^3D_1$ dominated state with a small admixture of $2^3S_1$\cite{Kuang:1989ub,Ding:1991vu,Rosner:2001nm,He:2008xb} as,  $|\psi(3770)\rangle=\cos\theta|1^3D_1\rangle+\sin\theta|2^3S_1\rangle$, where $\theta$ is the  mixing angle and we set it to be $-20^{\circ}$ in the calculation\footnote{
Refs.~\cite{Kuang:1989ub} reported the $S$-$D$ mixing angles $\theta=10^\circ$ and $-30^\circ$. Subsequent work in Ref.~\cite{Ding:1991vu} yielded multiple solutions $\pm 19^\circ$, $10^\circ$, $-30^\circ$, $13^\circ$, and $-26^\circ$; however, the $\pm 19^\circ$ results were derived by completely neglecting the $D$-wave contribution, while the $-30^\circ$ solution is experimentally disfavored. Ref.~\cite{Rosner:2001nm} extracted the $S$-$D$ mixing angles $\theta=(12\pm2)^\circ$ and $(-27\pm2)^\circ$. Nevertheless, the branching fraction $\mathcal{B}(\psi(3770)\to\rho\pi)$ predicted with $\theta=12^\circ$ significantly exceeds the latest experimental upper limit $\mathcal{B}(\psi(3770)\to\rho\pi)<5\times10^{-6}$~\cite{ParticleDataGroup:2024cfk}. Adopting Eqs.~(3) and (4) from Ref.~\cite{Rosner:2001nm} together with the parameter set utilized in this work, including $R_{2S}(0)$, $R''_{1D}(0)$, $m_c$, $M_{\psi(3770)}$, $M_{\psi(3686)}$, $\Gamma(\psi(3770)\to e^+e^-)$, and $\Gamma(\psi(3686)\to e^+e^-)$, we refit the experimental ratio
\[\frac{M_{\psi(3770)}^2\Gamma(\psi(3770)\to e^+e^-)}{M_{\psi(3686)}^2\Gamma(\psi(3686)\to e^+e^-)}=\left|\frac{0.892\sin\theta+0.062\cos\theta}{0.892\cos\theta-0.062\sin\theta}\right|^2=0.117,\]
and obtain two viable mixing-angle solutions, $\theta=12^\circ$ and $\theta=-20^\circ$. Of these, $\theta=12^\circ$ is disfavored by the above $\mathcal{B}(\psi(3770)\to\rho\pi)$ bound, so we adopt $\theta=-20^\circ$ for our main calculations.
}. 
The cross section of $\psi(3770)$ may be expressed as \begin{eqnarray}
\sigma_{\psi(3770)}=\cos^2\theta\,\sigma_{1^3D_1}+\sin^2\theta\,\sigma_{2^3S_1}+2\sin\theta\cos\theta\,\sigma_{SD},
\label{eq:mixsigma}
\end{eqnarray}
where the last term stands for the $S$-$D$ interference contribution between the $1^3D_1$ and $2^3S_1$ components, and the relevant LDME is $\frac{R_{2S}(0)}{\sqrt{4\pi}}\sqrt{\frac{15}{8\pi}}R''_{1D}(0)$.
	The $t$-channel $Z^0$-fragmentation contribution is negligible for all processes due to suppression from the $Z^0$ propagator.
From the upper-middle panel of Fig.\ref{ccc}, for $\chi_{c1}$ production at high energies ($\sqrt{s}>500$ GeV), the $t$-channel $Z^0$-fragmentation contribution ($R\simeq-0.5$) crosses over the $s$-channel counterpart ($R=-\frac{11}{10}$). Consequently, the magnitude of the relativistic correction decreases with increasing collision energy, and the K-factor ($\frac{\sigma_{\text{NLO}}}{\sigma_{\text{LO}}}$) gradually approaches $1-0.5v^2\simeq 0.9$ as the energy grows (see the blue line in the lower-right panel of Fig.\ref{ccc}).
	For $1^{--}$ heavy‑quarkonia ($J/\psi$, $\Upsilon$, $\psi_{1}$, $\Upsilon_1$), the $t$-channel photon‑fragmentation contribution dominates over its $s$-channel counterpart, especially for charmonium. The ratio of the $t$-channel (ISR) photon‑fragmentation cross section to the $s$-channel $Z^0$-propagated cross section at the $Z^0$ pole, namely $\big(\frac{\sigma_{t-\gamma}}{\sigma_{s-Z}}\big)_{\sqrt{s}=m_Z}$, 
takes values of about $194$, $2700$, $166$, $20$,  and $229$ for $J/\psi$, $\psi_{1}$, $\psi(3770)$, $\Upsilon$, and $\Upsilon_{1}$ production, respectively.		
	
	\begin{table}
		\caption{The SDC ratio R at the high energy limit ($\sqrt{s}\gg m_Z$) for ${}^3L_1$ heavy quarkonium production. Here we define $\xi=m_Q/m_Z, \zeta=\Gamma_Z/m_Z$. The last two columns show the total cross sections (the sum of s-channel and t-channel production cross sections, and including all cross-terms) at $\sqrt{s}=m_Z$. }
		\begin{tabular}{|c| |c|c|c|c|c| |c|c|}
			\hline
			~&\multicolumn{2}{c|}{s-channel}&\multicolumn{2}{c| }{t-channel}&Total &\multicolumn{2}{c|}{$\sigma_{\text{Tot}}$ (units:$ fb$) }\\
			\hline
			\hline
			~&$\gamma$-propagator&$Z^0$-propagator&$\gamma$-fragmentation&$Z^0$-fragmentation&&LO&NLO$(v^2)$\\
			\hline
			\hline
			$J/\psi$&~&$-\frac{1}{6}$&$-\frac{11}{6}$&$\frac{-176\xi^4+40\xi^2+1+\zeta^2}{6(16\xi^4-8\xi^2+1+\zeta^2)}\simeq0.17$&$-1.83$&1796 &1043 \\		
			\hline
			$\chi_{c1}$&$-\frac{11}{10}$&$-\frac{11}{10}$&~&$\frac{-80\xi^4+24\xi^2-1-\zeta^2}{2(16\xi^4-8\xi^2+1+\zeta^2)}\simeq-0.50$&$-0.50$&0.16 & 0.12\\
			\hline
			$\psi_{1}$&~&$-\frac{6}{7}$&$-3$&$\frac{-48\xi^4+16\xi^2-1-\zeta^2}{16\xi^4-8\xi^2+1+\zeta^2}\simeq-1.0$&$-3.0$& 2.26& 0.70\\
			\hline
			$\Upsilon$&~&$-\frac{1}{6}$&$-\frac{11}{6}$& $\frac{-176\xi^4+40\xi^2+1+\zeta^2}{6(16\xi^4-8\xi^2+1+\zeta^2)}\simeq0.19$&$-1.83$&94.6 & 78.1\\
			\hline
			$\chi_{b1}$&$-\frac{11}{10}$&$-\frac{11}{10}$&~&$\frac{-80\xi^4+24\xi^2-1-\zeta^2}{2(16\xi^4-8\xi^2+1+\zeta^2)}\simeq-0.48$&$-0.48$&0.089 & 0.079\\
			\hline
			$\Upsilon_{1}$&~&$-\frac{6}{7}$&$-3$&$\frac{-48\xi^4+16\xi^2-1-\zeta^2}{16\xi^4-8\xi^2+1+\zeta^2}\simeq-0.97$&$-3.0$&0.024 & 0.017\\
			\hline
		\end{tabular}
		\label{3L1Rratio}
	\end{table}
	The ratios of the relativistic SDCs to the LO SDCs in the high-energy limit for the ${}^3L_1$ heavy quarkonia are listed in Table~\ref{3L1Rratio}. Within this limit, the $t$-channel $Z^0$-fragmentation dominates the cross section for the ${}^3P_1$ state, and the $t$-channel $\gamma^*$-fragmentation dominates for the ${}^3S_1$ and ${}^3D_1$ states, as seen in Figs.~(\ref{ccc},~\ref{bbb}). This behavior is also reflected by the $R$-ratio values of the total cross sections given in the sixth column of Table~\ref{3L1Rratio}, as compared with those of individual channels. The total cross sections at LO and NLO in the $v^2$ expansion for these ${}^3L_1$ quarkonia are also provided in the last two columns.

For comparison, we also adopt the vector-meson-dominance (VMD) method~\cite{Bodwin:2006yd,Bodwin:2007ga,Trunin:2024ibx} for processes dominated by $t$-channel $\gamma^*$-fragmentation. This approach effectively incorporates all $\alpha_s$ and $v^2$ corrections. 
	Specifically, we have
	\bea
	\frac{d\sigma^{\text{VMD}}_{H\gamma}}{d\cos\theta}&=&\frac{e^2g_V^2}{M^4}\frac{d\sigma_{\gamma^*\gamma}}{d\cos\theta}
	\eea
where $M$ is the physical mass of the meson $H$ and the effective coupling constant $g_V$ is extracted from the leptonic decay width of the vector meson via
\bea
\Gamma[H\rightarrow e^+e^-]&=& \frac{4\pi\alpha^2g_V^2}{3M^3}
\eea
With the coupling $g_V$ fixed, the full VMD differential cross section is given below. We introduce the dimensionless kinematic variables $\omega=\sqrt{s}/M, \rho=m_e/M, \kappa=m_e/\sqrt{s}$.
	\bea
	&\frac{d\sigma^{\text{VMD}} }{d\cos\theta}=\frac{6\pi \alpha\Gamma[H\rightarrow e^+e^-]}{s^{3/2}\omega^3(\omega^2-1)(1-(1-4\kappa^2)\cos^2\theta)^2}\cr
	&\quad\times\{\omega^2[(\omega^2+1)^2-4\omega^2\cos^2\theta-(\omega^2-1)^2\cos^4\theta]+8\rho^2[\omega^2(\omega^2-2)\sin^2\theta+(\omega^2-1)^2\cos^4\theta]\}
	\label{VMD}
	\eea
The VMD results for $J/\psi$, $\Upsilon$, and $\psi(3770)$ differential cross sections are also shown in Fig.~\ref{ccc} and Fig.~\ref{bbb} for comparison. It can be seen that the relativistic corrections bring the LO NRQCD results toward the VMD predictions.

We now present a summary of the production processes for all quarkonia considered.
The numerical results are collected in Table~\ref{comparis}. In this table, the second and third columns give the LO and $\text{NLO}(v^2)$ cross sections, respectively.
The fourth and fifth columns present the results of resummed relativistic corrections from Ref.~\cite{Trunin:2024ibx} and NLO results including QCD corrections adapted from Refs.~\cite{Chen:2013itc,Sun:2014kva}, respectively.
We rescale these QCD corrections to match the input parameters adopted in the present work, including $\alpha$, $|R_S(0)|^2$, $|R'_P(0)|^2$ and the heavy-quark mass $m_Q$. In this rescaling procedure, we account for variations of the heavy-quark mass: shifting $m_c$ from $1.5\,\text{GeV}$ to $1.75\,\text{GeV}$ reduces P-wave cross sections by $37\%$; for bottomonium, adjusting $m_b$ from $4.9\,\text{GeV}$ to $4.71\,\text{GeV}$ increases the $\eta_b$ cross section by $4\%$ and the $\Upsilon$ cross section by $12\%$. These scaling factors, describing the mass-induced cross-section variations, are estimated via separate LO calculations.
With this rescaling, we further evaluate the $\text{NLO}(v^2,\alpha_s)$ results, listed in the sixth column. 
The VMD predictions for the t-channel dominated processes are given in the seventh column.

It can be seen that relativistic corrections reduce the LO cross sections for charmonium, exhibiting qualitatively similar behavior to the resummation results. 
For reference, we obtain
$\sigma^{(v^2)}/\sigma_{\text{LO}}=-42\%$, $-19\%$, $-30\%$, $-25\%$, $-16\%$, $-30\%$
for $J/\psi$, $\eta_c$, $\chi_{c0}$, $\chi_{c1}$, $\chi_{c2}$, $h_c$, respectively. 
The pronounced discrepancy for bottomonium production arises from the different input values of
$\langle\boldsymbol{p}^2\rangle_{b\bar{b}}=-0.19~\mathrm{GeV}^2$
adopted in Ref.~\cite{Trunin:2024ibx}, in contrast to
$\langle\boldsymbol{p}^2\rangle_{b\bar{b}}\simeq2.2~\mathrm{GeV}^2$
used in the present work. 
From the table, the combined relativistic and QCD corrections yield a prediction comparable to the VMD result for $\Upsilon$ production.

	\begin{table}
		\caption{Total cross sections for heavy-quarkonium production in the process $e^+e^-\rightarrow H+\gamma$ at $\sqrt{s}=m_Z$.
$\sigma_{\text{LO}}$ denotes the LO result in both $v^2$ and $\alpha_s$ expansions. $\sigma_{\text{NLO}(v^2)}$ includes NLO relativistic corrections. $\sigma_{\text{Resum}}$ is the resummed relativistic result from Ref.~\cite{Trunin:2024ibx}. $\sigma_{\text{NLO}(\alpha_s)}$ is the cross sections incorporating QCD corrections from Refs.~\cite{Chen:2013itc,Sun:2014kva}. $\sigma_{\text{NLO}(v^2,\alpha_s)}$ corresponds to the NLO result including both $v^2$ and $\alpha_s$ corrections evaluated with the input parameters adopted in the present work. $\sigma_{\text{VMD}}$ stands for the VMD prediction.
}
		\begin{tabular}{c c c c c c c}
			\hline
			H & $\sigma_{LO}$&  $\sigma_{NLO(v^2)}$& $\sigma_{\text{Resum}}$\cite{Trunin:2024ibx}& $\sigma_{NLO(\alpha_s)}$\cite{Chen:2013itc} &$\sigma_{NLO(v^2,\alpha_s)}$ & $ \sigma_{\text{VMD}}$ \\
			\hline
			$J/\psi$&1796&1043 & $700^{+223}_{-188}$&  & & 561\\	
			$\psi(2S)$&520& 303 &  &  & &198 \\	
			$\eta_c$&1.38& 1.11&$0.64^{+0.13}_{-0.13}$ &0.929  &1.37&  \\	
			$h_{c}$&0.53& 0.37&$0.29^{+0.22}_{-0.15}$ & 0.157&0.02 &  \\	
			$\chi_{c0}$&0.026& 0.018 &$0.014^{+0.011}_{-0.007}$& 0.023&0.0175 & \\
			$\chi_{c1}$&0.16&0.12  & $0.089^{+0.068}_{-0.047}$ &0.157 & 0.142 & \\
			$\chi_{c2}$&0.053&0.044 & $0.032^{+0.025}_{-0.017}$&0.010 & 0.005 & \\
			$\eta_{c2}$&0.0045&0.0026 & & &  &\\
			$\psi_{1}$&2.26&0.70  & & & &\\ 
				$\psi(3770)$&84.7&  45.8&  &  && 21.7\\	
			$\psi_{2}$&0.045& 0.028 & & & & \\	
			$\psi_{3}$&0.018& 0.013& & &  &\\
			\hline
			H & $\sigma_{LO}$&  $\sigma_{NLO(v^2)}$& $\sigma_{\text{Resum}}$\cite{Trunin:2024ibx}& $\sigma_{NLO(\alpha_s)}$\cite{Sun:2014kva} &$\sigma_{NLO(v^2,\alpha_s)}$  & $ \sigma_{\text{VMD}}$ \\
			\hline
			$\Upsilon$&94.6&78.1 &$71^{+8}_{-8}$ &36.51  & 51& 44.8 \\	
			$\eta_b$&2.15&1.97 &$1.58^{+0.13}_{-0.13}$&1.25 &  1.936  &\\	
			$h_{b}$&0.090& 0.078&$0.11^{+0.01}_{-0.01}$ &0.0414& 0.037  & \\	
			$\chi_{b0}$&0.014&  0.012&$0.017^{+0.002}_{-0.002}$ &0.0115 & 0.0126  &\\
			$\chi_{b1}$&0.089& 0.079 &$0.11^{+0.01}_{-0.01}$ &0.0683 & 0.0744 & \\
			$\chi_{b2}$&0.030& 0.028&$0.036^{+0.003}_{-0.003}$ &0.0117 & 0.012 & \\
			$\eta_{b2}$&0.0010& 0.0009& & & & \\
			$\Upsilon_{1}$&0.024& 0.017 & & &  &\\ 	
			$\Upsilon_{2}$&0.0033 &0.0028 & & &  &\\	
			$\Upsilon_{3}$&0.0013& 0.0012&& &  &\\
			\hline
		\end{tabular}
		\label{comparis}
	\end{table}

	\subsection{Events}

	Using the cross sections presented in Tables~\ref{TCS} and~\ref{3L1Rratio}, we can estimate the event yields at the CEPC and FCC‑ee, whose integrated luminosities are 
	$16~\text{ab}^{-1}$ (2 IPs in 2 yrs)\cite{CEPCStudyGroup:2018ghi} and $150~\text{ab}^{-1}$ (4 IPs in 4 yrs)\cite{Agapov:2022bhm}, respectively. The results are shown in Table~\ref{events}. It can be seen that the event yields for 
$\eta_Q, J/\psi, \Upsilon, \psi_{1}, \psi(2S)$ and $\psi(3770)$
	are all above 
	$10^4$. 
Therefore, future $Z$-factory experiments will provide a promising platform to study heavy quarkonium production in association with a photon, although significant backgrounds arise from the subprocesses $e^+e^-\to q+\bar{q}+\gamma$ ($q=u,d,s,c,b$)~\cite{Chung:2008km,Sun:2014kva}, together with feeddown contributions from higher excited states.

	Furthermore, using the decay branching fractions of the final state quarkonium \cite{ParticleDataGroup:2024cfk}, the reconstructed event yield can be estimated as
	\bea
	N = \sigma \times \mathcal{L} \times Br \times \epsilon,
	\eea
	where $\mathcal{L}$ is the integrated luminosity, $Br$ is the branching fraction for the decay mode of  $H$, and $\epsilon$ is the total detection efficiency. The reconstructed events are listed in Table~\ref{Revents}. 
With substantial negative corrections for $h_c$ and $\chi_{Q2}$ production, the expected event yields reduce to $\mathcal{O}(1)$ for $\chi_{Q2}$ and $\mathcal{O}(10^2)$ for $h_c$, respectively. The $h_c$ state should still be accessible at a future super Z factory.

	\begin{table}
		\caption{ 
The expected production event yields at CEPC and FCC-ee. Values outside and inside the parentheses correspond to the LO and NLO results, respectively. For $\eta_Q$, $h_Q$, $\chi_{Q0,1,2}$, and $\Upsilon$, the NLO results include both relativistic and one-loop QCD corrections; for all other states, only relativistic corrections are included.
}
		\begin{tabular}{|c|c|c||c|c|c|}
			\hline
			charmonium& CEPC ($16~\text{ab}^{-1}$)& FCC-ee($150~\text{ab}^{-1}$)& bottomonium&  CEPC($16~\text{ab}^{-1}$)&FCC-ee($150~\text{ab}^{-1}$) \\
			\hline
			\hline
			$N_{\eta_c+\gamma}$&  $2.2\times10^4(2.2\times10^4)$  &$2.1\times10^5(2.1\times10^5)$   &	$N_{\eta_b+\gamma}$&$3.4\times10^4(3.1\times10^4)$ &    $3.2\times10^5(2.9\times10^5)$	\\
			\hline
			$N_{J/\psi+\gamma}$& $2.9\times10^7(1.7\times10^7)$& $2.7\times10^8(1.6\times10^8)$ &  	 $N_{\Upsilon+\gamma}$& $1.5\times10^6(8.2\times10^5)$ &  $1.4\times10^7(7.7\times10^6)$ 	\\
			\hline
			$N_{h_c+\gamma}$& $8401(320)$ & $7.9\times10^4(3000)$ &  		$N_{h_b+\gamma}$ &  $1443(592)$&   $1.4\times10^4(5550)$ 	\\
			\hline
			$N_{\chi_{c0}+\gamma}$& $416(280)$ & $3902(2625)$  &  	$N_{\chi_{b0}+\gamma}$&  $217(202)$&  $2035(1890)$ 	\\
			\hline
			$N_{\chi_{c1}+\gamma}$&$2524(2272)$  & $2.4\times10^4(2.1\times10^4)$ &  	$N_{\chi_{b1}+\gamma}$& $1425(1190)$  &  $1.3\times10^4(1.1\times10^4)$  	\\
			\hline
			$N_{\chi_{c2}+\gamma}$& $843(80)$& $7908(750)$  &  	$N_{\chi_{b2}+\gamma}$& $483(192)$  &   $4533(1800)$	\\
			\hline
			$N_{\eta_{c2}+\gamma}$& $72(42)$ & $670(395)$ &  	$N_{\eta_{b2}+\gamma}$&   $17(14)$&  $156(128)$ 	\\
			\hline
			$N_{\psi_{1}+\gamma}$& $3.6\times10^4(1.1\times10^4)$& $3.4\times10^5(1.1\times10^5)$ &  		$N_{\Upsilon_{1}+\gamma}$&  $378(266)$  &  $3546(2493)$ 	\\
			\hline
			$N_{\psi_{2}+\gamma}$&$721(449)$ & $6756(4208)$ &  	$N_{\Upsilon_{2}+\gamma}$& $54(46)$  & $510(427)$   	\\
			\hline
			$N_{\psi_{3}+\gamma}$& $287(207)$& $2690(1939)$ &  		$N_{\Upsilon_{3}+\gamma}$&$21(18)$   &  $197(173)$	\\
			\hline
				$N_{\psi(2S)+\gamma}$&   $8.3\times10^6(4.8\times10^6)$   &$7.8\times10^7(4.5\times10^7)$     &  		 &   &  	\\
			\hline
		$N_{\psi(3770)+\gamma}$&  $1.4\times10^6(7.3\times10^5)$     &   $1.3\times10^7(6.9\times10^6)$  &  		 &   &  	\\
			\hline
		\end{tabular}
		\label{events}
	\end{table}
	
	\begin{table}
		\caption{
Number of reconstructed events at CEPC and FCC-ee. Values outside and inside the parentheses correspond to the LO and NLO results, respectively. For $\eta_Q$, $h_Q$, $\chi_{Q0,1,2}$, and $\Upsilon$, the NLO results include both relativistic and one-loop QCD corrections; for all other states, only relativistic corrections are included. A total detection efficiency of $\epsilon=20\%$ is assumed.
}
		\begin{tabular}{|c|c|c|c|c|}
			\hline
			process	& quarkonium decay mode& Br&  $N_{\text{CEPC}}$ & $N_{\text{FCC-ee}}$ \\
			\hline
			\hline
			${\eta_c+\gamma}$&  hadronic decay   &43\%  & 1899(1885)&17799(17673) \\
			\hline
			${J/\psi+\gamma}$& $l^+l^-(l=e,\mu)$ & 12\%  & $6.9\times10^5(4.0\times10^5)$& $6.5\times10^6(3.8\times10^6)$\\
			\hline
			${\psi(2S)+\gamma}$& $l^+l^-(l=e,\mu)$ & 1.6\%  & $2.7\times10^4(1.6\times10^4)$& $2.5\times10^5(1.5\times10^5)$\\
			\hline
			${h_c+\gamma}$& $\eta_c\gamma\rightarrow(\text{hadron decay})+\gamma$  & 21.5\%  &361(14) &3387(129) \\
			\hline
			${\chi_{c0}+\gamma}$& $2\pi^+2\pi^-$  & 2.2\%  &2(1) &17(12) \\
			\hline
			${\chi_{c1}+\gamma}$& $\gamma l^+l^-(\gamma J/\psi)$  & 4.1\%  & 21(19)&194(175) \\
			\hline
			${\chi_{c2}+\gamma}$& $\gamma l^+l^-(\gamma J/\psi)$ & 2.3\%  &4(0) &36(3) \\
			\hline
			${\eta_{c2}+\gamma}$& $\gamma\gamma K\bar{K}\pi$  & 1.5\%  & 0(0)&2(1) \\
			\hline
			${\Upsilon+\gamma}$& $l^+l^-(l=e,\mu)$  & 4.86\%  & $1.5\times10^4(7932)$&$1.4\times10^5(7.4\times10^4)$ \\
			\hline
			$\chi_{b0}+\gamma$& $\gamma l^+l^-(\gamma\Upsilon)$ &0.09 \%  &0(0) &0(0) \\
			\hline
			$\chi_{b1}+\gamma$& $\gamma l^+l^-(\gamma\Upsilon)$  & 1.71\%  & 5(4)&46(38) \\
			\hline
			$\chi_{b2}+\gamma$& $\gamma l^+l^-(\gamma\Upsilon)$  &0.87 \%  &1(0) &8(3) \\
			\hline
		\end{tabular}
		\label{Revents}
	\end{table}

	\subsection{Theoretical uncertainty}
The main theoretical uncertainty originates from the LDMEs \cite{Liao:2021ifc}.
As reported in Ref.~\cite{Liao:2014rca},
we have $|R_{(c\bar{c})S}(0)|^2\in(0.565, 2.458)$, $|R_{(c\bar{c})P}'(0)|^2\in(0.053, 0.329)$, $|R_{(c\bar{c})D}''(0)|^2\in(0.012, 0.099)$, $|R_{(b\bar{b})S}(0)|^2\in(3.071, 16.12)$, $|R_{(b\bar{b})P}'(0)|^2\in(1.111, 5.874)$, $|R_{(b\bar{b})D}''(0)|^2\in(0.421, 4.469)$, with units $\mathrm{GeV}^{3+2l}$.
Consequently, the uncertainties induced by the LDMEs could be as follows: 
for S‑wave charmonium production, from $-62.0\%$ to $65.5\%$; for P‑wave charmonium, from $-47.4\%$ to $226.7\%$; for D‑wave charmonium, from $-24.5\%$ to $522.6\%$; for S‑wave bottomonium, from $-66.6\%$ to $75.4\%$; for P‑wave bottomonium, from $-29.1\%$ to $275.0\%$; and for D‑wave bottomonium, from $-33.4\%$ to $606.6\%$.
	
	We take $m_c\pm0.1 \text{GeV}$ and $m_b\pm0.2 \text{GeV}$ to see the uncertainties caused by the heavy quark mass. The results are shown in Table~\ref{massuncert}, 
from which one sees that the uncertainties are about $-6\% \sim +7\%$ for $\eta_c$ production, $-18\% \sim +23\%$ for $J/\psi$ production, $-15\% \sim +19\%$ for $\psi(2S)$ and P-wave charmonia production, $-16\% \sim +21\%$ for $\psi(3770)$ production, $-30\% \sim +46\%$ for $\psi_{1}$ production, and $-23\% \sim +31\%$ for other D-wave charmonia production; $-4\% \sim +5\%$ for $\eta_b$ production, $-11\% \sim +13\%$ for $\Upsilon$ and P-wave bottomonia production, $-24\% \sim +29\%$ for $\Upsilon_{1}$ production, and $-18\% \sim +22\%$ for other D-wave bottomonia production.

To estimate the uncertainties caused by a possible deviation of the c.m.\ energy from the $Z^0$ pole, we vary $E_{\text{cm}}$ by $\pm0.5 \text{GeV}$. The results are presented in Table~\ref{Ecmdev}. A $\pm0.5 \text{GeV}$ shift in the collision energy yields about $14\%$ changes in the total cross sections for most charmonium and bottomonium production channels, while for $J/\psi$, $\psi(2S)$, $\psi_{1}$, $\psi(3770)$, $\Upsilon$, and $\Upsilon_1$, the variation is less than about $2\%$. This observation is consistent with Ref.~\cite{Sun:2014kva}.

	\begin{table}
		\caption{ Total cross sections (in fb) up to $\mathcal{O}(v^2)$ at $\sqrt{s}=m_Z$ for $m_c\pm0.1 \text{GeV}$ and  $m_b\pm0.2 \text{GeV}$. Here, we define $m_{c\pm}=m_c\pm0.1 \text{GeV}$, $m_{b\pm}=m_b\pm 0.2\text{GeV}$, and the $m_{c,b}$ means the central value of quark mass, as listed in Table~\ref{inmass}. For the ${}^3L_1(L=S,P,D)$ heavy quarkonia, the t-channel contributions are also included. }
		\begin{tabular}{|c||c|c|c||c|c|c|c|}
			\hline
			&$m_{c-}$&$m_c$ & $m_{c+}$& &$m_{b-}$ &$m_b$&$m_{b+}$ \\
			\hline
			\hline
			$\sigma_{\eta_c+\gamma}$& 1.195&1.115 &1.045 &	$\sigma_{\eta_b+\gamma}$&2.062 &1.972 &1.890\\
			\hline
			$\sigma_{J/\psi+\gamma}$&1281 &1043 &860& $\sigma_{\Upsilon+\gamma}$&88.5&78.1&69.4\\
			\hline
			$\sigma_{h_c+\gamma}$&0.439 &0.368 & 0.311&	$\sigma_{h_b+\gamma}$&0.089 &0.078 &0.070\\
			\hline
			$\sigma_{\chi_{c0}+\gamma}$& 0.022&0.018 &0.015 &$\sigma_{\chi_{b0}+\gamma}$&0.013 &0.012 &0.010\\
			\hline
			$\sigma_{\chi_{c1}+\gamma}$& 0.141&0.118 &0.100 &$\sigma_{\chi_{b1}+\gamma}$&0.090 &0.079 &0.071\\
			\hline
			$\sigma_{\chi_{c2}+\gamma}$&0.053 &0.044 &0.037 &$\sigma_{\chi_{b2}+\gamma}$&0.032&0.028&0.025\\
			\hline
			$\sigma_{\eta_{c2}+\gamma}$& 0.0035& 0.0026&0.0020 &$\sigma_{\eta_{b2}+\gamma}$&0.0010 &0.0009&0.0007\\
			\hline
			$\sigma_{\psi_{1}+\gamma}$&1.025 &0.702 &0.491 &	$\sigma_{\Upsilon_{1}+\gamma}$&0.022 &0.017 &0.013\\
			\hline
			$\sigma_{\psi_{2}+\gamma}$&0.037 &0.028  &0.022 &	$\sigma_{\Upsilon_{2}+\gamma}$&0.0035&0.0028&0.0024\\
			\hline
			$\sigma_{\psi_{3}+\gamma}$&0.017 &0.013 &0.010 &	$\sigma_{\Upsilon_{3}+\gamma}$&0.0014&0.0012 &0.0010\\
			\hline
		$\sigma_{\psi(2S)+\gamma}$& 357 & 303& 259 &	 & &  & \\
			\hline
		$\sigma_{\psi(3770)+\gamma}$&55.5  &45.8 & 38.2 &	 & &  & \\
			\hline
		\end{tabular}
		\label{massuncert}
	\end{table}

	\begin{table}
		\caption{Total cross sections (in fb) up to $\mathcal{O}(v^2)$ of different heavy quarkonium   production  under three collision energies $E_{cm}=m_Z-0.5\text{GeV}, m_Z, m_Z+0.5\text{GeV}$. For the ${}^3L_1(L=S,P,D)$ heavy quarkonia, the t-channel contributions are also included.}
		\begin{tabular}{|c|c|c|c| |c|c|c|c|}
			\hline
			charmonium &\multicolumn{3}{c| |}{$\sigma$}&bottomonium &\multicolumn{3}{c|}{$\sigma$}\\
			\hline
			\hline
			$\sqrt{s}$&$m_{Z}-0.5$ GeV&$ m_Z$&$m_{Z}+0.5$ GeV&$\sqrt{s}$&$m_{Z}-0.5$ GeV &$m_Z$&$m_{Z}+0.5$ GeV \\
			\hline
			\hline
			$\eta_c$&0.957 &1.115 & 0.970&$\eta_b$& 1.697&1.972 &1.702 \\
			\hline
			$J/\psi$&1052 &1043 &1031 & $\Upsilon$& 78.6&78.1 &76.4 \\
			\hline
			$h_{c}$&0.317 &0.368 &0.317 &	$h_{b}$ & 0.068&0.078 &0.067 \\
			\hline	
			$\chi_{c0}$& 0.016& 0.018&0.016 & 	$\chi_{b0}$& 0.010& 0.012& 0.010\\
			\hline	
			$\chi_{c1}$& 0.101&0.118 &0.103 &$\chi_{b1}$ &0.068 &0.079 &0.069 \\
			\hline
			$\chi_{c2}$&0.038 &0.044 &0.038 & $\chi_{b2}$& 0.024&0.028 &0.024 \\
			\hline
			$\eta_{c2}$&0.0023 &0.0026 &0.0023 & $\eta_{b2}$& 0.0007& 0.0009&0.0007 \\
			\hline
			$\psi_{1}$& 0.724&0.702 &0.680 &$\Upsilon_{1}$ &0.0168 &0.0166 & 0.0165\\
			\hline	
			$\psi_{2}$& 0.024& 0.028& 0.024& $\Upsilon_{2}$& 0.0025&0.0028 &0.0024 \\
			\hline	
			$\psi_{3}$&0.011 &0.013 &0.011 &$\Upsilon_{3}$ & 0.0010& 0.0012&0.0010 \\
			\hline
				$\psi(2S)$&305 &303 &299 &  &  &   &   \\
			\hline
				$\psi(3770)$&  46.1 & 45.8  &  45.2 &  &  &   &   \\
			\hline
		\end{tabular}
		\label{Ecmdev}
	\end{table}


	\section{Summary}
	\label{sumary}

	In this work, we study the $\mathcal{O}(v^2)$ relativistic corrections to heavy quarkonium plus photon production at the Z factory within the framework of NRQCD. S-, P-, and D-wave heavy quarkonium states are considered. Both the $s$-channel and $t$-channel processes are comprehensively analyzed. We find that the relativistic corrections are substantial for charmonium production, reducing the LO cross sections by roughly $20\%$–$70\%$, while they are mild for bottomonium production, with a reduction of $10\%$–$30\%$. We also compare our results with those obtained from the VMD model and from the resummed relativistic corrections~\cite{Trunin:2024ibx}. The relativistic corrections shift the LO cross section toward the VMD result. For charmonium, our findings agree with the resummation approach, whereas for bottomonium, discrepancies arise from the different input values of $\langle\boldsymbol{p}^2\rangle_{b\bar{b}}$.

Both the LO and NLO results suggest that the future super Z factory could serve as a favorable platform for investigating the $e^+e^-\rightarrow H(Q\bar{Q})+\gamma$ processes, for which the production event yields exceed $10^4$ for many heavy quarkonium states. 
For D‑wave states, our calculations indicate that the production event yields are non‑negligible for $\psi_{J}\,(J=1,2,3)$, $\psi(3770)$, and $\Upsilon_{1}$ at FCC‑ee, spanning roughly from a few times $10^{3}$ up to $10^{7}$. Thus, future $e^+e^-$ colliders may enable their observation. 
The theoretical uncertainties originating from LDMEs, heavy quark masses, and deviations of the collision energy from $m_Z$ are also discussed.

	The NNLO QCD corrections have been computed for the processes 
	$e^+e^-\rightarrow\eta_{Q}(\chi_{c0,1,2})+\gamma$ at BESIII and B-factory energies \cite{Chen:2017pyi,Yu:2020tri,Sang:2020fql,Li:2025pbt}. It would also be meaningful to extend such calculations to the Z factory energy scale. Furthermore, including the color-octet production channels could improve the reliability of the predictions.

	\section{Acknowledgements}  
    X.P. Wang would like to thank A. Trunin for helpful discussions. 
	This work was supported by the National Natural Science Foundation of China (Grant Nos. 11705078 and 12575087).

		\section{Appendix}
	
	We present the LO and relativistic-corrected SDCs, i.e., $F(n)$ and $G(n)$, in Eqs.~(\ref{eqb}--\ref{eqe}). We list the $\gamma$- and $Z^0$-propagated channels for $s$-channel production, while only the $\gamma$-fragmentation channel is displayed for $t$-channel ${}^3L_1$ production. Here, the subscript ``$\gamma/Z$'' denotes the corresponding $\gamma/Z$-propagated process. The remaining channels are omitted for brevity.
	
	For s-channel production, we consider
	
	$\eta_Q$ production,
    \bea
    F({}^1S_0)_{\gamma}&=&\frac{6144\pi^3\alpha^3 e_Q^4\big[16m_Q^4-8m_Q^2(s+t)+s^2+2st+2t^2\big]}{s(s-4m_Q^2)^2} \label{eqb}\\	 
    G({}^1S_0)_{\gamma}&=&-\frac{5120\pi^3\alpha^3 e_Q^4\big[16m_Q^4-8m_Q^2(s+t)+s^2+2st+2t^2\big]}{s\,m_Q^2(s-4m_Q^2)^2}\\
    F({}^1S_0)_{Z}&=&\frac{24\pi^3\alpha^3 e_Q^2 s\big[(1-4\sin^2\theta_w)^2+1\big](1-4|e_Q|\sin^2\theta_w)^2\big(16m_Q^4-8m_Q^2(s+t)+s^2+2st+2t^2\big)}
    {\cos^4\theta_w\sin^4\theta_w\,(s-4m_Q^2)^2\big[(s-m_Z^2)^2+m_Z^2\Gamma_Z^2\big]}\\
    G({}^1S_0)_{Z}&=&-\frac{20\pi^3\alpha^3 e_Q^2 s\big[(1-4\sin^2\theta_w)^2+1\big](1-4|e_Q|\sin^2\theta_w)^2\big(16m_Q^4-8m_Q^2(s+t)+s^2+2st+2t^2\big)}
    {\cos^4\theta_w\sin^4\theta_w\,m_Q^2(s-4m_Q^2)^2\big[(s-m_Z^2)^2+m_Z^2\Gamma_Z^2\big]}
    \eea
    
	$\psi/\Upsilon$ production,
    \bea
    F({}^3S_1)_{Z}&=&\frac{24\pi^3\alpha^3 e_Q^2\big[(1-4\sin^2\theta_w)^2+1\big]
    \big[16m_Q^4(s+4t)-8m_Q^2(s^2+3st+2t^2)+s(s^2+2st+2t^2)\big]}
    {\cos^4\theta_w\sin^4\theta_w\,(s-4m_Q^2)^2\big((s-m_Z^2)^2+m_Z^2\Gamma_Z^2\big)}\\
    G({}^3S_1)_{Z}&=&4\pi^3\alpha^3 e_Q^2\big[(1-4\sin^2\theta_w)^2+1\big]
    \big[192m_Q^6(3s+4t)-16m_Q^4(19s^2+10st+12t^2)+m_Q^2(44s^3-8st^2)\cr&-&s^2(s^2+2st+2t^2)\big]
    \big/\big[\cos^4\theta_w\sin^4\theta_w\,m_Q^2(s-4m_Q^2)^3\big((s-m_Z^2)^2+m_Z^2\Gamma_Z^2\big)\big]
    \eea
    
	$h_Q$ production,
    \bea
    F({}^1P_1)_{Z}&=&\frac{24\pi^3\alpha^3 e_Q^2\big[(1-4\sin^2\theta_w)^2+1\big]
    \big[16m_Q^4(s+4t)-8m_Q^2(s^2+3st+2t^2)+s(s^2+2st+2t^2)\big]}
    {\cos^4\theta_w\sin^4\theta_w\,m_Q^2(s-4m_Q^2)^2\big((s-m_Z^2)^2+m_Z^2\Gamma_Z^2\big)}\\
    G({}^1P_1)_{Z}&=&-\big\{12\pi^3\alpha^3 e_Q^2\big[(1-4\sin^2\theta_w)^2+1\big]
    \big[16m_Q^4(13s+44t)-8m_Q^2(13s^2+35st+22t^2)\cr&+&13s(s^2+2st+2t^2)\big]\big\}/
    \big\{5\cos^4\theta_w\sin^4\theta_w\,m_Q^4(s-4m_Q^2)^2\big[(s-m_Z^2)^2+m_Z^2\Gamma_Z^2\big]\big\}
    \eea
    
	$\chi_{Q0}$ production,
    \bea
    F({}^3P_0)_{\gamma}&=&\frac{2048\pi^3\alpha^3 e_Q^4 (s-12m_Q^2)^2\big[16m_Q^4-8m_Q^2(s+t)+s^2+2st+2t^2\big]}
    {m_Q^2 s (s-4m_Q^2)^4}\\
    G({}^3P_0)_{\gamma}&=&\frac{1024\pi^3\alpha^3 e_Q^4 \big[10560m_Q^6-2224m_Q^4 s+268m_Q^2 s^2-13s^3\big]
    \big(16m_Q^4-8m_Q^2(s+t)+s^2+2st+2t^2\big)}
    {5\,m_Q^4 s (s-4m_Q^2)^5}\\
    F({}^3P_0)_{Z}&=&\big\{8\pi^3\alpha^3 e_Q^2 s\big[(1-4\sin^2\theta_w)^2+1\big](1-4|e_Q|\sin^2\theta_w)^2 (s-12m_Q^2)^2
    \big[16m_Q^4-8m_Q^2(s+t)\cr&+&(s^2+2st+2t^2)\big]\big\}/
    \big\{\cos^4\theta_w\sin^4\theta_w\,m_Q^2 (s-4m_Q^2)^4\big[(s-m_Z^2)^2+m_Z^2\Gamma_Z^2\big]\big\}\\
    G({}^3P_0)_{Z}&=&\big\{4\pi^3\alpha^3 e_Q^2 s\big[(1-4\sin^2\theta_w)^2+1\big](1-4|e_Q|\sin^2\theta_w)^2
    \big(10560m_Q^6-2224m_Q^4 s+268m_Q^2 s^2-13s^3\big)\cr&&
    \big[16m_Q^4-8m_Q^2(s+t)+s^2+2st+2t^2\big]\big\}/
    \big\{5\cos^4\theta_w\sin^4\theta_w\,m_Q^4 (s-4m_Q^2)^5\big[(s-m_Z^2)^2+m_Z^2\Gamma_Z^2\big]\big\}
    \eea
    
	$\chi_{Q1}$ production,
    \bea
    F({}^3P_1)_{\gamma}&=&\frac{12288\pi^3\alpha^3 e_Q^4 \big(16m_Q^4(s+4t)-8m_Q^2(s^2+3st+2t^2)+s(s^2+2st+2t^2)\big)}
    {m_Q^2(s-4m_Q^2)^4}\\
    G({}^3P_1)_{\gamma}&=&\big\{6144\pi^3\alpha^3 e_Q^4 \big[64m_Q^6(31s+84t)-16m_Q^4(73s^2+150st+84t^2)+4m_Q^2s(53s^2+88st+66t^2)\cr&-&11s^2(s^2+2st+2t^2)\big]\big\}/
    \big\{5\,m_Q^4(s-4m_Q^2)^5\big\}\\
    F({}^3P_1)_{Z}&=&\big\{48\pi^3\alpha^3 e_Q^2 s^2 \big[(1-4\sin^2\theta_w)^2+1\big](1-4|e_Q|\sin^2\theta_w)^2
    \big[16m_Q^4(s+4t)-8m_Q^2(s^2+3st+2t^2)\cr&+&s(s^2+2st+2t^2)\big]\big\}/
    \big\{\cos^4\theta_w\sin^4\theta_w\,m_Q^2(s-4m_Q^2)^4\big[(s-m_Z^2)^2+m_Z^2\Gamma_Z^2\big]\big\}\\
    G({}^3P_1)_{Z}&=&\big\{24\pi^3\alpha^3 e_Q^2 s^2 \big[(1-4\sin^2\theta_w)^2+1\big](1-4|e_Q|\sin^2\theta_w)^2
    \big[64m_Q^6(31s+84t)-16m_Q^4(73s^2+150st+84t^2)\cr&+&4m_Q^2s(53s^2+88st+66t^2)-11s^2(s^2+2st+2t^2)\big]\big\}\cr&/&
    \big\{5\cos^4\theta_w\sin^4\theta_w\,m_Q^4(s-4m_Q^2)^5\big[(s-m_Z^2)^2+m_Z^2\Gamma_Z^2\big]\big\}
    \eea
    
	$\chi_{Q2}$ production,
    \bea
    F({}^3P_2)_{\gamma}&=&\big\{4096\pi^3\alpha^3 e_Q^4 \big[1536m_Q^8-768m_Q^6(s+t)+16m_Q^4(7s^2+24st+12t^2)-8m_Q^2s(s^2+7st+6t^2)\cr&+&s^2(s^2+2st+2t^2)\big]\big\}/
    \big\{m_Q^2 s (s-4m_Q^2)^4\big\}\\
    G({}^3P_2)_{\gamma}&=&\big\{2048\pi^3\alpha^3 e_Q^4 \big[92160m_Q^{10}-512m_Q^8(71s+90t)+64m_Q^6(33s^2+404st+180t^2)-112m_Q^4s(s^2+46st+32t^2)\cr&+&4m_Q^2s^2(33s^2+112st+98t^2)-7s^3(s^2+2st+2t^2)\big]\big\}/
    \big[5\,m_Q^4 s (s-4m_Q^2)^5\big]\\
    F({}^3P_2)_{Z}&=&\big\{16\pi^3\alpha^3 e_Q^2 s \big[(1-4\sin^2\theta_w)^2+1\big](1-4|e_Q|\sin^2\theta_w)^2
    \big[1536m_Q^8-768m_Q^6(s+t)\cr&+&16m_Q^4(7s^2+24st+12t^2)-8m_Q^2s(s^2+7st+6t^2)+s^2(s^2+2st+2t^2)\big]\big\}\cr&/&
    \big\{\cos^4\theta_w\sin^4\theta_w\,m_Q^2 (s-4m_Q^2)^4\big[(s-m_Z^2)^2+m_Z^2\Gamma_Z^2\big]\big\}\\
    G({}^3P_2)_{Z}&=&\big\{8\pi^3\alpha^3 e_Q^2 s \big[(1-4\sin^2\theta_w)^2+1\big](1-4|e_Q|\sin^2\theta_w)^2
    \big[92160m_Q^{10}-512m_Q^8(71s+90t)\cr&+&64m_Q^6(33s^2+404st+180t^2)-112m_Q^4s(s^2+46st+32t^2)+4m_Q^2s^2(33s^2+112st+98t^2)\cr&-&7s^3(s^2+2st+2t^2)\big]\big\}/
    \big\{5\cos^4\theta_w\sin^4\theta_w\,m_Q^4 (s-4m_Q^2)^5\big[(s-m_Z^2)^2+m_Z^2\Gamma_Z^2\big]\big\}
    \eea
    
	$\eta_{Q2}$ production,
    \bea
    F({}^1D_2)_{\gamma}&=&\frac{4096\pi^3\alpha^3 e_Q^4 \big[16m_Q^4-8m_Q^2(s+t)+s^2+2st+2t^2\big]}
    {m_Q^4 s (s-4m_Q^2)^2}\\
    G({}^1D_2)_{\gamma}&=&-\frac{51200\pi^3\alpha^3 e_Q^4 \big[16m_Q^4-8m_Q^2(s+t)+s^2+2st+2t^2\big]}
    {7\,m_Q^6 s (s-4m_Q^2)^2}\\
    F({}^1D_2)_{Z}&=&\frac{16\pi^3\alpha^3 e_Q^2 s \big[(1-4\sin^2\theta_w)^2+1\big](1-4|e_Q|\sin^2\theta_w)^2
    \big[16m_Q^4-8m_Q^2(s+t)+s^2+2st+2t^2\big]}
    {\cos^4\theta_w\sin^4\theta_w\,m_Q^4 (s-4m_Q^2)^2\big((s-m_Z^2)^2+m_Z^2\Gamma_Z^2\big)}\\
    G({}^1D_2)_{Z}&=&-\frac{200\pi^3\alpha^3 e_Q^2 s \big[(1-4\sin^2\theta_w)^2+1\big](1-4|e_Q|\sin^2\theta_w)^2
    \big[16m_Q^4-8m_Q^2(s+t)+s^2+2st+2t^2\big]}
    {7\cos^4\theta_w\sin^4\theta_w\,m_Q^6 (s-4m_Q^2)^2\big((s-m_Z^2)^2+m_Z^2\Gamma_Z^2\big)}
    \eea	
    
	$\psi_1/\Upsilon_{1}$ production,
   \bea
    F({}^3D_1)_{Z}&=&\big\{2\pi^3\alpha^3 e_Q^2 \big[(1-4\sin^2\theta_w)^2+1\big]
    \big[2304m_Q^8(9s+4t)-384m_Q^6(24s^2+5st+6t^2)\cr&+&32m_Q^4s(23s^2+77st-3t^2)+8m_Q^2s^2(8s^2-81st-80t^2)+s^3(s^2+2st+2t^2)\big]\big\}\cr&/&
    \big\{5\cos^4\theta_w\sin^4\theta_w\,m_Q^4(s-4m_Q^2)^4\big[(s-m_Z^2)^2+m_Z^2\Gamma_Z^2\big]\big\}\\
    G({}^3D_1)_{Z}&=&\big\{4\pi^3\alpha^3 e_Q^2 \big[(1-4\sin^2\theta_w)^2+1\big]
    \big[1536m_Q^{10}(423s+124t)-256m_Q^8(1176s^2+91st+186t^2)\cr&+&64m_Q^6s(424s^2+1613st-95t^2)+16m_Q^4s^2(139s^2-2303st-1780t^2)+m_Q^2(-74s^5+2404s^4t+2380s^3t^2)\cr&-&3s^4(s^2+2st+2t^2)\big]\big\}/
    \big\{35\cos^4\theta_w\sin^4\theta_w\,m_Q^6(s-4m_Q^2)^5\big[(s-m_Z^2)^2+m_Z^2\Gamma_Z^2\big]\big\}
    \eea
	
   the $1^3D_1$--$2^3S_1$ interference term,
    \bea
    F(1^3D_1,2^3S_1)_{Z}&=&\big\{16\pi^3\alpha^3 \big[(1-4\sin^2\theta_w)^2+1\big]
    \big[192m_c^6(3s+4t)-16m_c^4(17s^2+2st+12t^2)+4m_c^2s(7s^2-12st-10t^2)\cr&+&s^2(s^2+2st+2t^2)\big]\big\}/
    \big\{3\sqrt{15}\cos^4\theta_w\sin^4\theta_w\,m_c^2(4m_c^2-s)^3\big[(s-m_Z^2)^2+m_Z^2\Gamma_Z^2\big]\big\}\\
    G(1^3D_1,2^3S_1)_{Z}&=&-\big\{4\pi^3\alpha^3 \big[(1-4\sin^2\theta_w)^2+1\big]
    \big[768m_c^8(471s+332t)-384m_c^6(466s^2-151st+166t^2)\cr&+&32m_c^4s(655s^2-1043st-951t^2)+8m_c^2s^2(58s^2+135st+92t^2)-43s^3(s^2+2st+2t^2)\big]\big\}\cr&/&
    \big\{63\sqrt{15}\cos^4\theta_w\sin^4\theta_w\,m_c^4(s-4m_c^2)^4\big[(s-m_Z^2)^2+m_Z^2\Gamma_Z^2\big]\big\}
    \eea	
    
	$\psi_2/\Upsilon_{2}$ production,
   \bea
    F({}^3D_2)_{Z}&=&\big\{8\pi^3\alpha^3 e_Q^2 \big[(1-4\sin^2\theta_w)^2+1\big]
    \big[512m_Q^8(9s+2t)-256m_Q^6(3s+t)^2+16m_Q^4s(21s^2+36st+20t^2)\cr&-&8m_Q^2s^2(3s^2+11st+8t^2)+3s^3(s^2+2st+2t^2)\big]\big\}\cr&/&
    \big\{\cos^4\theta_w\sin^4\theta_w\,m_Q^4(s-4m_Q^2)^4\big[(s-m_Z^2)^2+m_Z^2\Gamma_Z^2\big]\big\}\\
    G({}^3D_2)_{Z}&=&-\big\{4\pi^3\alpha^3 e_Q^2 \big[(1-4\sin^2\theta_w)^2+1\big]
    \big[10240m_Q^{10}(51s+10t)-512m_Q^8(513s^2+274st+50t^2)\cr&+&64m_Q^6s(675s^2+1068st+448t^2)-16m_Q^4s^2(381s^2+1198st+620t^2)+4m_Q^2s^3(291s^2+716st+578t^2)\cr&-&69s^4(s^2+2st+2t^2)\big]\big\}/
    \big\{7\cos^4\theta_w\sin^4\theta_w\,m_Q^6(4m_Q^2-s)^5\big[(s-m_Z^2)^2+m_Z^2\Gamma_Z^2\big]\big\}
    \eea	
	
    $\psi_3/\Upsilon_{3}$ production,
    \bea
    F({}^3D_3)_{Z}&=&\frac{16\pi^3\alpha^3 e_Q^2 \big[(1-4\sin^2\theta_w)^2+1\big]
    \big[16m_Q^4(3s+8t)-8m_Q^2(3s^2+7st+4t^2)+3s(s^2+2st+2t^2)\big]}
    {5\cos^4\theta_w\sin^4\theta_w\,m_Q^4(s-4m_Q^2)^2\big[(s-m_Z^2)^2+m_Z^2\Gamma_Z^2\big]}\\
    G({}^3D_3)_{Z}&=&-\big\{8\pi^3\alpha^3 e_Q^2 \big[(1-4\sin^2\theta_w)^2+1\big]
    \big[64m_Q^6(99s+152t)-16m_Q^4(249s^2+438st+152t^2)\cr&+&4m_Q^2s(201s^2+388st+286t^2)-51s^2(s^2+2st+2t^2)\big]\big\}\cr&/&
    \big\{35\cos^4\theta_w\sin^4\theta_w\,m_Q^6(4m_Q^2-s)^3\big[(s-m_Z^2)^2+m_Z^2\Gamma_Z^2\big]\big\}
    \eea

	For t-channel production, we consider	
		
    $\psi/\Upsilon$ production,
    \bea
    F({}^3S_1)_{\gamma}&=&\frac{768\pi^3\alpha^3 e_Q^2 \big[32m_Q^4 -8m_Q^2(t+u)+t^2+u^2\big]}
    {m_Q^2\, t\, u}\label{eqe1}\\
    G({}^3S_1)_{\gamma}&=&-\frac{128\pi^3\alpha^3 e_Q^2 \big[1536m_Q^6 -224m_Q^4(t+u)-40m_Q^2(t+u)^2+11(t^3+t^2u+tu^2+u^3)\big]}
    {m_Q^4\, t\, u\, (t+u)}
    \eea
	
	$\psi_1/\Upsilon_{1}$ production,
   \bea
    F({}^3D_1)_{\gamma}&=&\frac{320\pi^3\alpha^3 e_Q^2 \big[32m_Q^4 -8m_Q^2(t+u)+t^2+u^2\big]}
    {m_Q^6\, t\, u}\\
    G({}^3D_1)_{\gamma}&=&-\frac{320\pi^3\alpha^3 e_Q^2 \big[256m_Q^6 -16m_Q^2(t+u)^2+3(t^3+t^2u+tu^2+u^3)\big]}
    {m_Q^8\, t\, u\, (t+u)}
    \eea

    the $1^3D_1$--$2^3S_1$ interference term,
    \bea
    F(1^3D_1,2^3S_1)_{\gamma}&=&-\frac{512\sqrt{15}\pi^3\alpha^3 \big[32m_c^4 -8m_c^2(t+u)+t^2+u^2\big]}
    {9\,m_c^4\, t\, u}\\
    G(1^3D_1,2^3S_1)_{\gamma}&=&\frac{128\sqrt{15}\pi^3\alpha^3 \big[3072m_c^6 -224m_c^4(t+u)-136m_c^2(t+u)^2+29(t^3+t^2u+tu^2+u^3)\big]}
    {27\,m_c^6\, t\, u\, (t+u)}
    \label{eqe}
    \eea

	\hspace{2cm}


\begin{thebibliography}{10}

\bibitem{Bodwin:1994jh}
Geoffrey~T. Bodwin, Eric Braaten, and G.~Peter Lepage.
\newblock {Rigorous QCD analysis of inclusive annihilation and production of
  heavy quarkonium}.
\newblock {\em Phys. Rev. D}, 51:1125--1171, 1995.
\newblock [Erratum: Phys.Rev.D 55, 5853 (1997)].

\bibitem{Belle:2002tfa}
Kazuo Abe et~al.
\newblock {Observation of double c anti-c production in e+ e- annihilation at
  s**(1/2) approximately 10.6-GeV}.
\newblock {\em Phys. Rev. Lett.}, 89:142001, 2002.

\bibitem{Belle:2004abn}
Kazuo Abe et~al.
\newblock {Study of double charmonium production in e+ e- annihilation at
  s**(1/2) {\textasciitilde} 10.6-GeV}.
\newblock {\em Phys. Rev. D}, 70:071102, 2004.

\bibitem{BaBar:2005nic}
Bernard Aubert et~al.
\newblock {Measurement of double charmonium production in $e^+e^-$
  annihilations at $\sqrt{s}=10.6$ GeV}.
\newblock {\em Phys. Rev. D}, 72:031101, 2005.

\bibitem{Zhang:2005cha}
Yu-Jie Zhang, Ying-jia Gao, and Kuang-Ta Chao.
\newblock {Next-to-leading order QCD correction to e+ e- ---{\ensuremath{>}} J
  / psi + eta(c) at s**(1/2) = 10.6-GeV}.
\newblock {\em Phys. Rev. Lett.}, 96:092001, 2006.

\bibitem{Gong:2007db}
Bin Gong and Jian-Xiong Wang.
\newblock {QCD corrections to $J/\psi$ plus $\eta_c$ production in $e^{+}
  e^{-}$ annihilation at $S^{(1/2)}$ = 10.6-GeV}.
\newblock {\em Phys. Rev. D}, 77:054028, 2008.

\bibitem{He:2007te}
Zhi-Guo He, Ying Fan, and Kuang-Ta Chao.
\newblock {Relativistic corrections to J/psi exclusive and inclusive double
  charm production at B factories}.
\newblock {\em Phys. Rev. D}, 75:074011, 2007.

\bibitem{Bodwin:2007ga}
Geoffrey~T. Bodwin, Jungil Lee, and Chaehyun Yu.
\newblock {Resummation of Relativistic Corrections to e+ e- ---{\ensuremath{>}}
  J/psi + eta(c)}.
\newblock {\em Phys. Rev. D}, 77:094018, 2008.

\bibitem{Trunin:2024ibx}
Anton Trunin.
\newblock {Resummation of relativistic corrections to heavy quarkonium
  $+\,\gamma$ production at $Z$ factory}.
\newblock 6 2024.

\bibitem{Wang:2025sbx}
Xiao-Peng Wang, Guang-Zhi Xu, and Kui-Yong Liu.
\newblock {Double heavy quarkonia production with color-octet channels at a Z
  factory and at the CEPC and FCC-ee}.
\newblock {\em Phys. Rev. D}, 111(11):116007, 2025.

\bibitem{Wang:2026gso}
Xiao-Peng Wang, Yi-Jie Li, Guang-Zhi Xu, and Kui-Yong Liu.
\newblock {Associated production of charmonia and bottomonia with color-octet
  channels at the Z factory, CEPC, and FCC-ee}.
\newblock {\em Phys. Rev. D}, 114(1):016016, 2026.

\bibitem{Chen:2013mjb}
Gu~Chen, Xing-Gang Wu, Zhan Sun, Sheng-Quan Wang, and Jian-Ming Shen.
\newblock {Exclusive charmonium production from $e^+ e^-$ annihilation round
  the $Z^0$ peak}.
\newblock {\em Phys. Rev. D}, 88:074021, 2013.

\bibitem{Sun:2014kva}
Zhan Sun, Xing-Gang Wu, Gu~Chen, Yang Ma, Hong-Hao Ma, and Huan-Yu Bi.
\newblock {Bottomonium production associated with a photon at a high luminosity
  $e^+e^-$ collider with the one-loop QCD correction}.
\newblock {\em Phys. Rev. D}, 89(7):074035, 2014.

\bibitem{Liao:2021ifc}
Qi-Li Liao, Jun Jiang, Peng-Cheng Lu, and Gu~Chen.
\newblock {Production of excited heavy quarkonia in
  e+e-{\textrightarrow}{\ensuremath{\gamma}}*/Z0{\textrightarrow}|(QQ{\textasciimacron})[n]{\ensuremath{\rangle}}+{\ensuremath{\gamma}}
  at super Z factory}.
\newblock {\em Phys. Rev. D}, 105(1):016026, 2022.

\bibitem{BESIII:2014uzr}
M.~Ablikim et~al.
\newblock {Evidence for $e^+e^-\to\gamma\chi_{c1, 2}$ at center-of-mass
  energies from 4.009 to 4.360 GeV}.
\newblock {\em Chin. Phys. C}, 39(4):041001, 2015.

\bibitem{BESIII:2017nty}
Medina Ablikim et~al.
\newblock {Evidence for $e^{+}e^{-} \to \gamma\eta_c(1S)$ at center-of-mass
  energies between 4.01 and 4.60 GeV}.
\newblock {\em Phys. Rev. D}, 96(5):051101, 2017.

\bibitem{BESIII:2021yal}
M.~Ablikim et~al.
\newblock {Measurement of $e^+e^-\to\gamma \chi_{c0,c1,c2}$ cross sections at
  center-of-mass energies between 3.77 and 4.60~GeV}.
\newblock {\em Phys. Rev. D}, 104(9):092001, 2021.

\bibitem{BaBar:2003waz}
Bernard Aubert et~al.
\newblock {$J/\psi$ production via initial state radiation in $e^+e^- \to \mu^+
  \mu^- \gamma$ at an $e^+e^-$ center-of-mass energy near 10.6-GeV}.
\newblock {\em Phys. Rev. D}, 69:011103, 2004.

\bibitem{Belle:2015hcs}
Y.~L. Han et~al.
\newblock {Measurement of $e^+e^- \to \gamma\chi_{cJ}$ via initial state
  radiation at Belle}.
\newblock {\em Phys. Rev. D}, 92(1):012011, 2015.

\bibitem{Belle:2018jqa}
S.~Jia et~al.
\newblock {Observation of $e^+e^- \to \gamma \chi_{c1}$ and search for $e^+e^-
  \to \gamma \chi_{c0}, \gamma \chi_{c2},$ and $\gamma\eta_c$ at $\sqrt{s}$
  near 10.6 GeV at Belle}.
\newblock {\em Phys. Rev. D}, 98(9):092015, 2018.

\bibitem{Chung:2008km}
Hee~Sok Chung, Jungil Lee, and Chaehyun Yu.
\newblock {Exclusive heavy quarkonium + gamma production from e+ e-
  annihilation into a virtual photon}.
\newblock {\em Phys. Rev. D}, 78:074022, 2008.

\bibitem{Li:2009ki}
Dan Li, Zhi-Guo He, and Kuang-Ta Chao.
\newblock {Search for C= charmonium and bottomonium states in e+ e-
  ---{\ensuremath{>}} gamma + X at B factories}.
\newblock {\em Phys. Rev. D}, 80:114014, 2009.

\bibitem{Sang:2009jc}
Wen-Long Sang and Yu-Qi Chen.
\newblock {Higher Order Corrections to the Cross Section of e+e-
  ---{\ensuremath{>}} Quarkonium + gamma}.
\newblock {\em Phys. Rev. D}, 81:034028, 2010.

\bibitem{Chao:2013cca}
Kuang-Ta Chao, Zhi-Guo He, Dan Li, and Ce~Meng.
\newblock {Search for $C=+$ charmonium states in $e^+e^-\to \gamma+~X$ at
  BEPCII/BESIII}.
\newblock 10 2013.

\bibitem{Chen:2017pyi}
Long-Bin Chen, Yi~Liang, and Cong-Feng Qiao.
\newblock {NNLO QCD corrections to $\gamma + \eta_c(\eta_b)$ exclusive
  production in electron-positron collision}.
\newblock {\em JHEP}, 01:091, 2018.

\bibitem{Yu:2020tri}
Huai-Min Yu, Wen-Long Sang, Xu-Dong Huang, Jun Zeng, Xing-Gang Wu, and
  Stanley~J. Brodsky.
\newblock {Scale-fixed predictions for $\gamma + \eta_c$ production in
  electron-positron collisions at NNLO in perturbative QCD}.
\newblock {\em JHEP}, 01:131, 2021.

\bibitem{Sang:2020fql}
Wen-Long Sang, Feng Feng, and Yu~Jia.
\newblock {Next-to-next-to-leading-order radiative corrections
  to$e^+e^-\to\chi_{cJ}+\gamma$ at B factory}.
\newblock {\em JHEP}, 10:098, 2020.

\bibitem{Li:2025pbt}
Cong Li, Wen-Long Sang, and Hong-Fei Zhang.
\newblock {The next-to-next-to-leading-order QCD corrections to
  e$^{+}$e$^{−}${\textrightarrow}
  {\ensuremath{\eta}}$_{c}$/{\ensuremath{\chi}}$_{cJ}$ + {\ensuremath{\gamma}}
  at B factories}.
\newblock {\em JHEP}, 05:163, 2026.

\bibitem{Fan:2012dy}
Ying Fan, Jungil Lee, and Chaehyun Yu.
\newblock {Resummation of relativistic corrections to exclusive productions of
  charmonia in $e^+e^-$ collisions}.
\newblock {\em Phys. Rev. D}, 87(9):094032, 2013.

\bibitem{Li:2013nna}
Yi-Jie Li, Guang-Zhi Xu, Kui-Yong Liu, and Yu-Jie Zhang.
\newblock {Search for $C=+$ charmonium and XYZ states in $e^+e^-\to \gamma+ H$
  at BESIII}.
\newblock {\em JHEP}, 01:022, 2014.

\bibitem{Xu:2014zra}
Guang-Zhi Xu, Yi-Jie Li, Kui-Yong Liu, and Yu-Jie Zhang.
\newblock {$\alpha_sv^2$ corrections to $\eta_c$ and $\chi_{cJ}$ production
  recoiled with a photon at $e^+e^-$ colliders}.
\newblock {\em JHEP}, 10:071, 2014.

\bibitem{Brambilla:2017kgw}
Nora Brambilla, Wen Chen, Yu~Jia, Vladyslav Shtabovenko, and Antonio Vairo.
\newblock {Relativistic corrections to exclusive $\chi_{cJ} + \gamma$
  production from $e^+ e^-$ annihilation}.
\newblock {\em Phys. Rev. D}, 97(9):096001, 2018.
\newblock [Erratum: Phys.Rev.D 101, 039903 (2020)].

\bibitem{Benayoun:1999hm}
Maurice Benayoun, S.~I. Eidelman, V.~N. Ivanchenko, and Z.~K. Silagadze.
\newblock {Spectroscopy at B factories using hard photon emission}.
\newblock {\em Mod. Phys. Lett. A}, 14:2605--2614, 1999.

\bibitem{Shifman:1980dk}
Mikhail~A. Shifman and Michael~I. Vysotsky.
\newblock {FORM-FACTORS OF HEAVY MESONS IN QCD}.
\newblock {\em Nucl. Phys. B}, 186:475--518, 1981.

\bibitem{Jia:2008ep}
Yu~Jia and Deshan Yang.
\newblock {Refactorizing NRQCD short-distance coefficients in exclusive
  quarkonium production}.
\newblock {\em Nucl. Phys. B}, 814:217--230, 2009.

\bibitem{Braguta:2010mf}
V.~V. Braguta.
\newblock {Exclusive C=+ charmonium production in $e^+ e^- \to H+\gamma$ at
  B-factories within light cone formalism}.
\newblock {\em Phys. Rev. D}, 82:074009, 2010.

\bibitem{Bhatnagar:2022wgg}
Shashank Bhatnagar and Vaishali Guleria.
\newblock {({\ensuremath{\gamma}}, {\ensuremath{\chi}}cJ(J = 0,1)) and
  ({\ensuremath{\gamma}}, {\ensuremath{\eta}}c) production in electron-positron
  annihilation at s=10.6 GeV and 4.6 GeV in the framework of Bethe-Salpeter
  equation}.
\newblock {\em Nucl. Phys. A}, 1041:122783, 2024.

\bibitem{Guleria:2024rac}
Vaishali Guleria and Shashank Bhatnagar.
\newblock {$\chi_{c1}/\eta_c$ and photon production in electron-positron
  annihilation at B-Factories}.
\newblock {\em DAE Symp. Nucl. Phys.}, 67:969--970, 2024.

\bibitem{Bhatnagar:2024ghr}
Shashank Bhatnagar and Vaishali Guleria.
\newblock {$\gamma$ and $\chi _{c0}$ production in electron-positron
  annihilation}.
\newblock {\em DAE Symp. Nucl. Phys.}, 67:953--954, 2024.

\bibitem{Bhatnagar:2025keo}
Shashank Bhatnagar.
\newblock {{\ensuremath{\gamma}} and J/{\ensuremath{\psi}} production through
  the initial state radiation process in electron-positron annihilation at B
  factories}.
\newblock {\em Phys. Rev. D}, 112(5):054011, 2025.

\bibitem{CEPCStudyGroup:2018ghi}
Mingyi Dong et~al.
\newblock {CEPC Conceptual Design Report: Volume 2 - Physics {\&} Detector}.
\newblock 11 2018.

\bibitem{Agapov:2022bhm}
I.~Agapov et~al.
\newblock {Future Circular Lepton Collider FCC-ee: Overview and Status}.
\newblock In {\em {Snowmass 2021}}, 3 2022.

\bibitem{ILC:2013jhg}
{The International Linear Collider Technical Design Report - Volume 2:
  Physics}.
\newblock 6 2013.

\bibitem{Ma.JP}
J.P. Ma and Z.X. Zhang.
\newblock {Z-Factory Physics. Preface}.
\newblock {\em Sci. China: Phys. Mech. Astro}, 53:1947, 2010.

\bibitem{Chang:2010am}
Chao-Hsi Chang, Jian-Xiong Wang, and Xing-Gang Wu.
\newblock {Production of a Heavy Quarkonium with a Photon or via ISR at $Z$
  Peak in $e^+e^-$ Collider}.
\newblock {\em Sci. China Phys. Mech. Astron.}, 53:2031--2036, 2010.

\bibitem{Chen:2013itc}
Gu~Chen, Xing-Gang Wu, Zhan Sun, Xu-Chang Zheng, and Jian-Ming Shen.
\newblock {Next-to-leading order QCD corrections for the charmonium production
  via the channel $e^+ e^- \to H(|c\bar{c}\rangle) + \gamma$ round the $Z^0$
  peak}.
\newblock {\em Phys. Rev. D}, 89(1):014006, 2014.

\bibitem{Chung:2019ota}
Hee~Sok Chung, June-Haak Ee, Daekyoung Kang, U-Rae Kim, Jungil Lee, and
  Xiang-Peng Wang.
\newblock {Pseudoscalar Quarkonium+gamma Production at NLL+NLO accuracy}.
\newblock {\em JHEP}, 10:162, 2019.

\bibitem{Guo:2017jvc}
Feng-Kun Guo, Christoph Hanhart, Ulf-G. Mei{\ss}ner, Qian Wang, Qiang Zhao, and
  Bing-Song Zou.
\newblock {Hadronic molecules}.
\newblock {\em Rev. Mod. Phys.}, 90(1):015004, 2018.
\newblock [Erratum: Rev.Mod.Phys. 94, 029901 (2022)].

\bibitem{Olsen:2017bmm}
Stephen~Lars Olsen, Tomasz Skwarnicki, and Daria Zieminska.
\newblock {Nonstandard heavy mesons and baryons: Experimental evidence}.
\newblock {\em Rev. Mod. Phys.}, 90(1):015003, 2018.

\bibitem{Brambilla:2019esw}
Nora Brambilla, Simon Eidelman, Christoph Hanhart, Alexey Nefediev, Cheng-Ping
  Shen, Christopher~E. Thomas, Antonio Vairo, and Chang-Zheng Yuan.
\newblock {The $XYZ$ states: experimental and theoretical status and
  perspectives}.
\newblock {\em Phys. Rept.}, 873:1--154, 2020.

\bibitem{Chen:2022asf}
Hua-Xing Chen, Wei Chen, Xiang Liu, Yan-Rui Liu, and Shi-Lin Zhu.
\newblock {An updated review of the new hadron states}.
\newblock {\em Rept. Prog. Phys.}, 86(2):026201, 2023.

\bibitem{Godfrey:1985xj}
S.~Godfrey and Nathan Isgur.
\newblock {Mesons in a Relativized Quark Model with Chromodynamics}.
\newblock {\em Phys. Rev. D}, 32:189--231, 1985.

\bibitem{Barnes:2005pb}
T.~Barnes, S.~Godfrey, and E.~S. Swanson.
\newblock {Higher charmonia}.
\newblock {\em Phys. Rev. D}, 72:054026, 2005.

\bibitem{Rapidis:1977cv}
Petros~A. Rapidis et~al.
\newblock {Observation of a Resonance in e+ e- Annihilation Just Above Charm
  Threshold}.
\newblock {\em Phys. Rev. Lett.}, 39:526, 1977.
\newblock [Erratum: Phys.Rev.Lett. 39, 974 (1977)].

\bibitem{Belle:2013ewt}
V.~Bhardwaj et~al.
\newblock {Evidence of a new narrow resonance decaying to $\chi_{c1}\gamma$ in
  $B \to \chi_{c1} \gamma K$}.
\newblock {\em Phys. Rev. Lett.}, 111(3):032001, 2013.

\bibitem{BESIII:2015iqd}
M.~Ablikim et~al.
\newblock {Observation of the $\psi(1^3D_2)$ state in
  $e^+e^-\to\pi^+\pi^-\gamma\chi_{c1}$ at BESIII}.
\newblock {\em Phys. Rev. Lett.}, 115(1):011803, 2015.

\bibitem{Deng:2016stx}
Wei-Jun Deng, Hui Liu, Long-Cheng Gui, and Xian-Hui Zhong.
\newblock {Charmonium spectrum and their electromagnetic transitions with
  higher multipole contributions}.
\newblock {\em Phys. Rev. D}, 95(3):034026, 2017.

\bibitem{Li:2022qhg}
Wei Li, Su-Yan Pei, Tianhong Wang, Ying-Long Wang, Tai-Fu Feng, and Guo-Li
  Wang.
\newblock {Electromagnetic decays of X(3823) as the {\ensuremath{\psi}}2(1D32)
  state and its radial excited states}.
\newblock {\em Phys. Rev. D}, 107(11):113002, 2023.

\bibitem{Li:2023zig}
Ning Li, Yan Gao, Feiyu Chen, Ying Chen, Xiangyu Jiang, Chunjiang Shi, and Wei
  Sun.
\newblock {Radiative transition decay width of
  {\ensuremath{\psi}}2(3823){\textrightarrow}{\ensuremath{\gamma}}{\ensuremath{\chi}}c1
  from lattice QCD}.
\newblock {\em Phys. Rev. D}, 109(1):014513, 2024.

\bibitem{LHCb:2019lnr}
Roel Aaij et~al.
\newblock {Near-threshold $D\bar{D}$ spectroscopy and observation of a new
  charmonium state}.
\newblock {\em JHEP}, 07:035, 2019.

\bibitem{Yu:2019kcr}
Guo~Liang Yu and Zhi~Gang Wang.
\newblock {Analysis of the $X(3842)$ as a D-wave charmonium meson}.
\newblock {\em Int. J. Mod. Phys. A}, 34(26):1950151, 2019.

\bibitem{Li:2023cpl}
Wei Li, Su-Yan Pei, Tianhong Wang, Tai-Fu Feng, and Guo-Li Wang.
\newblock {Decay property of the X(3842) as the {\ensuremath{\psi}}3(1D33)
  state}.
\newblock {\em Phys. Rev. D}, 109(3):036011, 2024.

\bibitem{Belle:2020esr}
K.~Chilikin et~al.
\newblock {First search for the $\eta_{c2}(1D)$ in $B$ decays at Belle}.
\newblock {\em JHEP}, 05:034, 2020.

\bibitem{Belle:2021cjk}
S.~Jia et~al.
\newblock {Search for the $\eta_{c2}(1D)$ in $e^+e^-\to\gamma\eta_{c2}(1D)$ at
  $\sqrt{s}$ near 10.6 GeV at Belle}.
\newblock {\em Phys. Rev. D}, 104:012012, 2021.

\bibitem{CLEO:2004npj}
G.~Bonvicini et~al.
\newblock {First observation of a Upsilon(1D) state}.
\newblock {\em Phys. Rev. D}, 70:032001, 2004.

\bibitem{BaBar:2010tqb}
P.~del Amo~Sanchez et~al.
\newblock {Observation of the $\Upsilon{1^3D_J}$ Bottomonium State through
  Decays to $\pi^+\pi^-\Upsilon{1S}$}.
\newblock {\em Phys. Rev. D}, 82:111102, 2010.

\bibitem{He:2008xb}
Zhi-Guo He, Ying Fan, and Kuang-Ta Chao.
\newblock {QCD prediction for the non-D anti-D annihilation decay of
  psi(3770)}.
\newblock {\em Phys. Rev. Lett.}, 101:112001, 2008.

\bibitem{Fan:2009cj}
Ying Fan, Zhi-Guo He, Yan-Qing Ma, and Kuang-Ta Chao.
\newblock {Predictions of Light Hadronic Decays of Heavy Quarkonium 1D(2)
  States in NRQCD}.
\newblock {\em Phys. Rev. D}, 80:014001, 2009.

\bibitem{Fan:2011aa}
Ying Fan, Jin-Zhao Li, Ce~Meng, and Kuang-Ta Chao.
\newblock {B-meson Semi-inclusive Decay to $2^{-+}$ Charmonium in NRQCD and
  X(3872)}.
\newblock {\em Phys. Rev. D}, 85:034032, 2012.

\bibitem{Chen:2015zta}
Long-Bin Chen, Jun Jiang, and Cong-Feng Qiao.
\newblock {NLO QCD Corrections to $\eta_c+h_c(1P)/\psi_{1,2}(1D)$ Production at
  B-factories}.
\newblock {\em Chin. Phys. C}, 39(10):103101, 2015.

\bibitem{Dong:2013qw}
Hai-Rong Dong, Feng Feng, and Yu~Jia.
\newblock {${\mathcal O}(\alpha_s)$ corrections to $e^+e^-\rightarrow
  J/\psi+\eta_{c2}(\chi_{c1}^\prime)$ at $B$ factories}.
\newblock {\em Phys. Rev. D}, 98(3):034005, 2018.

\bibitem{Yang:2020pyh}
Li~Yang, Wen-Long Sang, Hong-Fei Zhang, Yu-Dong Zhang, and Ming-Zhen Zhou.
\newblock {Study on $\eta_{c2}(\eta_{b2})$ electromagnetic decay into double
  photons}.
\newblock {\em Phys. Rev. D}, 103(3):034018, 2021.

\bibitem{Wang:2013nya}
Tianhong Wang, Guo-Li Wang, Wan-Li Ju, and Yue Jiang.
\newblock {Annihilation rate of $2^{-+}$ charmonium and bottomonium}.
\newblock {\em JHEP}, 03:110, 2013.

\bibitem{Wang:2016hbl}
Tianhong Wang, Hui-Feng Fu, Yue Jiang, Qiang Li, and Guo-Li Wang.
\newblock {Annihilation Rates of $^3D_2(2^{--})$ and $^3D_3(3^{--})$ Heavy
  Quarkonia}.
\newblock {\em Int. J. Mod. Phys. A}, 32(06n07):1750035, 2017.

\bibitem{Li:2016cou}
Qiang Li, Tianhong Wang, Yue Jiang, Han Yuan, and Guo-Li Wang.
\newblock {$D$-wave charmonia $\eta_{c2}$($1{^1D_2}$), $\psi_2$($1{^3D_2}$),
  and $\psi_3$($1{^3D_3}$) in $B_c$ decays}.
\newblock {\em Eur. Phys. J. C}, 76(8):454, 2016.

\bibitem{Du:2024gna}
Xin-Yao Du, Su-Yan Pei, Wei Li, Man Jia, Qiang Li, Tianhong Wang, Bo~Wang, and
  Guo-Li Wang.
\newblock {{\ensuremath{\eta}}c2(D21) and its electromagnetic decays}.
\newblock {\em Phys. Rev. D}, 110(11):113009, 2024.

\bibitem{Ko:1996xw}
Pyungwon Ko, Jungil Lee, and H.~S. Song.
\newblock {Color octet mechanism in $\gamma$ + $p \to J/\psi$ + x}.
\newblock {\em Phys. Rev. D}, 54:4312--4325, 1996.
\newblock [Erratum: Phys.Rev.D 60, 119902 (1999)].

\bibitem{He:2009bf}
Zhi-Guo He, Ying Fan, and Kuang-Ta Chao.
\newblock {NRQCD Predictions of D-Wave Quarkonia D-3(J) (J = 1,2,3) Decay into
  Light Hadrons at Order alpha**3 (S)}.
\newblock {\em Phys. Rev. D}, 81:074032, 2010.

\bibitem{Bodwin:2002cfe}
Geoffrey~T. Bodwin and Andrea Petrelli.
\newblock {Order-$v^4$ corrections to $S$-wave quarkonium decay}.
\newblock {\em Phys. Rev. D}, 66:094011, 2002.
\newblock [Erratum: Phys.Rev.D 87, 039902 (2013)].

\bibitem{Xu:2012am}
Guang-Zhi Xu, Yi-Jie Li, Kui-Yong Liu, and Yu-Jie Zhang.
\newblock {Relativistic Correction to Color Octet J/psi Production at Hadron
  Colliders}.
\newblock {\em Phys. Rev. D}, 86:094017, 2012.

\bibitem{ParticleDataGroup:2024cfk}
S.~Navas et~al.
\newblock {Review of particle physics}.
\newblock {\em Phys. Rev. D}, 110(3):030001, 2024.

\bibitem{A:2023bxv}
Christas~Mony A. and Rohit Dhir.
\newblock {Spectroscopy and annihilation decay widths of charmonium and
  bottomonium excitations}.
\newblock {\em J. Phys. G}, 51(11):115004, 2024.

\bibitem{Sun:2018rgx}
Zhan Sun, Xing-Gang Wu, Yang Ma, and Stanley~J. Brodsky.
\newblock {Exclusive production of $J/\psi+\eta_c$ at the $B$ factories Belle
  and Babar using the principle of maximum conformality}.
\newblock {\em Phys. Rev. D}, 98(9):094001, 2018.

\bibitem{Jiang:2018wmv}
YingZhao Jiang and Zhan Sun.
\newblock {Further studies on the exclusive productions of $J/\psi +\chi _{cJ}$
  ( $J=0,1,2$ ) via $e^+e^-$ annihilation at the $B$ factories}.
\newblock {\em Eur. Phys. J. C}, 78(11):892, 2018.

\bibitem{Xu:2015goa}
Guang-Zhi Xu, Gang Li, Yi-Jie Li, Kui-Yong Liu, and Yu-Jie Zhang.
\newblock {Interference effects on Higgs mass measurement in $e^+e^-\to
  H(\gamma\gamma) Z$ at CEPC}.
\newblock {\em Chin. Phys. C}, 40(3):033101, 2016.

\bibitem{Jia:2026onq}
Meng-Kun Jia, Yi-Jie Li, Guang-Zhi Xu, and Kui-Yong Liu.
\newblock {Resonance-continuum interference with a possible tensor $X(6900)$ in
  radiative double-charmonium production at $\sqrt{s}=10.58~{\rm GeV}$}.
\newblock 8 2026.

\bibitem{Kuang:1989ub}
Yu-Ping Kuang and Tung-Mow Yan.
\newblock {Hadronic Transitions of $D$ Wave Quarkonium and $\psi(3770) \to
  J/\psi \pi \pi$}.
\newblock {\em Phys. Rev. D}, 41:155, 1990.

\bibitem{Ding:1991vu}
Yi-Bing Ding, Dan-Hua Qin, and Kuang-Ta Chao.
\newblock {Electric dipole transitions of psi (3770) and S - D mixing between
  psi (3686) and psi (3770)}.
\newblock {\em Phys. Rev. D}, 44:3562--3566, 1991.

\bibitem{Rosner:2001nm}
Jonathan~L. Rosner.
\newblock {Charmless final states and S D wave mixing in the psi-prime- prime}.
\newblock {\em Phys. Rev. D}, 64:094002, 2001.

\bibitem{Bodwin:2006yd}
Geoffrey~T. Bodwin, Eric Braaten, Jungil Lee, and Chaehyun Yu.
\newblock {Exclusive two-vector-meson production from e+ e- annihilation}.
\newblock {\em Phys. Rev. D}, 74:074014, 2006.

\bibitem{Liao:2014rca}
Qi-Li Liao and Guo-Ya Xie.
\newblock {Heavy quarkonium wave functions at the origin and excited heavy
  quarkonium production via top quark decays at the LHC}.
\newblock {\em Phys. Rev. D}, 90(5):054007, 2014.

\end{thebibliography}

\end{document}